\documentclass[10pt]{article}
\pdfoutput=1

\usepackage{amsthm,amsfonts,amsmath,amssymb}
\usepackage{mathtools}
\usepackage{stackrel}
\usepackage{leftidx}
\usepackage{scalerel}

\usepackage{euscript}
\usepackage{pxfonts}

\usepackage{enumerate}

\usepackage{hyperref}
\usepackage{cleveref}

\usepackage{xcolor}
\usepackage{tcolorbox}
\usepackage{graphicx}
\usepackage{subfigure}

\usepackage[all,2cell,arrow,matrix]{xy} \UseAllTwocells \SilentMatrices
\usepackage{tikz-cd}
\usepackage{tikz}
\usetikzlibrary{arrows,arrows.meta,decorations.markings,decorations.pathreplacing,calc}

\usepackage{verbatim}
\usepackage{comment}

\usepackage{epstopdf} % for pdflatex

\usepackage{xspace}
\tikzset{->-/.style={decoration={markings,mark=at position #1 with {\arrow{stealth}}},postaction={decorate}},->-/.default=0.6}

\usepackage{sectsty}
\sectionfont{\Large}
\usepackage[nottoc]{tocbibind} % References in ToC

\usepackage{appendix}

 \usepackage{sectsty}
 \sectionfont{\large}
 \subsectionfont{\normalsize}

\newcommand\void[1]       {}

\newcommand\be            {\begin{equation}}
\newcommand\ee            {\end{equation}}
\newcommand\bnu           {\begin{enumerate}}
\newcommand\enu           {\end{enumerate}}

\newcommand\bea       {\begin{eqnarray}}
\newcommand\eea       {\end{eqnarray}}
\newcommand\nn         {\nonumber \\}

\newcommand\etb           {&\!\! \displaystyle}

\newcommand\eps           {\varepsilon}
\newcommand\Hom           {\mathrm{Hom}}
\newcommand\id            {{\rm id}}

\newcommand\one           {{\bf1}}

\newcommand\vect        {\mathrm{Vec}}
\newcommand\hilb         {\mathrm{Hilb}}
 \newcommand\rev         {\mathrm{rev}}
 
 \newcommand\rep        {\mathrm{Rep}}
 \newcommand\totimes  {\otimes_{\EuScript{D}}}
 \newcommand\op         {\mathrm{op}}

 \newcommand\tr             {\mathrm{Tr}}
 \newcommand\Mod {\mathrm{Mod}}

\newtheorem{thm}{Theorem}[section]
\newtheorem{pthm}[thm]{Theorem$^\mathrm{ph}$}
\newtheorem{plemma}[thm]{Lemma$^\mathrm{ph}$}

\newtheorem{defn}[thm]{Definition}
\newtheorem{prop}[thm]{Proposition}
\newtheorem{cor}[thm]{Corollary}
\newtheorem{rema}[thm]{Remark}

\newtheorem{expl}[thm]{Example}

\numberwithin{equation}{section}
\numberwithin{thm}{section}

\newcommand\Cb            {\mathbb{C}}

\newcommand\Zb            {\mathbb{Z}}

\newcommand\EA   {\EuScript{A}}
\newcommand\EB   {\EuScript{B}}
\newcommand\EC   {\EuScript{C}}
\newcommand\ED  {\EuScript{D}}
\newcommand\EE  {\EuScript{E}}
\newcommand\EF  {\EuScript{F}}
\newcommand\EH  {\EuScript{H}}
\newcommand\EM  {\EuScript{M}}
\newcommand\EN  {\EuScript{N}}

\newcommand\arXiv[1]{\href{http://arxiv.org/abs/#1}{\nolinkurl{arXiv:#1}}}

\begin{document}

\begin{center} \LARGE
%A mathematical theory of anyon condensation
Anyon condensation and tensor categories
\end{center}

\vskip 1em
\begin{center}
{\large 
Liang Kong$^{\, a,b,}$\footnote{emails: 
{\tt  kong.fan.liang@gmail.com}}}
\\[1em]
$^a$ {\it Institute for Advanced Study (Science Hall) \\ 
Tsinghua University, Beijing 100084, China}
\\[1em]
$^b$ {\it Department of Mathematics and Statistics\\
University of New Hampshire, Durham, NH 03824, USA}
\end{center}

\bigskip
\begin{abstract}
Instead of studying anyon condensation in concrete models, we take an abstract approach. Assume that a system of anyons, which form a modular tensor category $\ED$, is obtained via an anyon condensation from another system of anyons (i.e. another modular tensor category $\EC$). By a bootstrap analysis, we derive the relation between $\EC$ and $\ED$ from natural physical requirements. It turns out that the tensor unit of $\ED$ can be identified with a connected commutative separable algebra $A$ in $\EC$. The modular tensor category $\ED$ consists of all deconfined particles and can be identified with the category of local $A$-modules in $\EC$. If this condensation occurs in a 2d region in the $\EC$-phase, then it also produces a 1d gapped domain wall between the $\EC$-phase and the $\ED$-phase. The confined and deconfined particles accumulate on the wall and form a fusion category that is precisely the category of right $A$-modules in $\EC$. We also consider condensations that are confined to a 1d line. We show how to determine the algebra $A$ from physical macroscopic data. We provide examples of anyon condensation in the toric code model, Kitaev quantum double models and Levin-Wen types of lattice models and in some chiral topological phases. In the end, we briefly discuss Witt equivalence between 2d topological phases. 

We also attach to this paper an Erratum and Addendum to the original version of “Anyon condensation and tensor categories” published in [Nucl. Phys. B 886 (2014) 436-482]. 
\end{abstract}

\tableofcontents

\section{Introduction}

Anyon condensation is an important subject to study in the field of topological orders. In 2002, Bais, Schroers, Slingerland initiated a systematic study of anyon condensations based on the idea of Hopf symmetry breaking \cite{bss1,bss2}. The theory was further developed by Bais and Slingerland in an influential work \cite{bs}, and was followed and further developed by many peoples (see for example \cite{bm-d,bsh,bw,buss,br,levin,bjq} and references therein). In spite of many works in this direction, the fundamental mathematical structure that controls the anyon condensation has not being identified in its full generality. Recently, Kapustin and Saulina \cite{kas}, followed by Levin \cite{levin} and Barkeshli, Jian and Qi \cite{bjq}, successfully identified the notion of Lagrangian subgroup \cite{drgno} with gapped boundaries. But these studies are restricted to the abelian Chern-Simons models and Kitaev quantum double models associated to abelian groups. A general condensation theory is still not available. 

On the other hand, it has been known to physicists for a long time that a system of anyons can be described by a (unitary) modular tensor category (MTC). All necessary mathematical notions are reviewed in Appendix\,\ref{sec:mtc}. A physical introduction of MTC can be found in Appendix E in \cite{kitaev1}. 
It is clear that an anyon condensation should be controled by certain mathematical structures in a MTC. Mathematicians know how to obtain new (unitary) MTC from a given one $\EC$ since the seminal works of B\"{o}ckenhauer, Evans and Kawahigashi \cite{bek1,bek2,bek3} in 1999-2001 and that of Kirillov, Jr. and Ostrik \cite{ko} in 2001 (see also \cite{cor}). They proved that given a connected commutative separable algebra $A$ in $\EC$, the category $\EC_A^{loc}$ of local $A$-modules is a MTC. What this mathematical result suggests to us is obvious: an anyon condensation in a 2d $\EC$-phase is determined by a connected commutative separable algebra $A$ in the MTC $\EC$. In the spring of 2009, Alexei Kitaev told me this connection between anyon condensation and connected commutative separable algebra in a MTC. He also provided a brief physical proof based on many-body wave functions \cite{kitaev2}. This connection was announced independently by Alexei Davydov in an international workshop in Sydney in 2011 \cite{da3}. In that talk, he stated explicitly this connection and examples were also provided there, but he did not provide any explanation why this connection is physically reasonable. This connection was also known to Fuchs, Schweigert and Valentino and was briefly mentioned in Section\,4 in their work \cite{fsv}. From 2009 to 2013, the author had also explained this result to a few groups of people privately and publicly. Anyon condensation was also studied in \cite{bsw} in the framework of Kitaev quantum double models. But the general theory was not given there.

The condensed matter physics community has not fully embraced this connection yet. This delay is partially due to the abstractness of the category theory. But perhaps more important reason is that it is unclear why these mathematical structures are demanded by physics. Through the influence of the works \cite{kas,kk,bsw,fsv}, recently, many physicists start to realize the relevance of some mathematical literatures and expressed the willingness to understand these abstract structures in more physical way. The main goal of this paper is to provide a detailed explanation of how each ingredient of the complete mathematical structures emerge naturally from concrete and natural physical requirements. We hope that this analysis can convince physicists that the tensor-categorical language, although abstract, is a powerful and necessary language for anyon condensation. 

\medskip
An anyon condensation is a complicated physical process, to which the mathematical structure associated is highly non-trivial. This mathematical structure captures the universal and model-independent structures of anyon condensations. Working with examples does not always shed lights on the underlining universal structure because a concrete example usually contains too many accidental structures that are misleading. Ideally, we would like to take no more assumptions than what we absolutely need. This suggests us to take a {\bf bootstrap approach} towards anyon condensation. Namely, instead of studying concrete models directly, we start from an abstract setting in which only two MTC's $\EC$ and $\ED$, i.e. two systems of anyons before and after the condensation, are given (see Figure\,\ref{fig:set-up}). We want to work out all the necessary relations between $\EC$ and $\ED$. We show in details in Section\,\ref{sec:bs-1} how these relations in terms of abstract tensor-categorical structures emerge from natural physical requirements. 
The final result is summarized below (see Appendix for the definitions of various mathematical notions):

\begin{figure}[t] 
$$
\begin{tikzpicture}
\fill[blue!30] (-2,0) rectangle (2,3) ;
\fill[green!30] (0,1.5) circle (1);
%\draw[fill=white] (-0.1,0.9) rectangle (0.1,1.1) node[midway,right] {$\,\phi$} ;
\node at (-1.5,2.5) {$\EC$} ;
\node at (0,1.5) {$\ED$} ;
\node at (-1.1,1) {$\EE$} ;
\draw [blue, ultra thick] (0,1.5) circle [radius=1] ;
\end{tikzpicture}
$$
  \caption{{\small The set-up of bootstrap analysis: We consider an anyon condensation occurring in a 2d topological phase described by a (unitary) MTC $\EC$. The condensed phase is described by another MTC $\ED$, and the gapped excitations on the domain wall form a (unitary) spherical fusion category $\EE$. }}
  \label{fig:set-up}
\end{figure}

\medskip
\noindent {\bf Main results of bootstrap analysis} (Theorem$^\mathrm{ph}$\,\ref{thm:main-1}): If a system of anyons (i.e. an MTC $\ED$) is obtained from another system of anyons (i.e. an MTC $\EC$) via a 2d condensation, and if a 1d gapped domain wall between the $\EC$-phase and the $\ED$-phase is produced by this 2d condensation with the wall excitations given by a (unitary) spherical fusion category $\EE$ (see Figure\,\ref{fig:set-up}), then we must have
\bnu
\item The vacuum of $\ED$-phase can be identified with a connected commutative symmetric normalized-special Frobenius algebra $A$ in $\EC$. Moreover, $\ED$ consists of all deconfined particles and can be identified with the category $\EC_A^{loc}$ of local $A$-modules in $\EC$, i.e. $\ED = \EC_A^{loc}$. 

\item $\EE$ consists of all confined and deconfined particles, and can be identified with $\EC_A$, which is  the category of right $A$-modules in $\EC$.

\item Anyons in the $\EC$-phase can move onto the wall according to the central functor 
$-\otimes A: \EC \to \EC_A$ defined by $C\mapsto C\otimes A$ for all $C\in \EC$.
 
\item Anyons in the $\ED$-phase can move onto the wall according to the embedding $\EC_A^{loc} \hookrightarrow \EC_A$, then can move out to the $\ED$-side freely. 

\enu

\begin{rema} {\rm
We use Theorem$^\mathrm{ph}$, Lemma$^\mathrm{ph}$, Proposition$^\mathrm{ph}$ and Corollary$^\mathrm{ph}$ to summarize and highlight the physical results obtained from our bootstrap analysis, and to distinguish them from rigorous mathematical results. %For example, the above main result is summarized in Theorem$^\mathrm{ph}$\,\ref{thm:bs-wall-simple}. 
}
\end{rema}

\begin{rema} {\rm
The basic mathematical structures used in this work have already appeared in the seminal works \cite{bek1,bek2,bek3} by B\"{o}ckenhauer, Evans and Kawahigashi in 1999-2001. Moreover, they worked in the unitary setting, which is the most relevant case in physics. But they used the language of the subfactor theory instead of the tensor-categorical language. The following dictionary provided by Kawahigashi should be helpful. 
\begin{center}
\begin{tabular}
[c]{|c|c|}\hline
tensor-categorical language & subfactor language \\\hline\hline
connected comm. separable algebra $A$ & local Q-system  \\  \hline
category $\EC_A$ of $A$-modules & $\alpha$-induced system  \\ \hline
category $\EC_A^{loc}$ of local $A$-modules  & ambichiral system  \\ \hline
the bulk-to-wall map & $\alpha^\pm$-induction \\ \hline
boundary-bulk duality & quantum double of $\alpha$-induced system   \\ \hline
\end{tabular}
\end{center}
}
\end{rema}

\begin{rema} {\rm
For most physical applications, we need the assumption of unitarity \cite{kitaev1}. Since our theory works pretty well in the non-unitary cases, we only assume the MTC without unitarity in the main body of this paper. We put all results related to the unitary in Remarks.
}
\end{rema}

\begin{rema} {\rm
After the appearance of the 3rd version of this paper on arXiv, I was informed by Sander Bais that this work has some overlaps with Sebas Eli\"{e}ns' thesis \cite{eliens}, in which a commutative algebra object as Bose condensates was discussed (see \cite[Sec.\,6.2]{eliens}). See also their recent paper \cite{erb} joint with Romers. 
}
\end{rema}

\medskip
The layout of the paper is: in Section\,\ref{sec:bs-1}, we carry out this bootstrap analysis and derive our main results; in Section\,\ref{sec:find-A-B}, we discuss how to use physical macroscopic data to determine the condensation; in Section\,\ref{sec:example}, we provide examples; in Section\,\ref{sec:mu-we}, we discuss the Witt equivalence between 2d topological orders; Appendix contains the definitions of all tensor-categorical notions appeared in this work. At the end of this paper, we attach an Erratum and Addendum to the original version of “Anyon condensation and tensor categories” published in [Nucl. Phys. B 886 (2014) 436-482].

\bigskip
\noindent {\bf Acknowledgement}:  I thank Alexei Kitaev for sharing his unpublished works with me and Alexei Davydov for sending me the slides of his talk. I thank Ling-Yan Hung, Chao-Ming Jian and Yi-Zhuang You for motivating me to write up this paper. Their comments on the first version of this paper lead to clarification in Remark\,\ref{rema:contradiction}\,\&\,\ref{rema:confusion}, \ref{rema:confusion2}. I want to thank Michael M\"{u}ger and Dmitri Nikshych for clarifying the notion of unitary category, and thank J\"{u}rgen Fuchs and Christoph Schweigert for clarifying the connection to their works and for many suggestions for improvement. I want to thank Sander Bais and Joost Slingerland for clarifying their contributions to this subject, and Yasuyuki Kawahigashi for clarifying the connection to the subfactor theory. I thank Xiao-Liang Qi, Xiao-Gang Wen, Zhong Wang, Yong-shi Wu for helpful discussion and Zhi-Hao Zheng for finding a mistake. I would like to thank the referee for many important suggestions for improvement. The author is supported by Basic Research Young Scholars Program, Initiative Scientific Research Program at Tsinghua University, and NSFC under Grant No. 11071134.

\section{Bootstrap analysis}  \label{sec:bs-1}

\subsection{Anyons in a modular tensor category}
Let us start with a $2$d topological phase containing a system of anyonic excitations which form a MTC $\EC$ (see Definition\,\ref{def:mtc}).  In particular, the MTC $\EC$ is equipped with a tensor product $\otimes_\EC$ (or $\otimes$ for simplicity), a tensor unit $\one_\EC$ (or $\one$ for simplicity), an associator $\alpha_{X,Y,Z}: (X\otimes Y) \otimes Z \xrightarrow{\simeq} X \otimes (Y\otimes Z)$, 
%unit morphisms $\one \otimes X \xrightarrow{l_X} X \xleftarrow{r_X} X \otimes \one$ 
a braiding $c_{X,Y}: X\otimes Y \xrightarrow{\simeq} Y\otimes X$ and a twist $\theta_X: X \xrightarrow{\simeq} X$ for all $X,Y,Z \in \EC$. This is our initial data. Notice that we have intentionally ignored unit isomorphisms and duality maps from the data because the role they play in our presentation is implicit. We assume that an anyon condensation happens in a 2d region inside of a 2d phase $\EC$ as depicted in Fig.\,\ref{fig:set-up}, and the anyons in the condensed phase form another MTC $\ED$, which is equipped with a tensor product $\otimes_\ED$, a tensor unit $\one_\ED$, an associator $\alpha_{L,M,N}^\ED: (L\otimes_\ED M) \otimes_\ED N \xrightarrow{\simeq} L\otimes_\ED (M\otimes_\ED N)$, 
%unit isomophisms $\one_\ED \otimes X \xrightarrow{l_X} X \xleftarrow{r_X} X \otimes_\ED \one_\ED$ 
a braiding $c_{M, N}^\ED: M\otimes_\ED N \xrightarrow{\simeq} N\otimes_\ED M$ and a twist $\theta_M^\ED: M \xrightarrow{\simeq} M$ for all $L, M, N\in \ED$. This is our final data. The goal of this work is to work out all the necessary relations between $\EC$ and $\ED$ from natural physical requirements. 

\medskip
A simple object in $\EC$ is called a {\it simple anyon}. A generic object in $\EC$ is a direct sum of simple objects, and is called a {\it composite anyon}. We also use the term $X$-anyons. For example, three $X$-anyons means $X\otimes X \otimes X$.

\subsection{Physical data associated to a condensation}
We list some basic ingredients of the relation between $\EC$ and $\ED$ and some necessary physical data associated to an anyon condensation.
\bnu

\item A composite anyon in $\ED$ is necessarily made of (or a $\Zb_{\geq 0}$-linear combination of) simple anyons in $\EC$. The condensation process does not affect the ingredients of such a composite anyon. The ground state or other states in the condensed phase are  those states in $\EH$ that survives the condensation. Therefore, all objects in $\ED$ are automatically objects in $\EC$, the condensation simply induces the {\it identity condensation map} $M \xrightarrow{\id_M} M$ for all $M$ in $\ED$. In particular, the categorical vacuum wave function (or the tensor unit) $\one_\ED$ should be viewed as an object $A$ in $\EC$, i.e. $\one_\ED = A$.  In general, $A$ is a composite anyon in $\EC$ unless the condensation is trivial. The object $A$ should be viewed as a categorical ground-state wave function in the condensed phase. Not all anyons in $\EC$ survive the condensation. The MTC $\ED$ consists of only deconfined particles. The precisely definition of a deconfined particle emerges later as we move on.

\item Since all anyons in $\ED$ are made of simple anyons in $\EC$, all the possible fusion-splitting channels in the condensed $\ED$-phase must come from those in $\EC$. The information of these channels is encoded in the hom spaces. Therefore, we must have an embedding: 
$$
\hom_\ED(M, N) \hookrightarrow \hom_\EC(M, N), \quad\quad \forall M,N \in \ED.
$$
In other words, $\ED$ can be viewed as a (not full) subcategory of $\EC$.

\item The vacuum $\one=\one_\EC$ in $\EC$-phase should condense into the vacuum $\one_\ED$ in $\ED$. Mathematically, this means that there exists a morphism $\iota_A: \one \to A$ in $\EC$.

\item The difference between these two phases lies mainly in the way they fuse anyons. Therefore, we would like to know the difference and relation between $M \otimes N$ and $M\otimes_\ED N$ for any pair of anyons $M, N\in \ED$. A condensation is a process of selecting an energy-favorable subspace from the original Hilbert space $\EH$. In particular, the condensation process select $M\otimes_\ED N$ from $M\otimes N$ as an energy-favorable subspace. Therefore, we expect that there is an onto map, called {\it condensation map} in $\EC$:
\be  \label{eq:rho-MN}
\rho_{M,N}: M\otimes N \to M\otimes_\ED N. 
\ee
Moreover, we require that $M\otimes_D N$ lies in $M\otimes N$ in a canonical way (automatic in the unitary cases). 
By that we mean, there is a canonical morphism 
\be \label{eq:e-MN}
e_{M,N}: M\otimes_\ED N \to M\otimes N
\ee 
such that $\rho_{M,N} \circ e_{M,N} =\id_{M\otimes_\ED N}$.

Since $\one_\ED = A$, we must have $A\totimes A = A$ and $A\totimes M=M=M\totimes A$.  We denote the map $\rho_{A,A}: A\otimes A \to A=A\totimes A$ by $\mu_A$ and $e_{A,A}$ by $e_A$. 

Since $\EC$ is semisimple, $\mu_A$ and $e_A$ define a decomposition of $A\otimes A$: 
\be  \label{eq:decompose-AA}
A\otimes A = A \oplus X
\ee
where $X$ can be chosen to be the cokernel of $e_A$. By the mathematical definition of direct sum, it amounts to the existence of maps $e_X, r_X$, together with $e_A,\mu_A$, as shown in the following diagram:
$$
\xymatrix{ A \ar@<0.5ex>[r]^{e_A} & A\otimes A  \ar@<0.5ex>[l]^{\mu_A}  \ar@<0.5ex>[r]^{r_X} & X~, \ar@<0.5ex>[l]^{e_X} } 
$$
satisfying $\mu_A \circ e_A = \id_A$, $r_X \circ e_X=\id_X$, and
\be \label{eq:e-r}
\mu_A \circ e_X =0, \quad\quad r_X \circ e_A =0, \quad\quad e_A \circ \mu_A + e_X \circ r_X = \id_{A\otimes A}.
\ee

\enu

\begin{rema} \label{rema:unitary} {\rm
If $\EC$ is unitary (see Definition\,\ref{def:unitary}), then we can choose $\rho_{M,N}$ and $e_{M,N}$ to be a part of orthonormal basis such that $e_{M,N} = \rho_{M,N}^\ast$. 
}
\end{rema}

\begin{rema}  {\rm
All physical observables are encoded in the hom spaces. Very often, physicists like to understand a morphism $f: X\to Y$ by the canonically associated linear maps: $f_\ast: \hom_\EC(i, X) \to \hom_\EC(i, Y)$ defined by $g \mapsto f\circ g$ for all simple objects $i\in \EC$. These two points of view are equivalent. In this work, we treat $f$ as a physical observable and use it directly instead of $f_\ast$, and call $f$ as a morphism or a map. 
}
\end{rema}

\begin{rema} \label{rema:contradiction} {\rm
Since $\ED$ is a subcategory of $\EC$, if a simple object in $\EC$ ``survives the condensation", it seems that it should remain to be simple in $\ED$. It  was known, however, in physics that a simple object in $\EC$ can split after ``condensing" to the boundary. This superficial contradiction is actually a confusion in language. These two ``condensations" are referring to two different ways to compare two different categories. We explain this point in Remark\,\ref{rema:confusion} and \ref{rema:confusion2}. 
}
\end{rema}

\subsection{Vacuum in \texorpdfstring{$\ED$}{D} as an algebra in \texorpdfstring{$\EC$}{E}}
 In this subsection, we explore the properties of the condensation maps associated to $A$. 

\bnu

\item {\it Associativity of $\mu_A$}: if we condense three $A$-anyons\footnote{In physics, a condensation involves a large number of particles. It does not make any sense to say ``condense three anyons". A condensation in an anyon system is triggered by interaction among anyons. This interaction (e.g. adding a pair-wise interaction $1-\rho_{M,N}$ to the Hamiltonian) makes the subspace $M\otimes_\ED N$ of $M\otimes N$ more favorable in energy. We believe that a condensation in a region $R$ in the bulk can be realized by turning on the interaction in many small disjoint disks, each of which contains only a small number of anyons, and gradually enlarging the disk area such that the entire region $R$ is covered by the disks. By ``condensing three anyons", we mean turning on the interaction in a small disk containing only 3 anyons and projecting the local Hilbert space associated to the small disk onto the subspace of energy favorable states. A real condensation is a combination of such projections in a large quantity (in the thermodynamics limit). We use the terminology ``condense three anyons" here just for convenience. We use it and similar terms in many places later. On the other hand, to tell a complete story of anyon condensation, one would like to really write down a Hamiltonian system that can realize a given phase transition. It is an important problem in physics (see \cite{buss}), but beyond the scope of this paper. We hope that the gap between a complete physical theory of anyon condensation and the bootstrap study in this work can be filled in the near future.} 
in the bulk of $\EC$-phase, this process is independent of which pair of $A$ condense first. This independence leads to the following commutative diagram:
\be \label{diag:asso}
\xymatrix{
(A \otimes A) \otimes A  \ar[rr]^{\alpha_{A,A,A}} \ar[d]_{\mu_A1} &  & A \otimes (A\otimes A) \ar[d]^{1\mu_A}  \\
(A\otimes_\ED A) \otimes A \ar[d]_{\mu_A}  & & A\otimes (A\otimes_\ED A) \ar[d]^{\mu_A}  \\
(A \totimes A) \totimes A  \ar@{=}[r]  & A & A\totimes (A \totimes A)  \ar@{=}[l]
}
\ee
which means that $\mu_A: A\otimes A \to A$ is an associative multiplication.

\item {\it Unit properties}: The identity condensation map $\id_A: A\to A$ should be stable under a perturbation of the vacuum $\one$ in $\EC$. This leads to the following commutative diagrams: 
\be \label{diag:unit}
\xymatrix{
\one \otimes A \ar[r]^{\iota_A 1} \ar@{=}[d] & A\otimes A \ar[d]_{\mu_A} \\
A \ar[r]^{\id_A} & A
}
\quad\quad\quad 
\xymatrix{
A \otimes A  \ar[d]_{\mu_A} & A\otimes \one \ar@{=}[d] \ar[l]_{1\iota_A} \\
A  & A \ar[l]_{\id_A}
}
\ee
where the first diagram says that if we start with an $A$-anyon, then ``create" a vacuum $\one_\EC$ nearby, then condense it into $A$, then condense this $A$ further with the second $A$ into the new vacuum $A$, this process is physically not distinguishable with doing nothing (or the identity condensation map). %We can imagine that the process of the upper path is virtually happening all the time. The identity condensation map should be stable under such virtual processes. 
The meaning of the second commutative diagram in (\ref{diag:unit}) is similar. 

\item {\it Commutativity}: The condensation of two vacuums $A\otimes A$ is independent of whether we move one $A$-particle around the other $A$-particle along an arbitrary path before or after the condensation. This leads to the following commutative diagram:
\be \label{diag:comm}
\xymatrix{
A\otimes A  \ar[rr]^{c_{A,A}} \ar[d]_{\mu_A} & & A\otimes A  \ar[d]_{\mu_A} \\
A=A\totimes A \ar[rr]^{c_{A,A}^\ED=\id_A} & & A\totimes A =A
}
\ee

The commutative diagrams (\ref{diag:asso}), (\ref{diag:unit}) and (\ref{diag:comm}) are nothing but the defining properties of a commutative $\EC$-algebra for the triple $(A, \mu_A, \iota_A)$ (recall Definition\,\ref{def:alg}).

\begin{figure}[t] 
 \begin{picture}(250, 182)
   \put(0,-20){\scalebox{0.7}{\includegraphics{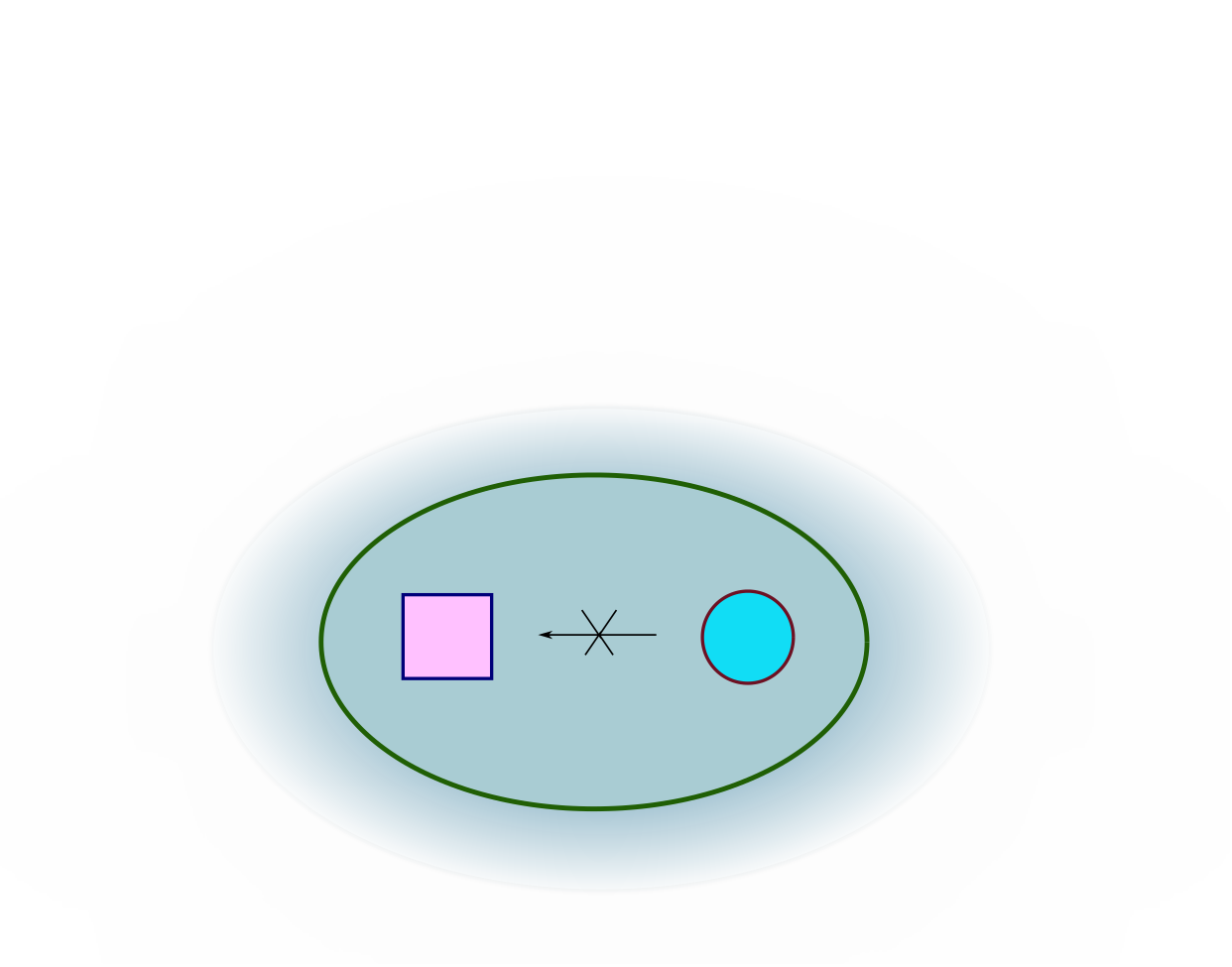}}}
   \put(0,-20){
     \setlength{\unitlength}{.75pt}\put(-18,-19){
     \put(345, 163)      { $ A $}
     \put(209,163)      { $X$}
     \put(265, 110)     { $A\otimes A$}
     \put(385, 220)     { $A$-cloud}
     }\setlength{\unitlength}{1pt}}
  \end{picture}
%  \centerline{\includegraphics[scale=0.7]{pic-stability-eps-converted-to}}
  \caption{{\small Stability of the vacuum $A$ in $A\otimes A$ under the action of $A$: The condensed vacuum $A$ in $\ED$ can be viewed as a canonical building block of $A\otimes A$. This information is encoded in the map $e_A: A\hookrightarrow A\otimes A$. When $A\otimes A$ is shrouded by (or simply a part of) an $A$-cloud (a tensor product $A^{\otimes n}$ for large $n$), $A$ in $A\otimes A$ should be stable under the action of $A$ on $A\otimes A$ from both sides. In other words, splitting $A$ into $X$ under such action is forbidden. Otherwise, $A$ is not stable. Therefore, we obtain that the map $e_A$ must be stable under the $A$-action on $A\otimes A$ from both sides. Mathematically, this condition says that two composed maps (\ref{eq:A-L-stable}) and (\ref{eq:A-R-stable}) are zero maps.}}
  \label{fig:stability-vacuum}
\end{figure}

\item {\it The stability of the vacuum $A$ in $A\otimes A$ under the $A$-action}\footnote{This stability is different from the usual stability of the vacuum under the small perturbations of Hamiltonian.}: The vacuum $A$, which lies in $A\otimes A$, i.e. $e_A: A \hookrightarrow A\otimes A$, should be stable under the screening of a cloud of vacuum (see Figure\,\ref{fig:stability-vacuum}). It implies that the $A$-action on $A\otimes A$ cannot create any splitting channels from $A$ to $X$. Otherwise the vacuum $A$ in an $A$-cloud can decay, which is physically unnatural. More precisely, this means that the following two composed maps
\be \label{eq:A-L-stable}
A \otimes A \xrightarrow{1e_A} A\otimes A\otimes A \xrightarrow{\mu_A1} A\otimes A \xrightarrow{r_X} X
\ee
and 
\be \label{eq:A-R-stable}
A \otimes A \xrightarrow{e_A1} A\otimes A\otimes A \xrightarrow{1\mu_A} A\otimes A \xrightarrow{r_X} X
\ee
must be zero maps. We use the conditions (\ref{eq:A-L-stable}) and (\ref{eq:A-R-stable}) to show that the algebra $A$ is separable (see Definition\,\ref{def:separable}) in the next paragraph. %Physicists can skip it. 

Notice that $A\otimes A$ is naturally an $A$-$A$-bimodule. By the associativity, the map $\mu_A$ is automatically an $A$-$A$-bimodule map. Using this fact, together with (\ref{eq:A-L-stable}) and (\ref{eq:A-R-stable}) being zero maps, it is easy to show that the map $e_A$ is an $A$-$A$-bimodule map. Moreover, using (\ref{eq:e-r}), it is easy to show that the following composed map:
$$
A\otimes X \xrightarrow{1 e_X} A\otimes A \otimes A \xrightarrow{\mu_A1}  A\otimes A \xrightarrow{r_X} X
$$
defines a left $A$-module structure on $X$. Similarly, the following composed map:
$$
X\otimes A \xrightarrow{e_X1} A\otimes A \otimes A \xrightarrow{1\mu_A}  A\otimes A \xrightarrow{r_X} X
$$
defines a right $A$-module structure on $X$. These two module structures are compatible so that $X$ is an $A$-$A$-bimodule. Using the fact that $e_A$ and $\mu_A$ are $A$-bimodule maps, it is easy to show that both $e_X$ and $r_X$ are $A$-$A$-bimodule maps. In other words, the decomposition (\ref{eq:decompose-AA}) is also a decomposition of $A$-$A$-bimodules. 

In mathematics, such an algebra $(A, \mu_A, \iota_A)$ is called {\it separable} (see Definition\,\ref{def:separable}). An important property of an separable algebra in $\EC$ is that both the category $\EC_A$ of $A$-modules in $\EC$ and the category $\EC_{A|A}$ of $A$-$A$-bimodules in $\EC$ are semisimple (see for example \cite{ko}). 

\item The algebra $(A, \mu_A, \iota_A)$ is connected, i.e. $\hom_\EC(\one, A) \simeq \Cb$ (see also Definition\,\ref{def:separable}): As we will show later that all objects $M$ in $\ED$ are $A$-modules and morphisms are $A$-module maps. Therefore, 
\be \label{eq:connect}
\mathbb{C} \simeq \hom_\ED (A, A) = \hom_A(A, A) \simeq \hom_\EC(\one, A)
\ee

\enu

Above bootstrap results can be summarized as follows:
\begin{plemma} \label{lemma:A=algebra}
$\ED$ is a subcategory of $\EC$. The vacuum $A=\one_\ED$ of the $\ED$-phase is a connected  commutative separable algebra in $\EC$. 
\end{plemma}

Since $\dim \hom_\EC(A, \one) =1$, we can choose a map $\epsilon_A: A \to \one$ such that 
$\epsilon_A \circ \iota_A = \dim A \cdot \id_A$. Here we assume that $\dim A\neq 0$, which is always true if $\EC$ is a unitary MTC. It is known that the pairing $(A\otimes A \xrightarrow{\mu_A} A \xrightarrow{\epsilon_A} \one)$ is non-degenerate. This implies (\cite[Lemma\,3.7]{tft1}) that $A$ has a canonical Frobenius algebra structure (see Definition\,\ref{def:Frob}). Moreover, by \cite[Cor.\,3.10]{tft1}, this Frobenius algebra is automatically symmetric; by \cite[Lemma.\,3.11]{tft1}, it is also normalized-special (see Definition\,\ref{def:Frob}). As a consequence, the coproduct $\Delta_A$ satisfies $\mu_A \circ \Delta_A = \id_A$ and $\Delta_A \circ \mu_A$ is a projector on $A\otimes A$. Moreover, $\Delta_A$ is an $A$-bimodule map because of the defining property of a Frobenius algebra. In other words, $\Delta_A$ give a splitting of the $A$-bimodule map $\mu_A: A\otimes A \to A$. Using results in \cite{tft1}, one can prove that 
\be
e_A=\Delta_A.
\ee
Therefore, $A$ has a natural structure of a connected commutative symmetric normalized-special Frobenius algebra in $\EC$ \cite{tft1}. %Conversely, the latter algebra can reproduce the original 2d-condensable algebra. In other words, these two concepts are equivalent. 

Alternatively, one can argue directly that $\Delta_A:=e_A: A\to A\otimes A$ gives a coassociative comultiplication because the vacuum $A$ lies in $A\otimes A\otimes A$ in a canonical way. Moreover, (\ref{eq:A-L-stable}) and (\ref{eq:A-R-stable}) imply the identities in (\ref{eq:Frob-property}). Since $A$ is connected, i.e. $\dim \hom_\EC(A, \one) =1$, for any $f\in \hom_\EC(A, \one)$, the map $(f\otimes 1) \circ e_A$ is a right $A$-module map and thus must be $c\cdot \id_A$ for some scalar $c\in \Cb$ because $A$ is a simple right $A$-module (proved later). Then we can choose the counit $\epsilon: A\to \one$ so that the counit condition hold. Again we obtain a structure of a commutative symmetric normalized-special Frobenius algebra on $A$.

\begin{rema} \label{rema:unitary-A} {\rm
When $\EC$ is unitary, we have $e_A=\mu_A^\ast$. Then the coassociativity follows from the associativity automatically. Choose the counit $\epsilon_A:=\iota_A^\ast$. Then the counit property is automatic. By \cite{tft1}, we have $\epsilon_A \circ \iota_A = \dim A \cdot \id_A$ automatically. A Frobenius algebra is called $\ast$-Frobenius algebra if $\Delta_A=\mu_A^\ast$. As a consequence, $A$ is a necessarily
commutative symmetric normalized-special $\ast$-Frobenius algebra in $\EC$. }
\end{rema}

To simplify our terminology, we introduce the following notion.
\begin{defn} {\rm
A 2d-condensable algebra $A$ in a MTC $\EC$ is a connected commutative symmetric normalized-special Frobenius algebra $\EC$. In the unitary case, a 2d-condensable algebra means a commutative symmetric normalized-special $\ast$-Frobenius algebra in $\EC$. 
}
\end{defn}

\begin{expl} {\rm
In the toric code model, the bulk anyons form a MTC $Z(\rep(\Zb_2))$, which is the Drinfeld center of the fusion category $\rep(\Zb_2)$. Then $1\oplus e$ and $1\oplus m$ are two examples of 2d-condensable algebra in $Z(\rep(\Zb_2))$. More examples of 2d-condensable algebras in other models are given in Section\,\ref{sec:example}.
}
\end{expl}

\smallskip
The condensation also must preserve the twists (a generalized notion of spin). This leads to the following conditions:
\be \label{diag:preserve-spin-A}
\xymatrix{ A \ar[r]^{\theta_A} \ar[d]_{\id_A} & A  \ar[d]^{\id_A} \\
A \ar[r]^{\theta_A^\ED} & A, 
}
\quad\quad\quad\quad
\xymatrix{
M \ar[r]^{\theta_M} \ar[d]_{\id_M} & M  \ar[d]^{\id_A} \\
M \ar[r]^{\theta_M^\ED} & M
}
\ee
for all $M\in \ED$. Since $A$ is the vacuum, $\theta_A^\ED =\id_A$. We must require that $\theta_A = \id_A$. In physical language, it means that $A$ must be a {\bf boson}. This condition turns out to be a redundant condition because a commutative Frobenius algebra $A$ is symmetric if and only if $\theta_A = \id_A$ \cite[Prop.\,2.25]{cor}. Therefore, we obtain
\begin{cor} \label{cor:boson}
A 2d-condensable algebra $A$ in $\EC$ is automatically a boson, i.e. $\theta_A = \id_A$. 
\end{cor}

\subsection{General deconfined particles}

The second diagram in (\ref{diag:preserve-spin-A}) simply means $\theta_M^\ED = \theta_M$. Before we discuss its meaning, we would like to first explore the properties of the condensation maps $\mu_M:=\rho_{A,M}: A\otimes M \to A\otimes_\ED M=M$ for all $M\in \ED$ and $e_{A,M}: A\otimes_\ED M \hookrightarrow A\otimes M$. 
%Since $A$ is special Frobenius, we can first choose $e_M: M \to A\otimes M$ to be 
%\be \label{eq:e_M}
%e_M: M \simeq \one_\EC \otimes M \xrightarrow{(1\mu_M) \circ \Delta_A \circ \iota_A} A \otimes M.
%\ee

\bnu

\item the pair $(M, \mu_M)$ is a left $A$-module:
\bnu
\item {\it Associativity}: as before, if we condense two $A$-anyons and an $M$-anyon, the process should not depends on which two of them condense first. This leads to the following commutative diagram:
\be  \label{diag:asso-module}
\xymatrix{
A\otimes (A \otimes M) \ar[rr]^{\alpha_{A,A,M}} \ar[d]_{1\mu_M}  & & (A \otimes A) \otimes M \ar[d]^{\mu_A1} \\
A\otimes (A \otimes_\ED M)  \ar[d]_{\mu_M} & & (A \otimes_\ED A) \otimes M  \ar[d]^{\mu_M}  \\
A\totimes (A\totimes M) \ar@{=}[r] & M & (A\totimes A) \totimes M \ar@{=}[l]
}
\ee
%By the universal property of $\otimes_A$, there is a morphism $A \otimes_A M \to A\otimes_\ED M$. On the other hand, we have $A\otimes_\ED M \cong M \cong A\otimes_A M$. By the universal property of $\otimes_A$, we obtain

 \item {\it Unit property}: Due to the similar physical reason behind the unit property of $A$, we have
\be \label{diag:unit-module}
\xymatrix{
\one \otimes M \ar[r]^{\iota_A 1} \ar@{=}[d] & A\otimes M \ar[d]^{\mu_M} \\
M \ar[r]^{\id_M} &  M
}
\ee

\enu
Above two commutativity diagrams (\ref{diag:asso-module}) and (\ref{diag:unit-module}) are the defining properties of a left $A$-module for the pair $(M, \mu_M)$ (see Definition\,\ref{def:module}).

\item Similarly, $M$ equipped with a right $A$-action $\rho_{M,A}: M\otimes A \to M$ is a right $A$-module. 

\item $(M, \mu_M)$ is a local $A$-module: condensation process is irrelevant to how you arrange the initial configuration of an $A$-anyon and an $M$-anyon. More precisely, it means that if you start with an arbitrary initial position of these two anyons, then move one around the other along a path, then condense them, it is equivalent to first condense them, then move them around the same path. Mathematically, it means that the condensation respects the braiding. Thus, we obtain the following commutative diagram: 
\be \label{diag:preserve-braiding}
\xymatrix{
A \otimes M \ar[r]^{c_{A,M}} \ar[d]^{\mu_M} & M \otimes A \ar[r]^{c_{M,A}} \ar[d]^{\rho_{M,A}} & A\otimes M \ar[d]^{\mu_M}  \\
A \totimes M \ar[r]^{c_{A,M}^\ED} & M \totimes A \ar[r]^{c_{M,A}^\ED}  & A\totimes M
}
\ee
where $A\totimes M =M=M\totimes A$ and $c_{M,A}^\ED = c_{A,M}^\ED = \id_M$ \cite[Prop.\,XIII.1.2]{kassel}. Therefore, we obtain
$$
\mu_M \circ c_{M,A} \circ c_{A,M} = \mu_M. 
$$
Such an $A$-module $M$ is called {\it local} (see Definition\,(\ref{def:local})). This commutative diagram also means that the left $A$-module structure determines the right $A$-module structure in a unique way via braiding, i.e.
$$
\rho_{M,A} = \mu_M \circ c_{M,A} = \mu_M \circ c_{A,M}^{-1}.
$$
For this reason, we also denote $\rho_{M,A}$ by $\mu_M$. 

\begin{rema} \label{rema:deconfined} {\rm
The notion of a local $A$-module provides the precise mathematical definition of a ``deconfined particle'' used in physics literature. Automatically, a right A-module that is not local is precisely a ``confined particle''.
}
\end{rema}

\item {\it Stability of a condensed anyon $M$ in $A\otimes M$ and $M\otimes A$ under the $A$-action}: Similar to the stability of the vacuum $A$, we require that the condensed particle $M$ in $A\otimes M$, i.e. $e_{A,M}: M \hookrightarrow A\otimes M$ is stable under the screening of a cloud of the vacuum $A$. It implies that left A-action on $A\otimes M$ cannot create any splitting channels from the subobject $M$ to other complementary subobjects in $A\otimes M$. Similar to that of the stability of the vacuum $A$, we obtain that $e_{A,M}$ is a left $A$-module map. 
By the locality of $M$, the map $e_{M,A}: M \hookrightarrow M\otimes A$ is automatically a right $A$-module map and an $A$-$A$-bimodule map.

\item {\it Compatibility among $e_A$, $e_{A,M}$ and $e_{M,A}$}: Consider two physical processes described by the two paths in the first of the following two diagrams: 
\be  \label{diag:e-A-eAM-eMA}
\xymatrix{
M \ar[d]_{e_{A,M}} \ar[r]^{e_{A,M}} & A \otimes M \ar[d]^{e_A 1}  \\
A\otimes M  & A\otimes A \otimes M \ar[l]_{1\mu_M}
}
\quad\quad\quad
\xymatrix{
M \ar[d]_{e_{M,A}} \ar[r]^{e_{M,A}} & M \otimes A \ar[d]^{1e_A}  \\
M\otimes A  & M\otimes A \otimes A \ar[l]_{1\mu_M}
}
\ee
Notice that the physical processes described by the composed map $A\otimes M \xrightarrow{e_A1} A\otimes A \otimes M \xrightarrow{1\mu_M} A\otimes M$ can be viewed as something virtually happening all the time. Of course, one can embed $A$ into more $A$-anyons (or an $A$-cloud) and then fuse them with $M$ until the last $A$. It is a natural physics requirement that $e_{A,M}$ must be stable under such virtual processes. Therefore, we conclude that the first diagram in (\ref{diag:e-A-eAM-eMA}) is commutative. Similarly, we can convince ourselves that the commutativity of the second diagram in (\ref{diag:e-A-eAM-eMA}) is also a physical requirement. 

Using the Frobenius property of $\Delta_A=e_A$ and the identities: $\mu_M \circ e_{A,M}=\id_M$ and $\mu_M\circ e_{M,A}=\id_M$, and their graphic expressions (see Section\,\ref{app:module-cat}), we obtain the following identities: 
\be  \label{eq:e_M-e_A}
  e_{A,M}~=~ 
  \raisebox{-27pt}{
  \begin{picture}(65,55)
   \put(8,8){\scalebox{.75}{\includegraphics{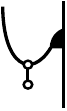}}}
   \put(8,8){
     \setlength{\unitlength}{.75pt}\put(-30,-38){
     \put(55, 28)  {\scriptsize $ M $}
     \put(55,94)  {\scriptsize $ M $}
     \put(26, 91)  {\scriptsize $ A $}
     }\setlength{\unitlength}{1pt}}
  \end{picture}}~,
  \quad\quad\quad
 e_{M,A}~=~ 
  \raisebox{-27pt}{
  \begin{picture}(65,55)
   \put(8,8){\scalebox{.75}{\includegraphics{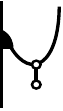}}}
   \put(8,8){
     \setlength{\unitlength}{.75pt}\put(-30,-38){
     \put(26, 28)  {\scriptsize $ M $}
     \put(26,94)  {\scriptsize $ M $}
     \put(55, 91)  {\scriptsize $ A $}
     }\setlength{\unitlength}{1pt}}
  \end{picture}} 
\ee

\begin{rema} \label{rema:unitary-AM} {\rm
When $\EC$ is unitary, then identities (\ref{eq:e_M-e_A}), together with $e_A=\mu_A^\ast$ and $\epsilon_A = \iota_A^\ast$, implies that $e_{A,M} = \mu_M^\ast$ and $e_{M,A} = \rho_{M,A}^\ast$. 
}
\end{rema}

\item Morphisms in $\ED$ are $A$-module homomorphisms: The morphisms in $\ED$ determine the fusion-splitting channels in the phase $\ED$. These fusion-splitting process must come from those fusion-splitting process (or morphisms) in $\EC$ and survived the condensation process. In particular, it means that such a morphism must remain intact after the screening by a cloud of the vacuum $A$. In other words, we have the following commutative diagrams: 
\be \label{diag:screening-map}
\xymatrix{
A\otimes M \ar[rr]^{1f}  & &  A\otimes N  \ar[d]_{\mu_N}  \\
A \totimes M =M  \ar[rr]^f \ar[u]_{e_{A,M}} & & N= A\totimes N.
}
\ee
for all $f\in \hom_\ED(M,N)$. By the fact that both $e_{A,M}$ and $\mu_N$ are left $A$-module maps, it is clear that the commutativity of the diagram (\ref{diag:screening-map}) is equivalent to the condition that $f$ is an $A$-module map, i.e. $\hom_\ED(M, N) = \hom_A(M,N)$. Mathematically, it means that the embedding $\ED \hookrightarrow \EC_A^{loc}$ is fully faithfully. This fact implies, in particular, the identity (\ref{eq:connect}). 

We would also like to point out that the upper path in diagram (\ref{diag:screening-map}) defines a {\it screening map} $\text{Sc}_A: \hom_\EC(M, N) \to \hom_{\EC_A^{loc}}(M, N)$ given by\footnote{In the special case $M=N$, assuming that $A$ is commutative symmetric special Frobenius, this screening map was given as the $Q_M$-operator defined in equation (3.35) in \cite{cor} (see also (\ref{fig:screening-op})). But we cannot use the $Q_M$-operator directly here because we want to apply the result to prove equation (\ref{eq:connect}), which was further used to prove that $A$ is a special symmetric commutative Frobenius algebra.} 
\be \label{eq:screening-op}
\text{Sc}_A: f \to \mu_M \circ (1f) \circ e_{A,M}.
\ee
An $A$-module map is automatically an $A$-$A$-bimodule map. This screening map $\text{Sc}_A$ is very natural from physical point of view because a fusion-splitting channel in $\EC$-phase screened by a cloud of the vacuum $A$ is automatically a fusion-splitting channel in $\ED$-phase. Using (\ref{eq:e_M-e_A}) and the locality of $M$ as $A$-module, the screening map defined in (\ref{eq:screening-op}) can be equivalently defined graphically as follows:
\be \label{fig:screening-op}
\mathrm{Sc}_A(g)   ~=~ \quad 
\raisebox{-57pt}{
  \begin{picture}(70, 122)
   \put(0,8){\scalebox{1}{\includegraphics{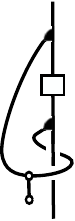}}}
   \put(0,8){
     \setlength{\unitlength}{.75pt}\put(-18,-19){
     \put( 16, 105)      {\scriptsize $ A $}
     \put(49,105)      {\scriptsize $g$}
     \put(30, 56)      {\scriptsize $ \Delta_A $}
     \put(70,56)       {\scriptsize $A$}
     \put(47, 10)      {\scriptsize $M$}
     \put(47,164)      {\scriptsize $N$}
     }\setlength{\unitlength}{1pt}}
  \end{picture}}
\ee
Using the normalized-specialness of the Frobenius algebra $A$, it is also easy to see that the screening map $\text{Sc}_A$ is a projector, i.e. $\text{Sc}_A \circ \text{Sc}_A = \text{Sc}_A$, and surjective. We adapt the superficially new definition, which appeared in \cite{ko}\cite{cor}, not only because it looks pictorially more like a screening of $M$ by a cloud of the vacuum $A$, but also because the new definition has other applications which does not work for the definition (\ref{eq:screening-op}). For example, if $M$ is a non-local left $A$-module and $g=\id_M: M\to M$, the screening operator defined in (\ref{fig:screening-op}) is actually a projection onto the largest local sub-$A$-module of $M$ \cite{ko}\cite{cor}. 

%Many morphisms we defined before are examples of morphisms in $\ED$, such as the condensation maps $\id_M$, $\rho_{M,N}$, $\mu_A$, $\mu_M$ and the embedding maps $e_A$, $e_{A,M}$, $e_{M,A}$ for $M, N\in \ED$. 
An important example of morphisms in $\ED$ is $\theta_M^\ED = \theta_M$ (recall the second diagram in (\ref{diag:preserve-spin-A})). Actually, for a left $A$-module $M$, the condition that $\theta_M\in \hom_A(M, M)$ is equivalent to the condition that $M$ is a local $A$-module \cite[Prop.\,3.17]{cor}. 

\begin{rema} \label{rema:unitary-Sc} {\rm
When $\EC$ is unitary, by Remark\,\ref{rema:unitary-A}, it is easy to see that $\text{Sc}_A \circ \ast = \ast \circ \text{Sc}_A$. This implies that $\EC_A^{loc}$ is a $\ast$-category (see Definition\,\ref{def:unitary}). 
}
\end{rema}

\item $\totimes=\otimes_A$ (see Definition\,\ref{def:tensor-A} for the definition of $\otimes_A$): Notice first that the condensation cannot distinguish the following two condensations: $(N\otimes A)\otimes_\ED M$ and $N\otimes (A\otimes_\ED M)$. Namely, we must have $(N\otimes A)\otimes_\ED M \simeq N\otimes (A\otimes_\ED M)$. The rest argument is a little mathematical. Notice that the canonical map $f_{N,M}: N\otimes_A M \to 
N\otimes_\ED M$ must be an epimorphism because $\rho_{N,M}$ is an epimorphism. It is enough to show that the kernel of $f_{N,M}$ is zero. Since $N$ can always be realized as a submodule of $N\otimes A$ (recall (\ref{eq:e_M-e_A})), it is enough to prove that the map $f_{N\otimes A, M}: (N \otimes A) \otimes_A M \to (N\otimes A)\otimes_\ED M$ is an isomorphism. On the one hand, $f_{N\otimes A, M}$ is an onto map. On the other hand, the domain is isomorphic to the codomain as objects:
$$
(N\otimes A) \otimes_A M \simeq N \otimes (A\otimes_A M) \simeq N\otimes (A\otimes_\ED M) 
\simeq (N\otimes A)\otimes_\ED M. 
$$
Therefore, $f_{N\otimes A, M}$ can only be an isomorphism. By the universal properties of $\otimes_A$, these isomorphisms $f_{N,M}$ defines an natural isomorphism between two tensor product functors $f: \otimes_A \xrightarrow{\simeq} \otimes_\ED$. Hence, we can take $\otimes_\ED = \otimes_A$. Moreover, for $f: M\to M'$ and $g: N\to N'$, it is easy to show that
\be \label{eq:f-A-g}
f\otimes_A g = \rho_{M',N'} \circ (f\otimes g) \circ e_{M,N}.
\ee
The associator $\alpha_{L,M,N}^A: L\otimes_A (M\otimes_A N) \to (L\otimes_A M) \otimes_A N$ is uniquely determined by $\alpha_{L,M,N}$ and the universal property of $\otimes_A$. Therefore, we must have $\alpha_{L,M,N}^\ED = \alpha_{L,M,N}^A$. More precisely, it can be expressed as follows:
\be \label{eq:alpha-D}
\alpha_{L,M,N}^A = \rho_{L\otimes_A M, N} \circ (\rho_{L,M}\otimes \id_N) \circ \alpha_{L,M,N} \circ 
e_{L, M\otimes_A N} \circ (\id_L \otimes_A e_{M,N}).
\ee

\begin{rema} \label{rema:unitary-alpha} {\rm
When $\EC$ is unitary, it is easy to show that $(f\otimes_A g)^\ast =
f^\ast \otimes_A g^\ast$ and $(\alpha_{L,M,N}^A)^\ast = (\alpha_{L,M,N}^A)^{-1}$. The unitarity of the unit morphisms in (\ref{eq:unitary-asso-unit}) is trivially true here. In other words, $\EC_A^{loc}$ is a monoidal $\ast$-category (see Definition\,\ref{def:bm-dagger}). 
}
\end{rema}

\enu

What we have shown so far is that $\ED$ must be a full sub-tensor category of the tensor category $\EC_A^{loc}$ of local $A$-modules in $\EC$. Moreover, there is a natural braiding in $\EC_A^{loc}$ \cite{bek2,ko}, defined by descending the braiding $c_{M,N}: M\otimes N \rightarrow N\otimes M$ to a braiding $c_{M,N}^A: M\otimes_A N \to N\otimes_A M$ via the universal property of $\otimes_A$:  
$$
\xymatrix{
M \otimes N \ar[r]^{c_{M,N}} \ar[d]^{\rho_{M,N}} & N\otimes M \ar[d]^{\rho_{N,M}} \\
M\otimes_A N \ar@{.>}[r]^{\exists ! c_{M,N}^A} & N\otimes_A M\, .
}
$$
On the other hand, above diagram is still commutative if we replace $c_{M,N}^A$ in above diagram by $c_{M,N}^\ED$ for the exact the same reason as those discussed above the diagram (\ref{diag:preserve-braiding}). By the universal property of $\otimes_A$, such $c_{M,N}^A$ is unique. Therefore, we must have $c_{M,N}^\ED = c_{M,N}^A$. We can express $c_{M,N}^A$ more explicit as follows:
\be  \label{eq:c-MAN}
c_{M,N}^\ED = c_{M,N}^A = \rho_{N, M} \circ c_{M, N} \circ e_{M,N}. 
\ee

\medskip
Above bootstrap results can be summarized as follows.
\begin{plemma} 
$\ED$ is a full braided monoidal subcategory of $\EC_A^{loc}$. 
\end{plemma}

\begin{rema} \label{rema:unitary-braiding} {\rm
When $\EC$ is unitary, the equation (\ref{eq:c-MAN}) implies that $(c_{M,N}^A)^\ast=(c_{M,N}^A)^{-1}$. In other words, $\EC_A^{loc}$ is a braided monoidal $\ast$-category (see Definition\,\ref{def:bm-dagger}), and $\ED$ is a braided monoidal $\ast$-subcategory of $\EC_A^{loc}$. 
}
\end{rema}

The category $\EC_A^{loc}$ is also rigid (see Definition\,\ref{def:rigidity}). The duality maps can be naturally defined. For example, if $M \in \EC_A^{loc}$, then the right dual $M^\vee$ in $\EC$ is automatically a local $A$-module. Moreover, 
the birth (or coevaluation) map $b_M^A: A \to M\otimes_A M^\vee$ is given by 
$$
A \xrightarrow{1 b_M} A \otimes M \otimes M^\vee \xrightarrow{\mu_M 1} M\otimes M^\vee \xrightarrow{\rho_{M,M}} M\otimes_A M^\vee. 
$$
and the death (or evaluation) map $d_M^A: M^\vee \otimes_A M \to A$ is given by
$$
M^\vee \otimes_A M \xrightarrow{(e_{M^\vee, M}) \iota_A} M^\vee \otimes M \otimes A \xrightarrow{11\Delta_A} M^\vee \otimes M \otimes A \otimes A \xrightarrow{1\rho_{M,A}1} M^\vee \otimes M \otimes A \xrightarrow{d_M 1} A
$$
where $\Delta_A=e_A$ and $e_{M^\vee, M}$ is defined in (\ref{eq:e-MN}) and it splits $\rho_{M^\vee, M}$. The duality maps in $\ED$ must coincide with the duality maps in $\EC_A^{loc}$ because $\otimes_\ED=\otimes_A$. The quantum dimensions in $\ED$ can be easily obtained from those in $\EC$ as follows:
$$
\dim_\ED M=\dim_\EC M/\dim_\EC A,
$$ 
and $\dim (\EC_A^{loc}) = \frac{\dim (\EC)}{\dim_\EC (A)}$ \cite{ko}\cite{cor}.  

\medskip
Although we have not completed our bootstrap analysis, as we will show later from our bootstrap study of domain wall, there is no additional relation between $\EC$ and $\ED$ that can tell us which objects in $\EC_A^{loc}$ shall be excluded in $\ED$ except the condition that $\ED$ is modular. In general, if a local $A$-module is excluded from $\ED$, there must be a principle, such as a symmetry constraint, to tell us why such exclusion happens. Since there is no such symmetry constraint in sight except the requirement of the modularity of $\ED$, we conclude that $\ED$ must be a maximum modular tensor subcategory in $\EC_A^{loc}$. 

On the other hand, we recall an important mathematical theorem proved in \cite[Thm.\,4.2]{bek2} (in unitary setting) and \cite[Thm.\,4.5]{ko} (see also \cite[Prop.\,3.21]{cor}). 
\begin{thm}  \label{thm:modular}
If $A$ is a 2d-condensable algebra in a MTC $\EC$, then the category $\EC_A^{loc}$ of local $A$-modules in $\EC$ is also modular. 
\end{thm}

\begin{rema} {\rm
If $\EC$ is a unitary MTC, since $\theta_M^\ED=\theta_M$, we have $(\theta_M^\ED)^\ast = \theta_M^\ast = \theta_M^{-1}=(\theta_M^\ED)^{-1}$. In other words, $\EC_A^{loc}$ is a ribbon $\ast$-category, hence, a unitary MTC \cite[Thm.\,4.2]{bek2}.   
}
\end{rema}

In particular, let $\lambda, \gamma$ be two simple objects in $\EC_A^{loc}$, the $s$-matrix in $\EC_A^{loc}$ is given by \cite{bek2}\cite[Eq.\,(4.3)]{ko} (see also \cite[Eq.\,(3.56)]{cor}): 
\be  
 s_{\lambda,\gamma}^A   ~=~  \frac{1}{\dim A} \quad
\raisebox{-40pt}{
  \begin{picture}(110, 85)
   \put(0,8){\scalebox{.9}{\includegraphics{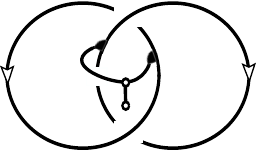}}}
   \put(0,8){
     \setlength{\unitlength}{.75pt}\put(-18,-19){
     \put( 168, 65)       {\scriptsize $ \gamma $}
     \put( 9, 62)      {\scriptsize $ \lambda $}
     \put( 53, 69)      {\scriptsize $ A $}
     \put( 80,67)       {\scriptsize $\Delta_A$}
     }\setlength{\unitlength}{1pt}}
  \end{picture}}
~~.
\ee

Therefore, we draw our conclusion: 
\begin{pthm} \label{thm:bs-D}
If a system of anyons, described by a (unitary) MTC $\ED$, is obtained via a condensation from another system of anyons given by a (unitary) MTC $\EC$, then $\ED$ consists of all deconfined particles. They form a (unitary) MTC $\EC_A^{loc}$, i.e. $\ED=\EC_A^{loc}$, where $A$ is a 2d-condensable algebra in $\EC$ and is the vacuum in $\EC_A^{loc}$. 
\end{pthm}

We have thus completed our bootstrap analysis on the condensed phase $\ED$.

%\section{Bootstrap analysis II: domain wall}  \label{sec:2d condensation}

\subsection{2d Condensations} 
If the domain wall between the $\EC$-phase and the $\ED$-phase is gapped, it gives an anomalous 1d topological phase. The wall excitations can be fused but not braided with each other. As a result, they form a unitary fusion category $\EE$. Moreover, we require that a pair of particle and its antiparticle can be annihilated or created from the vacuum, and the vacuum degeneracy is trivial\footnote{If this condition is not satisfied, the associated 1d topological phase is not stable \cite{wen-private}.}, i.e. $\hom_\EE(\one_\EE, \one_\EE)\simeq \Cb$. Therefore,  $\EE$ must be a unitary fusion category, which has a unique spherical structure \cite{kitaev1}\cite{eno02}. In this subsection, we discuss a special case, in which the vacuum of $\EE$ is precisely given by $A$. 

It was known to physicists that when a condensation occurs in a 2d (spatical dimension) region, called a {\it 2d condensation}, the condensed 2d $\ED$-phases consists of all the deconfined particles. Recall Remark\,\ref{rema:deconfined}, the mathematical definition of a deconfined particle is precisely a local $A$-module in $\EC$. At the same time, a confined particle can be defined by a ``non-local'' right $A$-module in $\EC$.
It was also intuitively clear to physicists that all the confined particles naturally accumulate on the domain wall and are confined to the wall. Moreover, deconfined particles in $\ED$ can move onto the wall and can then move back to $\ED$ freely. Therefore, the wall $\EE$ consists of both deconfined particles and confined ones, i.e. $\EE=\EC_A$ as categories. 

Note that a particle $X$ on the wall is naturally equipped with two-side actions of $A$. In particular, the left $A$-action on $X$ should be the one canonically induced from the right $A$-action as illustrated in Figure\,\ref{fig:left-right-A-module}. Mathematically, this statement simply says that the following diagram 
\be \label{diag:AX-c-XA}
\xymatrix{
A\otimes X \ar[r]  \ar[d]_{c_{X,A}^{-1}} &  X \\
X\otimes A \ar[ru] & 
}
\ee
is commutative. Therefore, an object in $\EC_A$ is automatically equipped with an $A$-$A$-bimodule structure. Then it is clear that $\otimes_A$ provides a tensor product structure on $\EC_A$ with tensor unit $A$. It turns out that $\EC_A$ is automatically a (unitary) spherical fusion category. This immediately leads to the main result of this paper. 

\begin{rema} \label{rem:braiding-convention} {\rm
Our braiding convention used in Diagram (\ref{diag:AX-c-XA}) is that particles on the $\EE$-walls always sit on the top when we braid them with particles in the $\ED$-phase. Similarly, particles in $\EC$-phase always sit on the top when we braid them with particles on the $\EE$-wall. 
}
\end{rema}

\begin{pthm} \label{thm:main-1}
If a system of anyons, described by a (unitary) modular tensor category $\ED$, is obtained via a 2d condensation from another system of anyons given by a (unitary) modular tensor category $\EC$, and if this 2d condensation also produces a gapped domain wall between $\EC$ and $\ED$ consisting of all confined and deconfined particles, then we have the following results. 
\bnu
\item The vacuum in $\ED$ can be identified with a connected commutative normalized-special Frobenius algebra $A$ in $\EC$. The $\ED$-phase consists of all deconfined particles, which form the (unitary) modular tensor category $\EC_A^{loc}$, i.e. $\ED = \EC_A^{loc}$.
\item The confined and deconfined particles naturally accumulate on the domain wall and form the (unitary) spherical fusion categories $\EC_A$.
%\item there is an algebraic homomorphism $\iota_B^A: A \hookrightarrow B$ in $\EC$ such that $B$ is an algebra over $A$;
\item The bulk-to-wall map from the $\EC$-side is given by the monoidal functor (or better a central functor see Remark\,\ref{rem:central}): 
\be  \label{eq:L-tensor-A}
-\otimes A: \EC \to \EC_A, \quad\mbox{defined by}\quad C\mapsto C\otimes A, \quad\quad \forall C\in \EC. 
\ee 
\item The bulk-to-wall map from the $\ED$-side is given by the canonical embedding $\EC_A^{loc}\hookrightarrow \EC_A$, and deconfined particles in $\EC_A$ can move out of the wall to the $\ED$-side freely. 

\enu
\end{pthm}

\begin{rema} \label{rem:monoidal} {\rm
Note that $C\mapsto C\otimes A$ is just the mathematical way to say that, as the particle approaching the wall, it approaches the $A$-cloud of vacuum. Both bulk-to-wall maps are necessarily monoidal functors. It is the mathematical translation of the fact that fusing two bulk particles in the bulk first then moving onto the wall is distinguishable to moving onto the wall first then fusing them along the wall. %The functors $-\otimes A$ and $\EC_A^{loc}\hookrightarrow \EC_A$ are automatically monoidal. 
}
\end{rema}

\begin{rema} \label{rem:central} {\rm
The functor $-\otimes A: \EC \to \EC_A$ is more than a monoidal functor. It is actually a central functor \cite{dmno} (see Definition\,\ref{def:central}). This is not an accidental fact for $-\otimes A$ but a natural physical requirement for all bulk-to-wall maps $L:\EC \to \EE$ as illustrated schematically in the two diagrams in equation\,(3.4) in \cite{fsv}. We briefly recall the argument below. When an anyon $C$ in $\EC$-phase move to the $\EE$-wall closely enough, it can be viewed as a particle on the wall, and $C$ can be braided with a wall excitation $X$ in a unique way. Namely, it can exchange positions with $X$ as long as the path of $C$ is in $\EC$-bulk and the path of $X$ is restricted on the wall. This braiding is only a half-braiding. It implies that $L$ is a central functor, i.e. $L$ factors as $(\EC \to Z(\EE) \xrightarrow{\text{forget}} \EE)$. This physical requirement is automatically satisfied for $L=-\otimes A$. Indeed, mathematically, there is a natural half braiding given by
\be \label{eq:half-braiding-A}
(C\otimes A)\otimes_A X \simeq C\otimes X \xrightarrow{c_{C,X}} X\otimes C \simeq X \otimes_A (C\otimes A)
%X \otimes_B (C\otimes B) \simeq X\otimes C \xrightarrow{c_{C,X}^{-1}} C\otimes X \simeq (C\otimes B)\otimes_B X
\ee
for $C\in \EC, X\in \EE$, satisfying all the coherence conditions of a central functor. Notice that the braiding is chosen according to the convention in Remark\,\ref{rem:braiding-convention}. Similarly, the bulk-to-wall map $\overline{\EC_A^{loc}} \hookrightarrow \EC_A$, where $\overline{\EC_A^{loc}}$ is the same fusion category as $\EC_A^{loc}$ but with braidings replaced by anti-braidings, is also a central functor.  
}
\end{rema}

\begin{figure}
$$
\begin{tikzpicture}
\fill[blue!30] (-2,0) rectangle (2,3) ;
\fill[green!30] (0,1.5) circle (1);
\draw[fill=black] (-0.04,2.45) rectangle (0.05,2.55) node[midway,above] {\footnotesize $X$} ;
\node at (-1.5,2.5) {$\EC$} ;
\node at (0,1.2) {$\ED$} ;
\node at (-1.1,1) {$\EE$} ;
\draw [blue, ultra thick] (0,1.5) circle [radius=1] ;
\draw[densely dashed,-stealth] (0,2) .. controls (0.2,2) and (0.3,2.3) ..(0.07,2.45) ;
\draw[densely dashed,-stealth] (0,2) .. controls (-0.2,2) and (-0.3,2.3) .. (-0.07,2.45) ;
\node at (0.2,1.8) {\scriptsize $A=\one_\ED$} ;
\end{tikzpicture}
$$
\caption{This picture show that the left $A$-action on a particle $X$ on the wall should be compatible with the right $A$-action.}
 \label{fig:left-right-A-module}
 \end{figure}

By \cite{DMNO13}, we have $Z(\EC_A) =\EC\boxtimes \overline{\EC_A^{loc}}$, where $Z(\EC_A)$ denotes the Drinfeld center of $\EC_A$. When $\EC_A^{loc}$ is trivial, i.e. $\EC_A^{loc}=\vect$, where $\vect$ denotes the category of finite dimensional vector spaces, we reobtain the boundary-bulk relation: $\EC=Z(\EE)$ for a gapped boundary $\EE=\EC_A$. 

\begin{rema} {\rm
The boundary-bulk relation, i.e. ``the bulk is the center of a boundary'' was first discovered in \cite{KK12} based on the Levin-Wen type of lattice models with gapped boundaries. It was noticed model-independently in \cite{FSV13} that the bulk MTC necessarily factors through the Drinfeld center of a boundary fusion category. If we agree that a gapped boundary of a 2d topological order is necessarily the result of a 2d condensation, then Theorem${}^{\mathrm{ph}}$\,\ref{thm:main-1} provided the first complete model-independent proof of the boundary-bulk relation for 2d topological orders with gapped boundaries. %A more general boundary-bulk relation for both gapped/gapless boundaries and for potentially gapless quantum liquid bulk phases in all dimensions was later proposed and proved in \cite{KWZ15,KWZ17}. 
}
\end{rema}

\subsection{1d Condensations} \label{sec:1d-condensation}

In this subsection, we show that more general types of gapped domain wall between $\EC$ and $\ED$ can occur. They can have vacuums that are different from $A$. Now we assume that the vacuum of the $\EE$-wall is $B$. As we will see later, this general case can be all obtained from a 2d condensation followed by a 1d condensation.  

\smallskip
All objects in $\EE$ should come from objects in $\EC$. These objects partially survive the condensation but are confined to live on the 1d wall. It is reasonable to view $\EE$ as a 1d phase condensed from $\EC$ such that the particles in it can only live on a 1d line. Once we take this point of view, many basic building blocks in $\EE$ can be analyzed similar to those in $\ED$. 
\bnu
\item $\EE$ is a subcategory of $\EC$. If $X\in \EE$, we must have the identity condensation map $\id_X: X \to X$. The vacuum $\one_\EE$ in $\EE$ can be viewed as an object $B$ in $\EC$. From the bootstrap point of view, it seems unnatural to take $B=A$ as a priori. %We would prefer to start our bootstrap study with minimum assumptions. We see later that bootstrap study tell us some relation between $A$ and $B$. 

\item We must have an embedding: $\hom_\EE(M, N) \hookrightarrow \hom_\EC(M, N)$
\item The vacuum $\one=\one_\EC$ should condense into $B$. Namely, there are a morphism $\iota_B: \one \to B$ in $\EC$ and a multiplication $\mu_B: B\otimes B \to B$. They endow $B$ with a structure of an algebra in $\EC$. 

On the other side, the vacuum $\one_\ED=A$ in $\ED$-phase should also fuse into the vacuum on the wall when we move the vacuum $A$ close to the wall. Therefore, we have an morphism $\iota_B^A: A \to B$ in $\EC$. The physical intuition immediately suggests that the following diagrams:
\be
\xymatrix{
A\otimes A \ar[r]^{\mu_A} \ar[d]_{\iota_B^A \iota_B^A} & A \ar[d]^{\iota_B^A} \\
B\otimes B \ar[r]^{\mu_B}  & B
}
\ee
is commutative. It is also natural that $\iota_B^A \circ \iota_A = \iota_B$. Therefore, $\iota_B^A: A \to B$ is an algebra homomorphism. Since $A$ is simple $A$-$A$-bimodule, it implies that $\iota_B^A$ is an embedding. 
%As a consequence, the wall excitations $(\EC_A)_{B|B}$ can be embedded into the category $\EC_A$. 

\item {\it $B$ is an algebra over $A$} (see Definition\,\ref{def:B-over-A}): Consider a $\ED$-vacuum $A$ and an $\EE$-vacuum $B$, then let $A$ fuse into the wall, then fuse with $B$ from either left or right side. Physically, two possible paths give the same fusion process $A\otimes B\to B$. This leads to the following conditions:  
\be  \label{diag:B-over-A}
\xymatrix{
A\otimes B \ar[r]^{\iota_B^A1} \ar[d]_{c_{B,A}^{-1}} & B\otimes B \ar[r]^{\mu_B} & B \\
B\otimes A \ar[r]^{1\iota_B^A} & B\otimes B \ar[ur]_{\mu_B} & 
}
\ee
Such algebra $B$ is called an algebra over $A$. It is equivalent to say that $B$ is an algebra in the monoidal category $\EC_A$ of right $A$-modules. 

Moreover, every object $X$ in $\EE$ is equipped with a structure of $A$-$A$-module structure such that the left $A$-action is compatible with the right $A$-action as illustrated in Figure\,(\ref{fig:left-right-A-module}). Mathematically, it means that in the following diagram
\be \label{diag:AXA}
\xymatrix{
A\otimes X \ar[r]  \ar[d]_{c_{X,A}^{-1}} &  X \\
X\otimes A \ar[ru] & 
}
\ee
is commutative. In other words, an object $X$ in $\EE$ is automatically an object in $\EC_A$. 

\enu

These are the basic data associated to the particles on the wall. We explore their properties below. 

\begin{itemize}
\item {\it As an algebra in $\EC_A$}: Consider the process of condensing three $B$-anyons, this process is independent of which pair of $B$ condenses first. This leads to a the same commutative diagram as (\ref{diag:asso}) but with $A$ replaced by $B$. Moreover, the multiplication map $\mu_B: B\otimes B\to B$ should also be compatible with $A$-actions. By the universal property of $\otimes_A$, it means that $\mu_B$ should factors through as follows: 
\be
\xymatrix{ B\otimes B \ar[r]^{\otimes_A}  \ar[dr]_{\mu_B} & B\otimes_A B \ar[d]^{\mu_B^A} \\
& B 
}
\ee
As a consequence, $\mu_B^A$ and $\iota_B^A$ endow $B$ with a structure of an algebra in $\EC_A$.

\item {\it Stability of the vacuum $B$ in $B\otimes_A B$ under the $B$-action}: By the same argument of the stability of the vacuum $A$, we obtain the stability of the vacuum $B$ which implies that both maps in (\ref{eq:A-L-stable}) and (\ref{eq:A-R-stable}) with $A$ replaced by $B$ are zero maps in $\EC_A$. This further implies that $e_B$ is a $B$-$B$-bimodule map in $\EC$. Moreover, since $e_B$ is necessarily compatible with $A$-actions, we obtain a new morphism $e_B^A:=(B\to B\otimes B \to B\otimes_A B)$, which is necessarily a $B$-$B$-bimodule map in $\EC_A$. Therefore, $B$ is an separable algebra in $\EC_A$. As a consequence, the category $(\EC_A)_B$ of $B$-modules in $\EC_A$ and the category $(\EC_A)_{B|B}$ of $B$-$B$-bimodules in $\EC_A$ are both semisimple. 

\item {\it Connectivity of $B$}: A disconnected separable algebra decomposes into direct sum of connected separable algebras. If $B$ is disconnected in $\EC_A$, the category $(\EC_A)_{B|B}$ is a multifusion category. As we will show later that $\EE=(\EC_A)_{B|B}$. Therefore, $B$ must be connected, i.e. $\dim \hom_{\EC_A}(A, B) =1$. Since $\hom_\EC(\one,B)=\hom_{\EC_A}(A, B)=\Cb$, $B$ is also connected in $\EC$. 

\end{itemize}

By similar arguments above Remark\,\ref{rema:unitary-A}, it is easy to see that $B$ is a connected symmetric normalized-special Frobenius algebra in $\EC_A$ with the multiplication $\mu_B^A$, the unit $\iota_B^A$ and the comultiplication $\Delta_B^A :=e_B^A$.

\begin{rema} {\rm
In the unitary case, $\EC_A$ is a unitary fusion category, and $B$ is necessary a connected symmetric normalized-special $\ast$-Frobenius algebra in $\EC_A$ (i.e. $\Delta_B^A=(\mu_B^A)^\ast$). 
}
\end{rema}

Note that $B$ is not necessarily commutative. But its other properties are similar to that of $A$. It defines a a new type of condensation but confined to the 1d domain wall. It motivates us to introduce the following simplified terminology. 
\begin{defn} {\rm
In a (unitary) fusion category $\EuScript{A}$, a {\it 1d-condensable algebra} is a connected symmetric normalized-special Frobenius ($\ast$-Frobenius) algebra. 
}
\end{defn}

Above bootstrap results can be summarized as follows. 
\begin{plemma} \label{lemma:B=algebra}
The vacuum $\one_\EE$ on the wall can be viewed as a 1d-condensable algebra $B$ in $\EC_A$.  
\end{plemma}

The remaining condensation data are given below. 
\bnu
\item[$5$] For any $X, Y\in \EE$, There should be a condensation map: $\rho_{X,Y}^\EE: X\otimes Y \to X\otimes_\EE Y$, which is necessarily factors through $\rho_{X, Y}^A: X \otimes_A Y \to X \otimes_\EE Y$, i.e. 
$$
\xymatrix{
X\otimes Y \ar[r] \ar[dr]_{\rho_{X,Y}^\EE} & X\otimes_A Y \ar[d]^{\rho_{X,Y}^A} \\
& X\otimes_\EE Y
}
$$ 
Similarly, we have a canonical embedding $e_{X,Y}^\EE: X\otimes_\EE Y \to X\otimes_A Y$ such that
$$
\rho_{X, Y}^A \circ e_{X, Y}^A =\id_{X\otimes_\EE Y}. 
$$
Since $\EC$ is semisimple, we can have a decomposition: $X\otimes_A Y = X\otimes_\EE Y \oplus U$ for some $U\in \EC$. In other words, we have $e_{X,Y}^{U|\EE}: U \to X\otimes_A Y$ and $r_{X,Y}^{U|\EE}: X\otimes Y \to U$ such that 
\be
r_{X,Y}^{U|A} \circ e_{X,Y}^{U|A} =\id_U, \quad\quad \rho_{X,Y}^A \circ e_{X,Y}^{U|A}=0, \quad\quad  r_{X,Y}^{U|A} \circ  e_{X,Y}^\EE=0
\ee
\be
e_{X,Y}^A \circ \rho_{X, Y}^A + e_{X,Y}^{U|A} \circ r_{X,Y}^{U|A} =\id_{X\otimes Y}. 
\ee
In particular, we define $\mu_B^A:=\rho_{B,B}^A$ and $e_B^A:=e_{B,B}^A$. 

\enu

\begin{rema} {\rm
In the unitary setting, we can choose $e_{X,Y}^A=(\rho_{X,Y}^A)^\ast$. 
}
\end{rema}

Now we would like to explore the properties of $\mu_X^L: =\rho_{B, X}^\EE: B\otimes_A X \to X=B\otimes_\EE X$ and $\mu_X^R:= \rho_{X, B}^\EE: X\otimes_A B \to X=X\otimes_\EE B$ for all $X\in \EE$. It is clear that the $B$-actions on $X$ is necessarily compatible with the $A$-actions. In other words, $\mu_X^L$ and $\mu_X^R$ are necessarily $A$-module maps, i.e. morphisms in $\EC_A$. 
\bnu
\item The pair $(X, \mu_X^L)$ gives a left $B$-module in $\EC_A$: the proof is entirely similar to that of left $A$-module. 
\item Similarly, $X$ equipped with a right $B$-action $\mu_X^R: X\otimes_A B \to X$ is a right $B$-module in $\EC_A$. 
\item The triple $(X, \mu_X^L, \mu_X^R)$ defines a $B$-$B$-bimodule in $\EC_A$. This follows from the following commutative diagram:
$$
\xymatrix{
(B\otimes_A X) \otimes_A B \ar[rr]^\simeq \ar[d]_{\mu_X^L1} &  & B\otimes_A (X \otimes_A B) \ar[d]^{1\mu_X^R} \\
(B\otimes_\EE X) \otimes_A B \ar[d]_{\mu_{B\otimes_\EE X}^R} & & B\otimes_A (X \otimes_\EE B)
\ar[d]^{\mu_{X\otimes_\EE B}} \\
(B\otimes_\EE X) \otimes_\EE B \ar@{=}[r] & X \ar@{=}[r] & B\otimes_\EE (X \otimes_\EE B),
}
$$
the physical meaning of which is obvious. 

\item {\it Stability of $X$ in $B\otimes_A X$ and $X\otimes_A B$ under the $B$-action}: Notice first that, by the associativity, it is automatically true that the map $\mu_X^L$ is a left $B$-module map and $\mu_X^R$ a right $B$-module map for all $X \in \EE$. Similar to the previously discussed stabilities, we obtain that $e_{B, X}^A$ is a left $B$-module map and $e_{X,B}^A$ a right $B$-module map.

\item {\it Compatibility among $e_B^A$, $e_{B,X}^A$ and $e_{X,B}^A$}: Similar to the compatibility among $e_A$, $e_{A,M}$ and $e_{M,A}$, we can show that the same diagrams (\ref{diag:e-A-eAM-eMA}) but with all $A$ replaced by $B$ and all $M$ by $X$ are commutative due to the same physical requirements. As a consequence, using the Frobenius properties of $\Delta_B^A$, we obtain the following identities: 
\be  \label{eq:e_M-e_B}
  e_{B,X}^A~=~ 
  \raisebox{-24pt}{
  \begin{picture}(65,55)
   \put(8,8){\scalebox{.75}{\includegraphics{pic-eAM-eps-converted-to}}}
   \put(8,8){
     \setlength{\unitlength}{.75pt}\put(-30,-38){
     \put(55, 28)  {\scriptsize $ X $}
     \put(55,94)  {\scriptsize $ X $}
     \put(26, 91)  {\scriptsize $ B $}
     }\setlength{\unitlength}{1pt}}
  \end{picture}}~,
  \quad\quad\quad
 e_{X,B}^A~=~ 
  \raisebox{-24pt}{
  \begin{picture}(65,55)
   \put(8,8){\scalebox{.75}{\includegraphics{pic-eMA-eps-converted-to}}}
   \put(8,8){
     \setlength{\unitlength}{.75pt}\put(-30,-38){
     \put(26, 28)  {\scriptsize $ X $}
     \put(26,94)  {\scriptsize $ X $}
     \put(55, 91)  {\scriptsize $ B $}
     }\setlength{\unitlength}{1pt}}
  \end{picture}} ~.
\ee

\item Morphisms in $\EE$ are $B$-$B$-bimodule maps in $\EC_A$: A morphism $f: X\to Y$ in $\EE$ should be stable under the screening of the vacuum $B$ from both sides. In other words, we should have the following two commutative diagrams: 
\be  \label{diag:screen-map-B-L}
\xymatrix{
B\otimes_A X \ar[rr]^{1f} &  & B\otimes_A Y  \ar[d]^{\mu_Y^L} \\
B\otimes_\EE X = X \ar[rr]^f \ar[u]_{e_{X,B}^A} & & Y = B\otimes_\EE Y.
}
\ee
\be \label{diag:screen-map-B-R}
\xymatrix{
X\otimes_A B \ar[rr]^{f1} & & Y\otimes_A B  \ar[d]^{\mu_Y^R} \\
X\otimes_\EE B = X \ar[rr]^f \ar[u]_{e_{B,X}^A} && Y = Y\otimes_\EE B.
}
\ee
Since $e_{B,X}^A, \mu_Y^L$ are left $B$-module maps and $e_{X,B}^\EE, \mu_Y^R$ are right $B$-module maps in $\EC_A$, we obtain that $f$ is a $B$-$B$-bimodule map if and only if diagrams in (\ref{diag:screen-map-B-L}) and (\ref{diag:screen-map-B-R}) are commutative. Namely, we have $\hom_\EE(X,Y) = \hom_{(\EC_A)_{B|B}}(X,Y)$, where $\hom_{(\EC_A)_{B|B}}(X,Y)$ denotes the set of $B$-$B$-bimodule maps from $X$ to $Y$ in $\EC_A$. 

Similar to (\ref{fig:screening-op}), we have a {\it screening map}: $\text{Sc}_B: \hom_{\EC_A}(X,Y) \to \hom_{(\EC_A)_{B|B}}(X, Y)$ defined by, for $g\in \hom_{\EC_A}(X,Y)$,
\be \label{fig:screening-map-B}
\mathrm{Sc}_B(g) ~:=~ \quad 
\raisebox{-57pt}{
  \begin{picture}(70, 122)
   \put(0,8){\scalebox{1}{\includegraphics{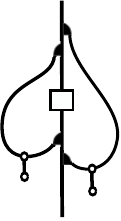}}}
   \put(0,8){
     \setlength{\unitlength}{.75pt}\put(-18,-19){
     \put( 16, 105)      {\scriptsize $ B $}
     \put( 85, 105)      {\scriptsize $B$}
     \put(55,95)      {\scriptsize $g$}
     \put(28, 68)      {\scriptsize $ \Delta_B $}
     \put(67, 62)       {\scriptsize $ \Delta_B$}
     \put(52, 10)      {\scriptsize $X$}
     \put(55,164)      {\scriptsize $Y$}
     }\setlength{\unitlength}{1pt}}
  \end{picture}}
\ee
Using the normalized-specialness of the Frobenius algebra $B$, it is easy to see that $\text{Sc}_B$ is a projector, i.e. $\text{Sc}_B \circ \text{Sc}_B = \text{Sc}_B$. 

\item $\otimes_\EE = \otimes_B$: the category $(\EC_A)_{B|B}$ of $B$-$B$-bimodules is a tensor category with tensor product $\otimes_B$. We have $B\otimes_B X = X = B\otimes_\EE X$. Moreover, the condensation cannot distinguish the following condensations: $(X\otimes_A B)\otimes_\EE Y$ and $X \otimes_A (B\otimes_\EE Y)$. By the same argument for $\otimes_\ED =\otimes_A$, we obtain that $\otimes_\EE = \otimes_B$. 

\enu
Using \cite[Lem.\,4.1]{defect-tft}, one can show that $(\EC_A)_{B|B}$ is a spherical multi-fusion category. As a consequence, we must have $\EE=(\EC_A)_{B|B}$. Moreover, $B$ is necessarily a connected algebra in $\EC_A$, i.e. $\dim \hom_{\EC_A}(A,B)=1$, in order for $(\EC_A)_{B|B}$ being a fusion category. 

\begin{pthm}
$\EE = (\EC_A)_{B|B}$ as spherical fusion categories. 
\end{pthm}

\begin{rema} {\rm
If $\EC$ is unitary, we choose $e_{X,Y}^A=(\rho_{X,Y}^A)^\ast$ and $\epsilon_B^A := (\iota_B^A)^\ast$. Then $B$ is automatically a connected symmetric normalized-special $\ast$-Frobenius algebra. Similar to algebra $A$, we can show that $\text{Sc}_B$ commutes with $\ast$. We obtain that $(\EC_A)_{B|B}$ is a $\ast$-category. It is a routine to check that $(\EC_A)_{B|B}$ is a unitary fusion category, which has a unique spherical structure \cite{kitaev1,eno02}. 
}
\end{rema}

Actually, what we have proven is more general than the bootstrap setting in this subsection. 
\begin{pthm} \label{thm:1d-condensation}
Consider a 1d anomalous topological order, such as a gapped domain wall between two 2d topological orders. Particles in it form a (unitary) fusion category $\EA$. If a 1d condensation occurs in this 1d phase, then the vacuum of the new phase is given by a 1d-condensable algebra $P$ in $\EA$ and particles in the new phase form a new (unitary) fusion category $\EA_{P|P}$ (of $P$-$P$-bimodules in $\EA$) with tensor product $\otimes_P$ and the tensor unit $P$. 
\end{pthm}

\begin{rema} {\rm
In the setting of Theorem${}^{\mathrm{ph}}$\,\ref{thm:1d-condensation}, if the 1d condensation occurs in a 1d region in the $\EA$-phase, it also produces a 0d domain wall between the $\EA$-phase and the $\EA_{P|P}$-phase. The mathematical description of this 0d domain wall is given by a pair $(\EA_P, Y)$, where $\EA_P$ is the category of right $P$-modules in $\EA$ and $Y$ is a distinguished object in $\EA_P$. Moreover, a particle $a\in \EA$ moves onto the 0d wall according to $a \mapsto a\otimes Y$, and a particle $Q\in \EA_{P|P}$ moves onto the 0d wall according to $Q\mapsto Y\otimes_P Q$. We leave it as an exercise. 
}
\end{rema}

Now we summarize all results from our bootstrap analysis as follows. 
\begin{pthm} \label{thm:main-2}
If a system of anyons, described by a (unitary) MTC $\ED$, is obtained from another system of anyons given by a (unitary) MTC $\EC$ via a 2d condensation, and if a 1d gapped domain wall is produced as a result of this 2d condensation and possibly an additional 1d condensation, then the following statements are true. 
\bnu
\item The vacuum in $\ED$ is given by a 2d-condensable algebra $A$ in $\EC$ and $\ED = \EC_A^{loc}$ as (unitary) MTC's.
\item The vacuum on the wall is given by a 1d-condensable algebra $B$ in $\EC_A$, and particles on the wall form the (unitary) spherical fusion category $(\EC_A)_{B|B}$.
%\item there is an algebraic homomorphism $\iota_B^A: A \hookrightarrow B$ in $\EC$ such that $B$ is an algebra over $A$;
\item The bulk-to-wall map from $\EC$-side is given by the central functor 
\be  \label{eq:L-tensor-B-1}
-\otimes B: \EC \to (\EC_A)_{B|B}, \quad\mbox{defined by}\quad C\mapsto C\otimes B, \quad\quad \forall C\in \EC. 
\ee 
\item The bulk-to-wall map from $\ED$-side is given by the central functor
\be
B\otimes_A -: \overline{\EC_A^{loc}}\to (\EC_A)_{B|B}, \quad\mbox{defined by}\quad M\mapsto B\otimes_A M, \quad \forall M \in \EC_A^{loc}.
\ee
\enu
\end{pthm}

\begin{rema} {\rm
The physical intuitions behind the functors $-\otimes B$ and $B\otimes_A -$ are the same as the intuition behind the functor $-\otimes A$ in Theorem${}^{\mathrm{ph}}$\,\ref{thm:main-1} (recall Remark\,\ref{rem:monoidal}). Mathematically, both functors $-\otimes B$ and $B\otimes_A -: \overline{\EC_A^{loc}} \hookrightarrow \EC_A$ are indeed monoidal and central (recall Remark\,\ref{rem:central}). In particular, the ``centralness'' of the functor $-\otimes B: \EC \to (\EC_A)_{B|B}$ is defined by the following half-braidings
\be \label{eq:half-braiding-B}
(C\otimes B)\otimes_B X \simeq C\otimes X \xrightarrow{c_{C,X}} X\otimes C \simeq X \otimes_B (C\otimes B), \quad\quad \forall C\in \EC, X\in (\EC_A)_{B|B}. 
%X \otimes_B (C\otimes B) \simeq X\otimes C \xrightarrow{c_{C,X}^{-1}} C\otimes X \simeq (C\otimes B)\otimes_B X
\ee

}
\end{rema}

\begin{rema} {\rm
For a 1d-condensable algebra $B$ in $\EC_A$, $(\EC_A)_{B|B}$ is Morita equivalent to $\EC_A$ \cite{sc} (see Definition\,\ref{def:morita-cat}) and we have $Z((\EC_A)_{B|B}) \simeq Z(\EC_A)$ as MTC's \cite{mueger1}. Therefore, the results given in Theorem\,\ref{thm:main-2} also demonstrate the boundary-bulk relation, i.e. $\EC \boxtimes \overline{\EC_A^{loc}} \simeq Z((\EC_A)_{B|B})$. 
}
\end{rema}

\begin{figure}[t] 
$$
 \begin{picture}(180, 182)
   \put(-60,-10){\scalebox{1.5}{\includegraphics{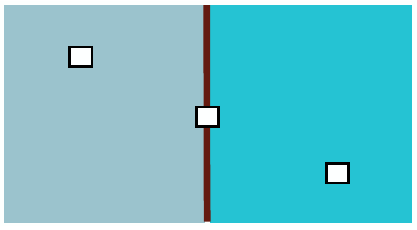}}}
   \put(-60,-10){
     \setlength{\unitlength}{.75pt}\put(-18,-19){
     \put(195,243)      { $\EE=(\EC_A)_{B|B} $-wall}
     \put(87, 180)     { $ C $}
     \put(25, 160)     { a bulk excitation} 
     \put(332,67)      { $ M $}
     \put(288, 47)     { a bulk excitation} 
     \put(208, 122)   { $ X $}
     \put(230,122)    { a wall excitation}
     \put(285, 215)     { $\ED=\EC_A^{loc}$-phase}
     \put(40, 40)     { $\EC$-phase}
     \put(135, 180)     { $\xrightarrow{-\otimes B}$ }
     \put(255, 68)      { $\xleftarrow{B\otimes_A -}$}
     }\setlength{\unitlength}{1pt}}
  \end{picture}
$$  
  \caption{This picture illustrate the results in Theorem${}^{\mathrm{ph}}$\,\ref{thm:main-2}.}
  \label{fig:bs-result}
\end{figure}

\section{Determining the condensation from physical data} \label{sec:find-A-B}

If we only have the abstract data of the initial phase $\EC$, the condensed phase $\ED$ and $\EE$-domain wall, how can we determine the 2d-condensable algebra $A$ and the algebra $B$? Are such algebras $A$ and $B$ unique? If not, is it possible to add more macroscopic and physically detectable information so that we can determine $A$ and $B$ uniquely? We would like to answer these questions in this section. 

\subsection{Gapped boundaries} \label{sec:boundary}
We would like to first consider a special case when $\ED$-phase is trivial, i.e. $\ED=\vect$ (or $\hilb$ if we assume unitarity). In other words, we have a gapped boundary given by $\EE$. 

Let us first look at a simple example: the toric code model. In this case, the bulk excitations are given by the MTC $Z(\rep(\Zb_2))$, which is the monoidal center of the unitary fusion category $\rep(\Zb_2)$ of the representations of the $\Zb_2$-group. It contains four simple anyons $1,e,m,\epsilon$. There are two different types of boundary: the smooth boundary and the rough boundary \cite{bk}\cite{kk}. The boundary excitations in both cases are given by the same unitary fusion category $\rep(\Zb_2)$. The difference between these two types of boundary lies in how bulk anyons condense when they approach the boundary, i.e. the bulk-to-boundary map. This information is macroscopic and physically detectable. In the smooth  boundary case, $m$-particles are condensed but $e$-particles are confined on the boundary; in the rough boundary case, $m$-particles are confined on the boundary but $e$ particles are condensed. As we will show in Section\,\ref{sec:toric-code}, in the smooth boundary case, the associated 2d-condensable algebra is $1 \oplus m$; in the rough boundary case, the associated 2d-condensable algebra is $1 \oplus e$. In other words, the condensation can be uniquely fixed by specifying the bulk-to-boundary map, which is a monoidal functor $F: Z(\rep(\Zb_2)) \to \rep(\Zb_2)$. If the bulk-to-boundary map is not given as a data, then the compatible 2d condensations in general are not unique. This phenomenon carries on to the most general cases.

\smallskip
In general, for a bulk phase given by a MTC $\EC$ and a gapped boundary phase given by $\EE$, even if the bulk-to-boundary map is not given, we must have a braided monoidal equivalence $\EC \simeq Z(\EE)$. Therefore, there exists a (possibly not unique) monoidal functor given by $L: \EC \simeq Z(\EE) \xrightarrow{\text{forget}} \EE$. Let $L^\vee$ be the right adjoint of $L$ (see Definition\,\ref{def:right-adjoint}). By \cite{eno,dmno}, $L^\vee(\one_\EE)$ has a natural structure of a 2d-condensable algebra in $\EC$. Moreover, it is a Lagrangian algebra (see Definition\,\ref{def:lag}). By Theorem\,\ref{thm:lag}, we have $Z(\EC_{L^\vee(\one_\EE)}^{loc})\simeq \vect$. Moreover, since $L^\vee(X)$ is naturally a right $L^\vee(\one_\EE)$-module, we obtain a functor $L^\vee: \EE \xrightarrow{} \EC_{L^\vee(\one_\EE)}$, which was proved to be a monoidal equivalence \cite{eno}. Therefore, we can certainly realize an $\EE$-boundary via a 2d condensation but not necessarily in a unique way because $L$ is in general not unique. 

In general, for given $\EC$ and $\EE$, it is possible to have more than one bulk-to-boundary maps. Indeed, if $\beta: \EC \to \EC$ is a non-trivial braided equivalence and $L:\EC\to \EE$ is a bulk-to-boundary map, then $L \circ \beta: \EC \to \EE$ gives a different bulk-to-boundary map. Then $\beta^{-1}(L^\vee(\one_\EE))$ is also a Lagrangian algebra. The condensation of $\beta(L^\vee(\one_\EE))$ gives exactly the same boundary excitations, i.e. 
$$
\EC_{\beta(L^\vee(\one_\EE))}\simeq \EC_{L^\vee(\one_\EE)} \simeq \EE. 
$$
In the case of toric code model, the bulk excitation $Z(\rep(\Zb_2))$ has a $\Zb_2$ automorphism group. The non-trivial automorphism is called electric-magnetic duality which exchanges an $e$-particle with an $m$-particle (see also \cite{bcka} for more general dualities). Therefore, any one of two bulk-to-boundary maps in the toric code model (discussed before) can be obtained from the other by applying the electric-magnetic duality.

The bulk-to-boundary map $L: \EC \to \EE$ is a physically detectable data. Once it is given, then the associated 2d condensation is fixed. Both the gapped boundary $\EE$ and the bulk-to-boundary map can be recovered from the Lagrangian algebra $L^\vee(\one_\EE)$ in $\EC$ as shown by the following commutative diagrams: 
\be \label{diag:L-A}
\raisebox{20pt}{
\xymatrix{
\EC \ar[rr]^{-\otimes  L^\vee(\one_\EE)} \ar[rd]_L &  & \EC_{L^\vee(\one_\EE)} \ar[dl]^{\simeq}  \\
& \EE & }}
\ee
It says that not only the boundary excitations $\EE$ coincide with $\EC_{L^\vee(\one_\EE)}$, but also the associated bulk-to-boundary maps coincide. We summarize these results below.
\begin{pthm}
Given a 2d topological order $\EC$ together with a gapped boundary phase $\EE$ and a given bulk-to-boundary map, i.e. a central functor $L: \EC \to \EE$, there is a unique 2d condensation of $\EC$ determined by the Lagrangian algebra $L^\vee(\one_\EE)$. More precisely, we have $\EE \simeq \EC_{L^\vee(\one_\EE)}$ and $L$ coincides with the functor $-\otimes L^\vee(\one_\EE): \EC \to \EC_{L^\vee(\one_\EE)}$. In other words, the gapped boundaries of a 2d $\EC$-phase one-to-one correspond to the Lagrangian algebras in $\EC$.
\end{pthm}

\begin{expl} {\rm
Consider the Ising topological order with anyon $1,\psi, \sigma$ and the fusion rules:
\be \label{eq:ising}
\sigma \otimes \sigma = 1 \oplus \psi, \quad \sigma \otimes \epsilon = \sigma, \quad
\psi \otimes \psi = 1. 
\ee
We use $\text{Ising}$ to denote the corresponding unitary MTC. By double folding the  Ising topological phase along a line, we obtain a double layered system $\text{Ising}\boxtimes \overline{\text{Ising}}$ with a gapped boundary, boundary excitations on which are given by the unitary fusion category $\text{Ising}$. The bulk-to-wall map is given by the usual fusion product functor $\text{Ising}\boxtimes \overline{\text{Ising}} \xrightarrow{L=\otimes} \text{Ising}$. We have
\be  \label{eq:ising2-lagrange}
L^\vee(\one_{\text{Ising}}) = (1 \boxtimes 1) \oplus (\psi \boxtimes \psi) \oplus (\sigma \boxtimes \sigma),
\ee
and $(\text{Ising}\boxtimes \overline{\text{Ising}})_{L^\vee(\one_{\text{Ising}})} \simeq \text{Ising}$ as fusion categories. The algebraic structure on $L^\vee(\one_{\text{Ising}})$ is guaranteed by abstract nonsenses \cite{eno,cardy,dmno}. But an explicit construction in terms of chosen bases of hom spaces is available in literature (see for example \cite[Prop.\,4.1]{mueger}\cite[Lem.\,6.19]{cor}\cite[Prop.\,2.25]{cardy}). 
}
\end{expl}

\begin{rema} {\rm
In the Abelian Chern-Simons theory based on the MTC $\mathcal{C}(G,q)$, where $G$ is a finite abelian group and $q$ a non-degenerate quadratic form, there is a one-to-one correspondence between Lagrangian algebras in the category $\mathcal{C}(G,q)$ and Lagrangian subgroups of $G$ \cite[Thm.\,5.5]{fsv}. So in this case, we recover the main result in \cite{kas,levin,bjq}. 
}
\end{rema}

\begin{rema} {\rm
By the folding trick, a domain wall between a $\EC$-phase and a $\ED$-phase can be viewed as a boundary of a $\EC\boxtimes \overline{\ED}$-phase. Therefore, the 1d gapped domain walls are classified by Lagrangian algebras in $\EC\boxtimes \overline{\ED}$. In the case $\EC=\ED$, by \cite{morita,cardy}\cite[Prop.\,4.8]{dmno}, such domain walls are equivalently classified by indecomposable semisimple $\EC$-modules.
}
\end{rema}

\begin{rema} \label{rema:confusion} {\rm
It was known that a simple anyon in $\EC$ can split into two particles on the boundary \cite{bs,buss}. For example, in the case of an $(\text{Ising} \boxtimes \text{Ising})$-bulk with an $\text{Ising}$-boundary, the simple anyon $\sigma \boxtimes \sigma$ in the bulk is mapped to $\sigma \otimes \sigma = \one \oplus \psi$ (see (\ref{eq:ising})) on the boundary.
This phenomenon has no contradiction to the fact that $\EE$ can be viewed as a subcategory of $\EC$ (recall Remark\,\ref{rema:contradiction}). To compare $\EC$ with $\EE$, we need first map one to the other. But there are many ways to do this. To view $\EE$ as a subcategory of $\EC$, we use the forgetful functor in the first diagram in (\ref{diag:L-A}); to see that a simple anyon in $\EC$ can split into two particles on the boundary, we use the bulk-to-boundary functor $-\otimes L^\vee(\one_\EE)$ in the second diagram in (\ref{diag:L-A}). Note that these two functors are adjoints of each other. For example, in the topological order $\EC=\text{Ising}\boxtimes \overline{\text{Ising}}$, the condensation of $L^\vee(\one_{\text{Ising}})$ in (\ref{eq:ising2-lagrange}) produces the trivial phase $\ED =  \hilb$ and a gapped boundary $\text{Ising}$. The simple anyon $\sigma\boxtimes \sigma$ in $\EC$ maps to the gapped boundary via the functor $-\otimes L^\vee(\one_\EE)$ and becomes 
$$
(\sigma\boxtimes \sigma) \otimes A = ((\sigma \otimes \sigma)\boxtimes 1) \otimes A  = ((1 \oplus \psi)\boxtimes 1) \otimes A = A\oplus ((\psi \boxtimes 1) \otimes A),
$$ 
which is decomposable on the boundary, or equivalently, via the functor $L$ and becomes $L(\sigma \boxtimes \sigma) = \sigma \otimes \sigma = 1 \oplus \psi$. 
}
\end{rema}

If we allow a 2d condensation and a 1d condensation on the boundary to realize the same data: $\EC\xrightarrow{L} \EE\hookleftarrow \vect$. Note that $A$ can still be fixed uniquely as $L^\vee(\one_\EE)$. However, this data is not enough to fix $B$. As you can see from (\ref{diag:L-A}), any $L: \EC \to \EE$ is equivalent to the standard functor $-\otimes L^\vee(\one_\EE): \EC \to \EC_{L^\vee(\one_\EE)}$. In general, $B$ can be noncommutative thus cannot be $L^\vee(\one_\EE)$. It turns out that $L^\vee(\one_\EE)$ is only the left center $C_l(B)$ of $B$. We define this notion now. Let $Y$ be a $B$-bimodule. The following map:
\be \label{eq:B-center}
P_Y =  \raisebox{-40pt}{
\begin{picture}(54,80)
   \put(0,8){\scalebox{.75}{\includegraphics{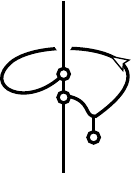}}}
   \put(0,8){
     \setlength{\unitlength}{.75pt}\put(-18,-19){
     \put(45, 10)  {\scriptsize $ Y $}
     \put(45,105)  {\scriptsize $ Y $}
     \put(78, 78)  {\scriptsize $ B $}
     \put(57,57)   {\scriptsize $B$}
     }\setlength{\unitlength}{1pt}}
  \end{picture}}\, ,
\ee
defines an idempotent from $Y$ to $Y$ \cite{tft1}. We define the left center of $Y$, denoted by $C_l(Y)$, to be the image of this map. In particular, for $Y=B$, we have $C_l(B) = \text{Im}\, P_B$. Choose a split $\iota_Y: C_l(Y) \to Y, r_Y: Y \to C_l(Y)$ such that $\iota_Y\circ r_Y=P_Y$ and $r_Y\circ \iota_Y = \id_{C_l(Y)}$. It is easy to show that the following two maps: 
for $f\in \hom_{B|B}(C\otimes B, Y)$ and $g \in \hom_\EC(C, C_l(Y))$, 
\bea
f & \mapsto & r_Y \circ f \circ (\id_C \otimes \iota_B), \nn
g & \mapsto & \mu_Y \circ (\iota_Y \otimes \id_B) \circ (g \otimes \id_B)
\eea
are well-defined and inverse to each other. Moreover, they define an natural isomorphism between the following two hom spaces:
\be
\hom_{(\EC_A)_{B|B}}( C\otimes B, Y) \simeq \hom_\EC(C, C_l(Y)).
\ee
Then it is clear from the Definition\,\ref{def:right-adjoint} of the right adjoint functor that we have $(-\otimes B)^\vee = C_l(-)$. Since $-\otimes B$ is dominant, by \cite[Lem.\,3.5]{dmno}, we have $(\EC_A)_{B|B} \simeq \EC_{C_l(B)}$ as fusion categories and the following commutative diagram.
$$
\xymatrix{
\EC \ar[rr]^{-\otimes C_l(B)} \ar[drr]_{-\otimes B} & & \EC_{C_l(B)} \ar[d]^{\simeq} \\
& & (\EC_A)_{B|B}
}
$$

Therefore, the bulk-to-boundary map $L: \EC \to \EE$ only determine the left center of $B$. Since there is no additional physically detectable data available to us. We can conclude that the algebra $B$ is not entirely physical. Only its left center, which is nothing but $A$ in this case, is physically detectable. Actually, in this case, the left center $C_l(B)$ coincides with the so-called full center of $B$ (\cite{unique}\cite{morita}). It is defined by $C_l(F^\vee(B))$, where $F: Z((\EC_A)_{B|B}) \to (\EC_A)_{B|B}$ is the forgetful functor, as an object in $Z((\EC_A)_{B|B}) \simeq \EC\boxtimes \vect = \EC$. The notion of a full center uniquely determines the Morita equivalent class of $B$ (see Def.\,\ref{def:morita-alg}) \cite[Thm.\,3.24]{morita}. Therefore, only the Morita class of $B$ is physical. Indeed, by definition, two 1d-condensable algebras $B_1$ and $B_2$ in the boundary fusion category $\EE$ are Morita equivalent if $\EE_{B_1} \simeq \EE_{B_2}$, which further implies $\EE_{B_1|B_1} \simeq \EE_{B_2|B_2}$. We won't be able to distinguish them by macroscopic physics.

\subsection{Non-trivial \texorpdfstring{$\ED$}{D}-phase}

We would like to answer the questions raised at the beginning of this section for the data 
$\EC \xrightarrow{L} \EE \hookleftarrow \ED$ for general $\ED$. If the condensation is purely 2d, by our bootstrap analysis, we have $\EE \simeq \EC_A \hookleftarrow \EC_A^{loc}\simeq \ED$ for
some algebra $A$ in $\EC$, and the functor $L: \EC \to \EC_A\simeq \EE$ is given by $-\otimes A$. Notice that the right adjoint of $-\otimes A$ is nothing but the forgetful functor $\EC_A \xrightarrow{\text{forget}} \EC$. This means that $A$ can be recovered from $(-\otimes A)^\vee(\one_\EE)$. 

Therefore, if we start from an abstract data $\EC \xrightarrow{L} \EE \hookleftarrow \ED$, one can immediately recover $A$ as $L^\vee(\one_\EE)$. By \cite[Lem.\,3.5]{dmno}, $L^\vee(\one_\EE)$ has a canonical structure of a 2d-condensable algebra. In this case, by \cite[Lem.\,3.5]{dmno}, we obtain that $\EE \simeq \EC_{L^\vee(\one_\EE)}$ as monoidal categories. Moreover, we have the following commutative diagram (recall (\ref{diag:L-A})):
$$
\xymatrix{
\EC  \ar[rr]^{-\otimes L^\vee(\one_\EE)} \ar[rd]_L & & \EC_{L^\vee(\one_\EE)} \ar[dl]^{(L^\vee)^{-1}}_{\simeq} \\
& \EE &
}
$$
Therefore, the bulk-to-wall map $L$ can be identified with the standard bulk-to-wall map $-\otimes L^\vee(\one_\EE)$. The category $\ED$ is just $\EC_{L^\vee(\one_\EE)}^{loc}$. %Since we have already assume that the condensation is 2d, i.e. $B=L^\vee(\one_\EE)$, the information $\EE \hookleftarrow \ED$ is redundant. %Actually, we will show in the next subsection that $L^\vee(\one_\EE)$ can also be recovered in a different way by using the data $\ED=\EC_{L^\vee(\one_\EE)}^{loc} \hookrightarrow \EE \simeq \EC_{L^\vee(\one_\EE)}$.

If we do not know whether the condensation associated to the physical data $\EC \xrightarrow{L} \EE \hookleftarrow \ED$ is purely 2d or a mixture of a 2d condensation and a 1d condensation, then, by the same arguments in Section\,\ref{sec:boundary}, $B$ cannot be uniquely determined by the physical data. But the Morita class of $B$ is uniquely determined.

\begin{rema} \label{rema:confusion2} {\rm
In physics literature (see for example \cite{bs,buss}), a simple anyon in the initial phase $\EC$ can be decomposable in the condensed phase $\ED=\EC_A^{loc}$. This phenomenon is similar to the one discussed in Remark\,\ref{rema:confusion}. To compare two categories, we need specify a functor between them. There is a forgetful functor $F: \EC_A^{loc} \hookrightarrow \EC$, which can be realized as the composition of the following two functors:
$$
F: \EC_A^{loc} \xrightarrow{R} \EC_A \xrightarrow{L^\vee = \, \mathrm{forget}} \EC.
$$
Its adjoint $F^\vee: \EC \to  \EC_A^{loc}$ is given by $F^\vee = R^\vee \circ L$. In other words, both functor $F$ and $F^\vee$ are the wall-tunneling maps between the $\EC$-phase and the $\ED$-phase. Note that even though there is no wall in the original setup, the 2d condensation choose the wall automatically. For a simple anyon $i$ in $\EC$, $F^\vee(i)$ is not simple in $\ED$ in general; for a simple anyon $M$ in $\EC$, $F(M)$ is not simple in $\EC$ in general. When physicists discuss the phenomenon of the splitting (in $\ED$-phase) of a simple anyon in $\EC$-phase, they applied the functor $F^\vee$ implicitly. 
For example, when $\EC=\text{Ising} \boxtimes \overline{\text{Ising}}$, take the 2d-condensable algebra $A = (1\boxtimes 1) \oplus (\psi \boxtimes \psi)$, which is a subalgebra of (\ref{eq:ising2-lagrange}) , then 
$$
L: \sigma \boxtimes \sigma \mapsto (\sigma \boxtimes \sigma) \otimes A = (\sigma \boxtimes \sigma) \oplus (\sigma \boxtimes \sigma).
$$
Note that $\sigma \boxtimes \sigma$ is a simple local $A$-module, thus can move freely into $\ED=\EC_A^{loc}$. Then we see that $F^\vee(\sigma \boxtimes \sigma)=(\sigma \boxtimes \sigma) \oplus (\sigma \boxtimes \sigma)$ is not simple, even though $\sigma \boxtimes \sigma$ is simple in $\EC$. Note that the splitting of $\sigma \boxtimes \sigma$ in the $\ED$-phase was studied in \cite{buss}, where the two summands in $F^\vee(\sigma \boxtimes \sigma)$ were denoted by $(\frac{1}{2}, \frac{1}{2})_0$ and $(\frac{1}{2}, \frac{1}{2})_1$, respectively.
}
\end{rema}

\section{Examples}  \label{sec:example}

In this section, we give some examples of 2d condensations in non-chiral and chiral topological phases. Recently, the gapped boundaries and domain walls have been studied intensively from various perspectives (see for example \cite{bsw,kk,fsv,levin,bjq,ww,kapustin,fsv2}).

\subsection{Toric code model}  \label{sec:toric-code}

In the toric code model \cite{kitaev}, the bulk excitations are given by the MTC $Z(\rep(\Zb_2))$, which is the monoidal center of the fusion category $\rep(\Zb_2)$. It contains four simple anyons $1,e,m,\epsilon$ with the following fusion rules:
$$
e\otimes e = m\otimes m = \epsilon \otimes \epsilon = 1, \quad e\otimes m = \epsilon. 
$$ 
There are two different types of boundary: a smooth boundary and a rough boundary \cite{bk}. Both boundaries have two boundary particles with the same fusion rule as $\rep(\Zb_2)$. However, these two boundaries are different and the difference can be detected by how bulk anyons approach the boundary. 

%They are characterized by two $\rep(\Zb_2)$-modules \cite{kk}. By \cite{kk}, the smooth boundary, viewed as a Levin-Wen type of lattice model, is defined by a boundary lattice associated to the $\rep(\Zb_2)$-module $\rep(\Zb_2)$ (see Definition\,\ref{def:C-mod}); for the rough boundary, the boundary lattice is defined by the $\rep(\Zb_2)$-module $\vect$.  The boundary excitations in these two cases are given, respectively, by $\text{Fun}_{\rep(\Zb_2)}(\rep(\Zb_2), \rep(\Zb_2))$ and $\text{Fun}_{\rep(\Zb_2)}(\vect, \vect)$, which are equivalent as fusion categories, i.e.
%$$ \text{Fun}_{\rep(\Zb_2)}(\rep(\Zb_2), \rep(\Zb_2)) \simeq \rep(\Zb_2) \simeq  \text{Fun}_{\rep(\Zb_2)}(\vect, \vect).$$ The difference of these two boundaries can be detected by how bulk anyons approach the boundary. 

In the case of smooth boundary, it was shown in \cite{bk} that when an $m$-anyon moves from the bulk to the boundary it simply disappeared or condensed. In this case, the associated 2d-condensable algebra $A_1$ is given by 
$$
A_1:= 1\oplus m.
$$ 
The boundary fusion category is given by $Z(\rep(\Zb_2))_{A_1}$, which is monoidally equivalent to $\rep(\Zb_2)$. According to Theorem\,\ref{thm:main-1}, the bulk-to-boundary map is given by the monoidal functor $-\otimes A_1: Z(\rep(\Zb_2)) \to Z(\rep(\Zb_2))_{A_1}$. Indeed, under this functor, we have
$$
1\mapsto 1\otimes (1\oplus m) = 1\oplus m, \quad\quad 
m \mapsto m \otimes (1 \oplus m) = 1\oplus m, 
$$
$$
e\mapsto e \otimes (1\oplus m) = e \oplus \epsilon, \quad\quad
\epsilon \mapsto \epsilon \otimes (1\oplus m) = \epsilon \oplus e. 
$$
Clearly, $m$ is mapped to the vacuum of the boundary. Even though the object $e\oplus \epsilon$ is not simple in $\rep(\Zb_2)$, it is the only simple right $A_1$-module other than $A_1$. Notice that $e\oplus \epsilon$ as an $A_1$-module is not local. So $1\oplus m$ is the only simple object in $Z(\rep(\Zb_2))_{A_1}^{loc}$, i.e. $Z(\rep(\Zb_2))_{A_1}^{loc} \simeq \vect$. $1\oplus m$ and $e\oplus \epsilon$ are the simple excitations on the smooth boundary. Their fusion products: 
$$
(1\oplus m) \otimes_{A_1} (e\oplus \epsilon) =  (e\oplus \epsilon) \otimes_{A_1} (1\oplus m) =
(e\oplus \epsilon),
$$
$$
(e\oplus \epsilon) \otimes_{A_1} (e\oplus \epsilon) = 1\oplus m
$$
coincide with those in $\rep(\Zb_2)$. Moreover, $Z(\rep(\Zb_2))_{A_1}\simeq \rep(\Zb_2)$
as fusion categories. 

The case of rough boundary is entirely similar. In this case, the 2d-condensable algebra is given by $A_2:= 1\oplus e$, which is Lagrangian, i.e. $Z(\rep(\Zb_2))_{A_2}^{loc} = \vect$. The boundary excitations are given by the fusion category
$Z(\rep(\Zb_2))_{A_2}$, which is also monoidally equivalent to $\rep(\Zb_2)$. The bulk-to-boundary map is given by the monoidal functor $-\otimes_{A_2}: Z(\rep(\Zb_2)) \to Z(\rep(\Zb_2))_{A_2}$, in which
$$
e \mapsto e\otimes (1\oplus e) = 1\oplus e, \quad\quad m \mapsto m\otimes (1\oplus e) = m\oplus \epsilon. 
$$
The boundary excitations and bulk-to-boundary maps associated to these two different types of boundaries are related by an EM-duality (i.e. exchanging $e$ with $m$).

\subsection{Levin-Wen types of lattice models}

The toric code model with boundaries is just one of a large family of Levin-Wen type of lattice models constructed in \cite{lw-mod}\cite{kk}. In these models, a bulk lattice is defined by a spherical fusion category $\EC$ and the boundary lattice is defined by a $\EC$-module $\EM$ (or ${}_\EC\EM$ if we want to make the $\EC$-action explicit) (see Definition\,\ref{def:C-mod}). In this case, the boundary excitations are given by the category $\EC_\EM^\vee:=\text{Fun}_\EC(\EM, \EM)^\rev$, where $\text{Fun}_\EC(\EM, \EM)^\rev$ denotes 
the category of $\EC$-module functors from $\EM$ to $\EM$ (see Definition\,\ref{def:mod-functor}) but with a reversed tensor product $\otimes^\rev$ defined by $x \otimes^\rev y := y\otimes x$ for $x,y\in \text{Fun}_\EC(\EM, \EM)$. The vacuum on the boundary is just the identity functor $\id_\EM: \EM \to \EM$. The bulk excitations are given by the monoidal center $Z(\EC)$ of $\EC$. The category $Z(\EC)$ is a MTC \cite{mueger} and can be identified with the category $\text{Fun}_{\EC|\EC}(\EC, \EC)$ of $\EC$-$\EC$-bimodule functors. In such a model, the bulk-to-boundary map $L_\EM: Z(\EC) \to \EC_\EM^\vee$ is given by the following central functor 
\be  \label{eq:btob-map}
L_\EM:\,\, (\EC \xrightarrow{\EF} \EC)   \mapsto (\EM \simeq \EC\boxtimes_\EC \EM \xrightarrow{\EF \boxtimes_\EC \id_\EM} \EC\boxtimes_\EC \EM \simeq \EM).
\ee

Let $L_\EM^\vee$ be the right adjoint of the $L_\EM$. Then $L_\EM^\vee(\id_\EM)$ is a Lagrangian algebra in $Z(\EC)$. For example, when $\EM=\EC$, we obtain a Lagrangian algebra in $Z(\EC)$:
$$
L_\EC^\vee(\id_\EC) = \oplus_i~  i \otimes i^\vee,
$$ 
where the direct sum runs over all simple objects $i$ in $\EC$. 
When $\EC$ is MTC, $Z(\EC) \simeq \EC \boxtimes \overline{\EC}$, then we can also write:
\be \label{eq:cc-mod-inv-CFT}
L_\EC^\vee(\id_\EC) = \oplus_i~  i \boxtimes i^\vee.
\ee
Moreover, the map $\EM \mapsto L_\EM^\vee(\id_\EM)$ defines a bijection between the set of indecomposable semisimple $\EC$-modules and that of Lagrangian algebras in $Z(\EC)$.

\begin{rema} {\rm
When $\EC$ is realized as the category of modules over a rational vertex operator algebra, Eq.\,(\ref{eq:cc-mod-inv-CFT}) coincides with the famous charge-conjugate modular invariant closed conformal field theory \cite{tft1,cardy}. More generally, a 2d-condensable algebra in $\EC\boxtimes \overline{\EC}$ is Lagrangian if and only if it is modular invariant in the sense of \cite[Thm.\,3.4]{cardy}. This fact might suggest something deeper in physics. 
}
\end{rema}

For general $\EM$, we have the following commutative diagram: 
\be \label{diag:L}
\xymatrix{
Z(\EC) \ar[rr]^{-\otimes L_\EM^\vee(\id_\EM)} \ar[dr]_{L_\EM} & & \EC_{L^\vee(\id_\EM)} \ar[dl]^\simeq \\
& \EC_\EM^\vee & 
}
\ee
The above diagram simply says that not only the boundary excitations $\EC_\EM^\vee$ coincide with the boundary excitations $\EC_{L^\vee(\id_\EM)}$ obtained from the condensation of $L^\vee(\id_\EM)$, their associated  bulk-to-boundary maps also coincide. Therefore, we conclude that the 1d boundary phase determined by an ${}_\EC\EM$-boundary lattice model can be obtained by condensing the algebra $L_\EM^\vee(\id_\EM)$ in the $Z(\EC)$-bulk. %It turns out that there is a one-to-one correspondence between Lagrangian algebras in $Z(\EC)$ and indecomposable semisimple $\EC$-modules \cite[Prop.\,4.8]{dmno}. Therefore, there is a unique 2d-condensable algebra $L_\EM^\vee(\id_\EM)$ in $Z(\EC)$  reproduce the gapped boundary phase defined by the ${}_\EC\EM$-boundary lattice model. 

%\begin{rema} {\rm These algebras $L_\EM^\vee(\id_\EM)\simeq Z(A)$ and $A$, together with an algebraic homomorphism $Z(A) \to A$, form a so-called Cardy algebra \cite{kong}, a notion which classifies open-closed rational conformal field theories (see also \cite{tft1}\cite{unique}\cite{cardy}). This connection between anyon condensation and rational CFT is not accidental and was partially clarified in \cite{kas2,levin} and in a mathematically rigorous way in \cite[Sec.\,6]{fsv} for abelian Chern-Simons theories, and was completely clarified in \cite{KZ18}. }\end{rema}

\subsection{Kitaev quantum-double models}

Kitaev quantum double models \cite{kitaev} cover a subset of non-chiral topological phases defined by Levin-Wen models. In this case, a complete classification of 2d condensations is known \cite{da2}. We discuss this classification in this subsection.

Let $G$ be a finite group with unit $e$. The bulk phase of a Kitaev quantum-double model is given by the unitary MTC $Z(\rep(G))$. If a 1d gapped boundary is given by the unitary fusion category $\rep(G)$, the bulk-to-boundary map is given by the forgetful functor $F: Z(\rep(G)) \to \rep(G)$. Then the associated condensation is given by a Lagrangian algebra $F^\vee(\Cb)$ where $\Cb$ is the trivial representation of $G$ and the tensor unit of $\rep(G)$. In this case, $F^\vee(\Cb)$ is given by the commutative algebra $\mathrm{Fun}(G)$. 

Lagrangian algebras in $Z(\rep(G))$ one-to-one correspond to indecomposable semisimple module categories over $\rep(G)$ \cite{dmno}, each of which is determined by a pair $(H,\omega)$, where $H$ is a subgroup and $\omega \in H^2(H, \Cb^\times)$ \cite{ostrik2}. 

There are more 2d-condensable algebras in $Z(\rep(G))$. They have been classified by Davydov in \cite{da2}. Because of its importance in the physical applications, we would like to spell out this classification explicitly. 

An explicit description of the category $Z(\rep(G))$ is given in \cite[Prop.\,3.1.1]{da2}. Its objects is a pair $(X,\rho_X)$, where $X$ is a $G$-graded vector spaces, i.e. $X= \oplus_{g\in G} X_g$, and $\rho_X: G \times X \to X$ is a compatible $G$-action, which means for $f, g\in G$ $(fg)(v) = f(g(v))$, $e(v)=v$ for all $v\in X$ and $f(X_g) = X_{fgf^{-1}}$. The tensor product of $(X, \rho_X)$ and $(Y,\rho_Y)$ is just usual tensor product of $G$-graded vector spaces with the $G$-action $\rho_{X\otimes Y}$ defined by $g(x\otimes y) = g(x) \otimes g(y)$ for $x\in X, y\in Y$. The tensor unit is $\Cb$ which is viewed as a $G$-grade vector space supported only on the unit $e$ and equipped with a trivial $G$-action. The braiding is given by 
$$
c_{X,Y} (x\otimes y) = f(y) \otimes x, \quad\quad\quad  x\in X_f, y\in Y, f\in G.
$$
The dual object $X^\vee = \oplus_{g\in G} (X^\vee)_g$ is given by
$$
(X^\vee)_g = (X_{g^{-1}})^\vee = \hom(X_{f^{-1}}, \Cb)
$$
with action $g(l)(x) = l(g^{-1}(x))$ for $l\in \hom(X_{f^{-1}}, \Cb), x\in X_{gf^{-1}g^{-1}}$. The twist is given by $\theta_X(x) = f^{-1}(x)$ for $x\in X_f$. The quantum dimension $\dim X$ is just the usual vector space dimension. 
 
 By \cite[Thm.\,3.5.1]{da2}, a 2d-condensable algebra $A=A(H,F,\gamma, \epsilon)$ is determined by a subgroup $H\subset G$, a normal subgroup $F$ in $H$, a cocycle $\gamma\in Z^2(F, \Cb^\times)$ and $\epsilon: H\times F \to \Cb^\times$ satisfying the following conditions: 
 $$
 \epsilon_{gh}(f) = \epsilon_g(hfh^{-1}) \epsilon_h(f), \quad\quad \forall g,h\in H, f\in F
 $$
 $$
 \gamma(f,g) \epsilon_h(fg) = \epsilon_h(f) \epsilon_h(g) \gamma(hff^{-1}, hgh^{-1}) \epsilon(f), \quad\quad
 \forall  h\in H, f,g\in F
 $$
 \be \label{eq:gamma-epsilon}
 \gamma(f,g) = \epsilon_f(g) \gamma(fgf^{-1},f), \quad\quad \forall f,g\in F. 
 \ee
 This algebra $A=A(H,F,\gamma, \epsilon)$ as a vector space is spanned by $a_{g,f}, g\in G, f\in F$, modulo the relations 
\be  \label{eq:a-ghf}
a_{gh,f} = \epsilon_h(f) a_{g,hfh^{-1}}, \quad\quad  \forall h\in H,
\ee
together with a $G$-grading $a_{g,f}\in A_{gfg^{-1}}$ and a $G$-action $h(a_{g,f}) = a_{hg,f}$. The multiplication is given by
$$
a_{g,f} a_{g',f'} = \delta_{g,g'} \, \gamma(f,f') \, a_{g,ff'}. 
$$
  
By \cite[Thm.\,3.5.3]{da2}, the algebra $A(H,F,\gamma, \epsilon)$ is Lagrangian if and only if $F=H$. In this case, $\epsilon$ is uniquely determined by $\gamma$ in (\ref{eq:gamma-epsilon}). Such algebra is determined by a pair $(H, \gamma)$ (see also \cite{ostrik2}). 

\smallskip
Among all of these algebras, a special class is very simple. Let $F$ be the trivial group. Both $\gamma$ and $\epsilon$ are trivial. In this case, by (\ref{eq:a-ghf}), $a_{gh,1}=a_{g,1}$. Therefore, the algebra is spanned by the coset $G/H$. Moreover, the $G$-action on $A=A[H]$, given by $f(a_{g,1}) = a_{fg,1}, \forall f,g\in G$, is an algebraic automorphism, i.e. $f(ab) = f(a) f(b)$ for $a,b\in A$. This algebra $A[H]$ is nothing but the function algebra on the coset $G/H$. In this case, the condensed phase $Z(\rep(G))_{A[H]}^{loc}$ defined by $A[H]$ is nothing but the unitary MTC of $Z(\rep_H)$ \cite{da2}, which can be realized by a quantum double model associated to the group $H$. Therefore, this condensation can be viewed as a symmetric broken process from gauge group $G$ to $H$ (see \cite{bss2}\cite{bs} for the idea of Hopf symmetry broken).

%\smallskip
%Commutative algebras in Drinfeld categories of abelian Lie algebras was studied in \cite{df}. 

%\subsection{Quantum-double models via Hopf algebra}

\subsection{Condensations in chiral topological phases} \label{sec:chiral}

If a topological phase, given by a MTC $\EC$, is chiral, it means that it does not admit a gapped boundary. Mathematically, it means that $\EC$ is not a monoidal center of any fusion category, or equivalently, there is no Lagrangian algebra in $\EC$. But $\EC$ can still have non-trivial 2d-condensable algebras. 

Many examples of 2d-condensable algebras in chiral topological orders can be constructed from conformal embedding of rational vertex operator algebras (VOA) \cite{voa}. Let $U$ and $V$ be two rational vertex operator algebras. The rationality means, in particular, that the 
category $\Mod_U$ of $U$-modules and the category $\Mod_V$ of $V$-modules are MTC's \cite{huang}. If $U \hookrightarrow V$ as a sub-VOA (preserve the Virasoro element), then $V$ is a finite extension of $U$ and can be viewed as an algebra in $\Mod_U$. Moreover, $V$ is a 2d-condensable algebra in $\Mod_U$ and $(\Mod_U)_V^{loc}\simeq \Mod_V$ \cite{hkl} (see also \cite[Thm.\,4.3,Remark\,4.4]{osvoa}). 
%Hence $[\rep_U]=[\rep_V]$. 
In other words, a topological phase associated to the MTC $\Mod_V$ can be obtained by condensing the 2d-condensable algebra $V$ in the topological phase associated to the MTC $\Mod_U$.

For example, let $V_{\hat{\mathfrak{g}},k}$ denotes the VOA associated to affine Lie algebra $\hat{\mathfrak{g}}$ at level $k$. A few well-known conformal embedding are:
$$
V_{\widehat{\mathfrak{sl}}_2,4} \hookrightarrow V_{\widehat{\mathfrak{sl}}_3,1}, \quad\quad
V_{\widehat{\mathfrak{sl}}_2,10} \hookrightarrow V_{\widehat{\mathfrak{sp}}_4,1}, \quad\quad
V_{\widehat{\mathfrak{sl}}_2,6} \otimes_\Cb V_{\widehat{\mathfrak{sl}}_2,6} \hookrightarrow V_{\widehat{\mathfrak{so}}_9,1}.
$$
$$
V_{\widehat{\mathfrak{su}}_m,n} \otimes_\Cb V_{\widehat{\mathfrak{su}}_n,m}
 \hookrightarrow V_{\widehat{\mathfrak{su}}_{mn},1}, \quad\quad
 V_{\widehat{\mathfrak{so}}_m,n} \otimes_\Cb V_{\widehat{\mathfrak{so}}_n,m}
 \hookrightarrow V_{\widehat{\mathfrak{so}}_{mn},1}.
$$
Examples of conformal embedding can be found in many places (see for example \cite[Appendix]{dmno}).

\section{Witt equivalence}  \label{sec:mu-we}

Anyon condensation provides a powerful tool to study the interrelations among topological phases. %In this section, we discuss the structure of multiple 2d condensations and 

\subsection{Completely anisotropic 2d phases}  \label{sec:multi-cond}

\begin{defn}  {\rm \cite{dmno}
A MTC is {\it completely anisotropic} if the only 2d-condensable algebra $A \in \EC$ is $A=\one_\EC$.
}
\end{defn}

Therefore, if a topological phase described by a completely anisotropic MTC, then it cannot be condensed further. We call such a topological phase completely anisotropic. 

\begin{expl} {\rm 
We give a few examples of completely anisotropic MTC's:  
\bnu
\item Fibonacci categories \cite{db}: simple objects are $1$ and $x$ with fusion rule $x \otimes x = 1\oplus x$. 
\item Tensor powers of Fibonacci categories \cite{db}. 
\item Ising model: simple objects are $1, \psi, \sigma$ with fusion rules given in (\ref{eq:ising}). It was proved in \cite{tft2} that only two simple special symmetric Frobenius algebra are $1$ and $1 \oplus \psi \simeq \sigma \otimes \sigma^\vee$. But it is easy to see that the algebra $\sigma \otimes \sigma^\vee$ is not commutative and not a boson. Therefore, the only 2d-condensable algebra is the trivial algebra $1$. 
\enu
}
\end{expl}

In general, a MTC $\EC$ might contain a lot of 2d-condensable algebras. Let $A$ be a 2d-condensable algebra in $\EC$. A commutative algebra $B$ over $A$ (recall Definition\,\ref{def:B-over-A}) is naturally a commutative algebra in $\EC_A^{loc}$. We have $\EC_B \simeq (\EC_A)_B$.

By \cite[Lem.\,4.3]{cor}\cite[Prop.\,2.3.3]{da2}, a commutative algebra over $A$  is separable (connected) if and only the corresponding algebra in $\EC_A^{loc}$ is separable (connected). Therefore, condensing a 2d-condensable algebra $B$ over $A$ in the $\EC$-phase can be obtained by the composition of two 2d condensations: first condensing $A$, then condensing $B$ in the condensed phase $\EC_A^{loc}$. In particular, we have $\EC_B^{loc} \simeq (\EC_A^{loc})_B^{loc}$ \cite{dmno}.

A maximum 2d-condensable algebra $A$ in $\EC$ creates a completely anisotropic topological phase $\EC_A^{loc}$. This completely anisotropic topological phase is trivial only if the original phase $\EC$ is non-chiral, i.e. admitting a gapped boundary. Examples of 2d condensations in chiral topological phases are discussed in Section\,\ref{sec:chiral}.  

\subsection{Witt equivalence} \label{sec:witt}

%In this subsection, we will drop the assumption that $\ED$ comes from $\EC$ via condensation. We would like to answer the question: is it possible to realize an arbitrary $\EC$-phase and an arbitrary $\ED$-phase bounded by a gapped domain wall via condensations?  It is equivalent to ask if any two phases connected by gapped domain walls can be achieved by condensations from a single phase. 

Two 2d topological orders $\EC$ and $\ED$ are called {\it Witt equivalent} if they can be connected by a gapped domain wall. This is a well-defined equivalence relation, which was first introduced by Kitaev in 2008 \cite{kitaev2}, and was introduced in mathematics for MTC's in \cite{dmno}, and was later translated back to physics in \cite[Sec.\,4]{fsv}. Mathematically, two MTC's $\EC$ and $\ED$ are Witt equivalence if there is a spherical fusion category $\EC$ such that 
\be \label{def:witt-eq}
\EC \boxtimes \overline{\ED} \simeq Z(\EE).
\ee
The equivalence classes of MTC's form a group, called the {\it Witt group} \cite{dmno}. It is an infinite group. The unit element is given by the Witt class $[\vect]$ of $\vect$. The multiplication of the group is given by the Deligne tensor product $\boxtimes$, which, in physics, amounts to stacking two topological orders. The inverse is given by $[\EC]^{-1} = [\overline{\EC}]$. In particular, a topological phase $\EC$ can have a gapped boundary or non-chiral if and only if $[\EC]=[\vect]$ (see \cite[Sec.\,3]{fsv} for a proof). An interesting result proved in \cite[Thm.\,5.13]{dmno} is that in each Witt class, there is a unique (up to braided equivalence) completely anisotropic MTC. A further study on Witt equivalence was carried out in \cite{dno}.

\begin{rema} {\rm
Theorem\,5.13 in \cite{dmno} is stated for non-degenerate braided fusion category, but can be generalized to (unitary) MTC's. %The proof of Theorem\,5.13 can be easily adapted to the (unitary) modular case because, for a 2d-condensable algebra $A$, the category $\EC_A^{loc}$ is automatically (unitary) spherical. 
}
\end{rema}

If two topological phases $\EC$ and $\ED$ are Witt equivalent, in general, one cannot obtain $\ED$ by a 2d condensation in $\EC$. But you can obtain both $\EC$ and $\ED$ from the new phase via two different 2d condensations. Indeed, by \cite[Cor.\,5.9]{dmno} (see also \cite[Sec.\,4]{fsv}), $\EC$ and $\ED$ are Witt equivalent if and only if there are spherical fusion categories $\EC_1$ and $\EC_2$ such that $\EC\boxtimes Z(\EC_1) \simeq \ED\boxtimes Z(\EC_2)$ as braided tensor categories. These two categories $\EC_1$ and $\EC_2$ can be determined as follows. If (\ref{def:witt-eq}) is true, multiplying both sides by $\ED$, we obtain 
$\EC \boxtimes (\overline{\ED} \boxtimes \ED) \simeq \ED \boxtimes Z(\EE)$, or equivalently, a braided monoidal equivalence 
$$
G: \EC\boxtimes Z(\ED) \simeq \ED \boxtimes Z(\EE)
$$
since $(\overline{\ED} \boxtimes \ED)\simeq Z(\ED)$ for $\ED$ being modular. One can start from a topological phase given by $\EA:=\ED \boxtimes Z(\EE)$, then condense two 2d-condensable algebras  $A_1$ and $A_2$ in $\EA$: 
$$
A_1:= G(\one_\EC \boxtimes F_\ED^\vee(\one_{\ED})), \quad\quad A_2:= \one_\ED \boxtimes F_\EE^\vee(\one_{\EE}), 
$$
where $F_\ED^\vee$ and $F_\EE^\vee$ are the right adjoint functors of the forgetful functors $F_\ED:Z(\ED) \to \ED$ and $F_\EE:Z(\EE) \to \EE$, respectively. After the condensations, we obtain two phases \cite[Prop.\,5.15]{dmno}:
$$
\EC \simeq \EA_{A_1}^{loc} \quad\quad \mbox{and} \quad\quad \ED \simeq \EA_{A_2}^{loc}~.
$$
Therefore, any pair of MTC's in the same Witt class can be obtain from a single 2d phases via two different 2d condensations.   
%As a consequence, we have shown that any two Witt equivalent topological phases can be obtained from a single phase via a two (potentially different) 2d condensations. 

\begin{rema} {\rm
By condensing $A_1$ and $A_2$ in the $\EA$-phase in two different 2d-regions, respectively, we obtain 
a very thick wall between the $\EC$-phase and the $\ED$-phase. According to our bootstrap analysis, the particles on the left side of the thick wall form the fusion category ${}_{A_1}\EA$; those on the right side of the wall form the fusion category $\EA_{A_2}$; in the middle of the thick wall is the original 2d phase $\EA$. Therefore, viewed from far away, this thick wall becomes a 1d wall defined by the fusion category ${}_{A_1}\EA \boxtimes_\EA \EA_{A_2}\simeq {}_{A_1}\EA_{A_2}$. The gapped domain walls between a $\EC$-phase and a $\ED$-phase are not unique of course. They are classified by Lagrangian algebras in $\EC\boxtimes \overline{\ED}$. 
}
\end{rema}

\begin{appendices}

\section{Appendix}

For the convenience of physics readers, we include in this appendix the mathematical definitions of various tensor-categorical notions appeared in this work. We do not spell out explicitly the coherence conditions used in some of these notions because they are usually lengthy and mysterious to the first time readers. 
For more details, readers should consult with reviews of this subject (see for example \cite{bakalov-kirillov,ce,mueger2,turaev,wang}).

\subsection{Modular tensor categories} \label{sec:mtc}
In this section, we review the definition of spherical fusion category and that of MTC. A beautiful introduction to the later notion from the point of view of anyons can be found in Appendix E in \cite{kitaev1}. 

\medskip
A monoidal category (or tensor category) is a category equipped with a tensor product $\otimes$ and a tensor unit $\one$ (or vacuum in physical language). The tensor product $\otimes$ is associative with the associativity isomorphisms:
\be  \label{eq:asso}
\alpha_{X,Y,Z}: \, X\otimes (Y \otimes Z) \xrightarrow{\simeq} (X\otimes Y) \otimes Z \quad\quad \forall X,Y,Z \in \EC,
\ee
which are required to satisfy the pentagon relations. The unit isomorphisms: 
\be
\one \otimes X \xrightarrow{l_X} X \xleftarrow{r_X} X\otimes \one
\ee
are required to satisfy the triangle relations. A braiding is a family of isomorphisms $c_{X,Y}: X\otimes Y \xrightarrow{\simeq} Y\otimes X$, satisfying the hexagon relations. 

\smallskip
\begin{defn}  \label{def:bm-functor} {\rm
A monoidal functor $F: \EC \to \ED$ between two monoidal categories $\EC$ and $\ED$ is a functor such that there are isomorphisms $F(X\otimes Y) \xrightarrow{\simeq} F(X) \otimes F(Y)$ (preserving the tensor products) and $F(\one) \xrightarrow {\simeq }\one$ (preserving the unit) satisfying some coherence properties. If both $\EC$ and $\ED$ are braided, $F$ is called braided monoidal if the following diagram:
$$
\xymatrix{
F(X\otimes Y) \ar[r]^{\simeq} \ar[d]_{F(c_{X,Y})} &  F(X)\otimes F(Y) \ar[d]^{c_{F(X), F(Y)}} \\
F(Y\otimes X) \ar[r]^{\simeq}  & F(Y) \otimes F(X)
}
$$
is commutative for all $X,Y\in \EC$. 
} 
\end{defn}

\begin{defn} \label{def:right-adjoint} {\rm
A right adjoint of a functor $F: \EC \to \ED$ between two categories is a functor $F^\vee: \ED \to \EC$ such that there are natural isomorphisms:  
$$
\hom_\ED(F(X), Y) \simeq \hom_\EC(X, F^\vee(Y)), \quad\quad \forall X\in \EC, Y\in \ED. 
$$ 
}
\end{defn}

\smallskip
A $\Cb$-linear category means that all hom spaces $\hom_\EC(A,B)$ for $A,B \in \EC$ are vector spaces over $\Cb$. $\EC$ is {\it semisimple} if every object in $\EC$ is a direct sum of simple objects. $\EC$ is called {\it finite} if there are only finite number of inequivalent simple objects. We denote the set of equivalence classes of simple objects in $\EC$ by $I$, elements in $I$ by $i,j,k,l\in I$. We have $|I|<\infty$. A simple unit means the unit $\one$ is in $I$. 

In a finite semisimple $\Cb$-linear category, it is possible to translate the associativity and unit isomorphisms to some very concrete data. The isomorphism (\ref{eq:asso}) can be recovered from the following isomorphisms:
$$
\hom_\EC( (i\otimes j) \otimes k, l) \xrightarrow{F} 
\hom_\EC( i\otimes (j \otimes k), l)
$$
In terms of the chosen basis, $F$ can be expressed by what is called fusion matrices in physics.  %A picture of this fusion matrix is given in equation (2.36) in \cite{tft1}.

\begin{defn} \label{def:rigidity} {\rm
A tensor category $\EC$ is called {\it rigid} if each $U\in  \EC$ has a left dual ${}^\vee U$ and a right dual $U^\vee$, together with the following duality maps:
\be
\begin{array}{llll}
  \raisebox{-8pt}{
  \begin{picture}(26,22)
   \put(0,6){\scalebox{.75}{\includegraphics{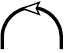}}}
   \put(0,6){
     \setlength{\unitlength}{.75pt}\put(-146,-155){
     \put(143,145)  {\scriptsize $ U^\vee $}
     \put(169,145)  {\scriptsize $ U $}
     }\setlength{\unitlength}{1pt}}
  \end{picture}}  
  \etb= d_U : U^\vee \otimes U \rightarrow \one
  ~~,\qquad &
  \raisebox{-8pt}{
  \begin{picture}(26,22)
   \put(0,6){\scalebox{.75}{\includegraphics{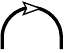}}}
   \put(0,6){
     \setlength{\unitlength}{.75pt}\put(-146,-155){
     \put(143,145)  {\scriptsize $ U $}
     \put(169,145)  {\scriptsize $ {}^\vee U $}
     }\setlength{\unitlength}{1pt}}
  \end{picture}}  
  \etb= \tilde d_U : U \otimes {}^\vee U \rightarrow \one
  ~~,
\\[2em]
  \raisebox{-8pt}{
  \begin{picture}(26,22)
   \put(0,0){\scalebox{.75}{\includegraphics{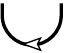}}}
   \put(0,0){
     \setlength{\unitlength}{.75pt}\put(-146,-155){
     \put(143,183)  {\scriptsize $ U $}
     \put(169,183)  {\scriptsize $ U^\vee $}
     }\setlength{\unitlength}{1pt}}
  \end{picture}}  
  \etb= b_U : \one \rightarrow U \otimes U^\vee
  ~~,
  &
  \raisebox{-8pt}{
  \begin{picture}(26,22)
   \put(0,0){\scalebox{.75}{\includegraphics{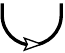}}}
   \put(0,0){
     \setlength{\unitlength}{.75pt}\put(-146,-155){
     \put(138,183)  {\scriptsize $ {}^\vee U $}
     \put(172,183)  {\scriptsize $ U $}
     }\setlength{\unitlength}{1pt}}
  \end{picture}}  
  \etb= \tilde b_U : \one \rightarrow {}^\vee U \otimes U
  ~,
\end{array}
\ee
where letter ``$b$" stands for ``birth" and ``$d$" for ``death", 
such that all the following conditions:
$$
  \raisebox{-18pt}{
  \begin{picture}(30,42)
   \put(0,6){\scalebox{.75}{\includegraphics{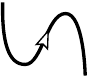}}}
   \put(0,6){
     \setlength{\unitlength}{.75pt}\put(-146,-155){
     \put(143,195)  {\scriptsize $ {}^\vee U $}
     \put(169,146)  {\scriptsize $ {}^\vee U $}
     }\setlength{\unitlength}{1pt}}
  \end{picture}}  ~=~ \id_{{}^\vee U} ~,
  \quad
  \raisebox{-18pt}{
  \begin{picture}(30,42)
   \put(0,6){\scalebox{.75}{\includegraphics{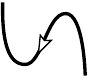}}}
   \put(0,6){
     \setlength{\unitlength}{.75pt}\put(-146,-155){
     \put(143,195)  {\scriptsize $ U $}
     \put(183,146)  {\scriptsize $ U $}
     }\setlength{\unitlength}{1pt}}
  \end{picture}}  ~=~ \id_{U} ~,
  \quad
  \raisebox{-18pt}{
  \begin{picture}(30,42)
   \put(0,6){\scalebox{.75}{\includegraphics{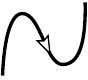}}}
   \put(0,6){
     \setlength{\unitlength}{.75pt}\put(-146,-155){
     \put(143,145)  {\scriptsize $ U $}
     \put(183,195)  {\scriptsize $ U $}
     }\setlength{\unitlength}{1pt}}
  \end{picture}}  ~=~ \id_{U} ~,
  \quad
  \raisebox{-18pt}{
  \begin{picture}(30,42)
   \put(0,6){\scalebox{.75}{\includegraphics{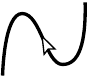}}}
   \put(0,6){
     \setlength{\unitlength}{.75pt}\put(-146,-155){
     \put(143,145)  {\scriptsize $ U^\vee $}
     \put(183,195)  {\scriptsize $ U^\vee $}
     }\setlength{\unitlength}{1pt}}
  \end{picture}}  ~=~ \id_{U^\vee} ~.
$$
are satisfied. $\EC$ is called {\it sovereign} if ${}^\vee U = U^\vee$ for all $U\in \EC$. 
} 
\end{defn}

\begin{defn} {\rm
A multi-fusion category is a finite semisimple $\Cb$-linear rigid tensor category $\EC$ with finite dimensional hom spaces. If the tensor unit in $\EC$ is simple, $\EC$ is called a fusion category. 
}
\end{defn}

Let $\EC$ be a rigid tensor category and $U\in \EC$ is an object. We naturally have $({}^\vee U)^\vee = U$ and ${}^\vee(U^\vee) = U$. If $a\in \hom_\EC(U, U^{\vee\vee})$, we define a left trace 
$$
\tr^L(a): \one \xrightarrow{b_U} U\otimes U^\vee \xrightarrow{a1} U^{\vee\vee} \otimes U^\vee \xrightarrow{d_{U^\vee}} \one. 
$$ 
If $a\in \hom_\EC(U, {}^{\vee\vee}U)$, we define a right trace:
$$
\tr^R(a): \one \xrightarrow{b_{{}^\vee U}} {}^\vee U\otimes U \xrightarrow{a1} {}^{\vee} U \otimes {}^{\vee\vee} U \xrightarrow{d_{{}^{\vee\vee} U}} \one. 
$$
\begin{defn} \label{def:spherical} {\rm
A {\it pivotal structure} on a rigid tensor category $\EC$ is an isomorphism $a: \id_\EC \to \vee\vee$, i.e. a collection of isomorphisms $a_U: U \xrightarrow{\simeq} U^{\vee\vee}$ natural in $U$ and satisfying $a_{U\otimes V} = a_U \otimes a_V$. In this case, we have ${}^{\vee\vee} U \simeq U \simeq U^{\vee\vee}$. $\EC$ is called {\it spherical} if $\tr^L(a_U) = \tr^R(a_U)$ for all $U \in \EC$. We set $\tr=\tr^{L/R}$ in this case. 
}
\end{defn}

If $\EC$ is spherical, we define quantum dimension for $U\in \EC$ by $\dim U :=\tr(a_U)$.

\begin{defn} \label{def:unitary} {\rm
A $\ast$-category $\EC$ is a $\Cb$-linear category equipped with a functor $\ast: \EC \to \EC^\op$ 
which acts trivially on the objects and is antilinear and involutive on morphisms, i.e. $\ast:\Hom(A,B) \to \Hom(B,A)$ is defined so that 
\be \label{eq:dagger}
(g \circ f)^\ast = f^\ast \circ g^\ast, \quad\quad (\lambda f)^\ast = \bar{\lambda} f^\ast,\quad\quad f^{\ast\ast} = f. 
\ee
for $f: U \to V$, $g: V \to W$, $h: X \to Y$, $\lambda \in \Cb^\times$.
A $\ast$-category is called {\it unitary} if $\ast$ satisfies the positive condition: $f\circ f^\ast =0$ implies $f=0$. 
}
\end{defn}

\begin{rema} {\rm
That $\ast$ preserves the identity maps follows from (\ref{eq:dagger}). More precisely, for $X\in \EC$, we have $\id_X = (\id_X \circ \id_X^\ast)^\ast = \id_X \circ \id_X^\ast = \id_X^\ast$. 
}
\end{rema}

A functor $F: \EC \to \ED$ between two $\ast$-categories is required to be adjoint preserving, i.e. $F(f^\ast) = F(f)^\ast$. 

\begin{defn}  \label{def:bm-dagger} {\rm
A monoidal $\ast$-category $\EC$ is a monoidal category such that $\ast$ is compatible with the monoidal structures, i.e.
\bea
&(g\otimes h)^\ast=g^\ast \otimes h^\ast,\quad\quad \forall g: V\to W, h: X \to Y, & \\
&\alpha_{X,Y,Z}^\ast=\alpha_{X,Y,Z}^{-1},\quad l_X^\ast=l_X^{-1},\quad r_X^\ast=r_X^{-1}. & \label{eq:unitary-asso-unit}
\eea
A braided monoidal $\ast$-category requires that $\ast$ is compatible with the braiding, i.e. $c_{X,Y}^\ast = c_{X,Y}^{-1}$ for all $X,Y$. 
}
\end{defn}

In a monoidal $\ast$-category $\EC$, if an object $X$ has a right dual $(U^\vee, b_U, d_U)$, it automatically has a left dual which is given by $(U^\vee, d_U^\ast, b_U^\ast)$. Similarly, a left dual automatically gives a right dual. For this reason, we adopt a symmetric notation for duality, we denote both the right and left duals as $\overline{U}$, i.e. $\overline{U}:=U^\vee = {}^\vee U$, and we set $\tilde{b}_U=d_U^\ast$ and $\tilde{d}_U = b_U^\ast$. A unitary fusion category has a unique pivotal structure which is spherical \cite{kitaev,eno02}. 

\begin{prop} \label{prop:positive-dim} {\rm \cite{eno02}}
For a unitary fusion category, the quantum dimensions of objects are real and positive.
\end{prop}

\begin{defn} \label{def:ribbon}{\rm
A ribbon category is a rigid braided tensor category $\EC$ with a twist $\theta_U: U \xrightarrow{\simeq} U$ such that the following conditions holds:
$$
\theta_\one =\id_\one, \quad \theta_{U^\vee} = (\theta_U)^\vee, \quad \theta_{U\otimes V} = c_{V,U}\circ c_{U,V} \circ (\theta_U \otimes \theta_V),
$$
where $(\theta_U)^\vee := (d_U\otimes \id_{U^\vee}) \circ (\id_{U^\vee} \otimes \theta_U \otimes \id_{U^\vee}) \circ (\id_{U^\vee} \otimes b_U) $. 
}
\end{defn}

If the category $\EC$ is ribbon, we can identify ${}^\vee U = U^\vee$, i.e. $\EC$ is sovereign. 

\begin{defn} \label{def:unitary-ribbon} {\rm 
A ribbon $\ast$-category $\EC$ is a braided monoidal $\ast$-category which is ribbon and $\theta_X^\ast = \theta_{X}^{-1}$ for all objects $X$. $\EC$ is unitary ribbon if $\ast$ is also positive. 
}
\end{defn}

\begin{defn} \label{def:mtc} {\rm
A (unitary) MTC is a $\Cb$-linear semisimple finite (unitary) ribbon category such that 
the matrix $[s_{i,j}]$ defined by
\be  
 s_{i,j}   ~=~  \quad
\raisebox{-30pt}{
  \begin{picture}(110,65)
   \put(0,8){\scalebox{.75}{\includegraphics{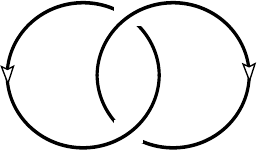}}}
   \put(0,8){
     \setlength{\unitlength}{.75pt}\put(-18,-19){
     \put( 98, 48)       {\scriptsize $ i $}
     \put( 50, 48)      {\scriptsize $ j $}
     }\setlength{\unitlength}{1pt}}
  \end{picture}}
\ee
is non-degenerate. 
}
\end{defn}
We have $s_{i,j}=s_{j,i}$ and $s_{0,i}=\dim i$. The dimension of $\EC$ is defined by $$\dim \EC = \sum_{i\in I} \,\, (\dim i)^2.$$

\subsection{Algebras in a MTC} \label{app:module-cat}

Let $\EC$ be a braided tensor category. 

\begin{defn} \label{def:alg} {\rm
An algebra in $\EC$ (or a $\EC$-algebra) is a triple $(A, \mu, \iota)$, where $A$ is an object in $\EC$, $m$ is a morphism $A\otimes A\to A$ and $\iota: \one \to A$ satisfying the following conditions: 
$$
\mu \circ (\mu \otimes \id_A) \circ \alpha_{A,A,A} = \mu \circ (\id_A \otimes \mu).
$$
$$
\mu \circ (\iota \otimes \id_A) = \id_A = \mu\circ (\id_A \otimes \iota). 
$$
The algebra $A$ is called commutative if $\mu = \mu \circ c_{A,A}$. 
}
\end{defn}
We denote the ingredients of an algebra graphically as follows: 
$$
  \mu = \raisebox{-20pt}{
  \begin{picture}(30,45)
   \put(0,6){\scalebox{.75}{\includegraphics{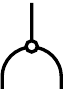}}}
   \put(0,6){
     \setlength{\unitlength}{.75pt}\put(-146,-155){
     \put(143,145)  {\scriptsize $ A $}
     \put(169,145)  {\scriptsize $ A $}
     \put(157,202)  {\scriptsize $ A $}
     }\setlength{\unitlength}{1pt}}
  \end{picture}}  
  ~~,\quad\quad\quad 
  \iota = \raisebox{-15pt}{
  \begin{picture}(10,30)
   \put(0,6){\scalebox{.75}{\includegraphics{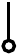}}}
   \put(0,6){
     \setlength{\unitlength}{.75pt}\put(-146,-155){
     \put(146,185)  {\scriptsize $ A $}
     }\setlength{\unitlength}{1pt}}
  \end{picture}}
$$

\begin{defn} \label{def:module} {\rm
A left module over an algebra $A=(A,\mu,\iota)$ is a pair $(M, \mu_M)$, where $M$ is an object in $\EC$ and $\mu_M: A\otimes M \to M$ such that 
$$
\mu_M^L \circ (\mu \otimes \id_M) \circ \alpha_{A,A,M} = \mu_M \circ (\id_A \otimes \mu_M)
$$
and $\mu_M \circ (\iota_A \otimes \id_M) =\id_M$. 
The definition of a right $A$-module $(M, \mu_M^R)$ is similar. An $A$-$B$-bimodule is a triple $(M, \mu_M^L, \mu_M^R)$ such that $(M, \mu_M^L)$ is a left $A$-module and $(M, \mu_M^R)$ is a right $B$-module such that 
$$
\mu_M^R \circ (\mu_M^L \otimes \id_B) \circ \alpha_{A,M,B} = \mu_M^L \circ (\id_A \otimes \mu_M^R). 
$$
}
\end{defn}
We denote the module structure graphically as follows: 
$$
  \mu_M^L = \raisebox{-20pt}{
  \begin{picture}(30,45)
   \put(0,6){\scalebox{.75}{\includegraphics{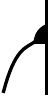}}}
   \put(0,6){
     \setlength{\unitlength}{.75pt}\put(-146,-155){
     \put(142,146)  {\scriptsize $ A $}
     \put(165,143)  {\scriptsize $ M $}
     \put(161,206)  {\scriptsize $ M $}
     }\setlength{\unitlength}{1pt}}
  \end{picture}}  
  ~~,\quad\quad\quad
  \mu_M^R = \raisebox{-20pt}{
  \begin{picture}(30,45)
   \put(0,6){\scalebox{.75}{\includegraphics{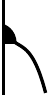}}}
   \put(0,6){
     \setlength{\unitlength}{.75pt}\put(-146,-155){
     \put(142,146)  {\scriptsize $ M $}
     \put(165,146)  {\scriptsize $ A $}
     \put(142,206)  {\scriptsize $ A $}
     }\setlength{\unitlength}{1pt}}
  \end{picture}}
$$

\begin{defn} \label{def:local}  {\rm
For a commutative algebra $A$, a module $(M, \mu_M)$ is called {\it local} if 
$\mu_M = \mu_M \circ c_{M, A} \circ c_{A,M}$.
}
\end{defn}

\begin{defn} \label{def:separable} {\rm
A $\EC$-algebra $(A, \mu, \iota)$ is called {\it separable} if $\mu: A\otimes A \to A$ splits as a morphism of $A$-$A$-bimodule. Namely, there is an $A$-$A$-bimodule map $e: A\to A\otimes A$ such that $\mu \circ e = \id_A$. 
A separable algebra is called {\it connected} if $\dim \hom_\EC(\one, A) = 1$. A commutative separable algebra is also called {\it \'{e}tale} algebra in \cite{dmno}. 
}
\end{defn}

If a $\EC$-algebra $A$ is separable, the category $\EC_A$ of $A$-module and the category $\EC_{A|A}$ of $A$-$A$-bimodules are both semisimple. In this paper, a connected separable commutative $\EC$-algebra is also called a 2d-condensable algebra for simplicity. 

\begin{defn} \label{def:morita-alg} {\rm
Two $\EC$-algebras $A$ and $B$ are called Morita equivalent if $\EC_A \cong \EC_B$. Or equivalently, there are an $A$-$B$-bimodule $M$ and a $B$-$A$-bimodule $N$ such that $M\otimes_B N\simeq A$ and $N\otimes_A M \simeq B$ as bimodules. 
}
\end{defn}

Let $\EC$ be a MTC. 
\begin{defn} \label{def:lag} {\rm \cite{drgno}
A connected commutative separable algebra $A$ is called {\it Lagrangian} if 
$(\dim A)^2 = \dim \EC$. 
}
\end{defn}

\begin{thm} \label{thm:lag} {\rm (see for example \cite{dmno})}
For a Lagrangian algebra $A$ in $\EC$, the category $\EC_A^{loc}$ of local $A$-modules is trivial, i.e. $\EC_A^{loc} \simeq \vect$. 
\end{thm}

\begin{defn} \label{def:coalg} {\rm 
A coalgebra is a triple $(A,\Delta,\eps)$ where $\Delta : A \rightarrow A \otimes A$ and $\eps : A \rightarrow \one$ obey the co-associativity: 
$$
\Delta \circ (\Delta \otimes \id_A) = \alpha_{A,A,A} \circ \Delta \circ (\id_A \otimes \Delta)
$$
and the counit condition $(\epsilon \otimes \id_A) \circ \Delta = \id_A = (\id_A \otimes \epsilon) \circ \Delta$. 
}
\end{defn}
We use the following graphical representation for the morphisms of a coalgebra,
\be
  \Delta = \raisebox{-20pt}{
  \begin{picture}(30,45)
   \put(0,6){\scalebox{.75}{\includegraphics{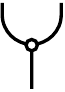}}}
   \put(0,6){
     \setlength{\unitlength}{.75pt}\put(-146,-155){
     \put(143,202)  {\scriptsize $ A $}
     \put(169,202)  {\scriptsize $ A $}
     \put(157,145)  {\scriptsize $ A $}
     }\setlength{\unitlength}{1pt}}
  \end{picture}}
  ~~,\quad\quad\quad
  \epsilon = \raisebox{-15pt}{
  \begin{picture}(10,30)
   \put(0,10){\scalebox{.75}{\includegraphics{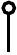}}}
   \put(0,10){
     \setlength{\unitlength}{.75pt}\put(-146,-155){
     \put(146,145)  {\scriptsize $ A $}
     }\setlength{\unitlength}{1pt}}
  \end{picture}}
  ~~.
\ee

\begin{defn} \label{def:Frob} {\rm 
A Frobenius algebra $A = (A,\mu,\iota,\Delta,\epsilon)$ is an algebra and a coalgebra such that the coproduct is an intertwiner of $A$-$A$-bimodules, i.e.
\be \label{eq:Frob-property}
(\id_A \otimes \mu) \circ (\Delta \otimes \id_A) = \Delta \circ \mu = (\mu \otimes \id_A) \circ (\id_A \otimes \Delta).
\ee
Let now $\EC$ be a sovereign tensor category.
A Frobenius algebra in $\EC$ is {\em symmetric} iff
\be
  \raisebox{-35pt}{
  \begin{picture}(50,75)
   \put(0,8){\scalebox{.75}{\includegraphics{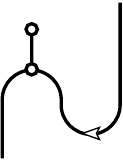}}}
   \put(0,8){
     \setlength{\unitlength}{.75pt}\put(-34,-37){
     \put(31, 28)  {\scriptsize $ A $}
     \put(87,117)  {\scriptsize $ A^\vee $}
     }\setlength{\unitlength}{1pt}}
  \end{picture}}
  ~=~
  \raisebox{-35pt}{
  \begin{picture}(50,75)
   \put(0,8){\scalebox{.75}{\includegraphics{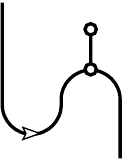}}}
   \put(0,8){
     \setlength{\unitlength}{.75pt}\put(-34,-37){
     \put(87, 28)  {\scriptsize $ A $}
     \put(31,117)  {\scriptsize $ A^\vee $}
     }\setlength{\unitlength}{1pt}}
  \end{picture}}
\ee
A Frobenius algebra is called {\it normalized-special} if 
$$
\mu \circ \Delta =\id_A \quad\quad\mbox{and}\quad\quad \epsilon \circ \iota = \dim (A) ~\id_\one.
$$ 
}
\end{defn}

\begin{defn} \label{def:tensor-A} {\rm
Let $A$ be a normalized-special Frobenius algebra in a MTC $\EC$ and let $M$ be a right $A$-module and $N$ be a left $A$-module. The tensor product $M \otimes_A N$ can be defined by the image of the following idempotent (or projector):
\be
  P_{\otimes A} ~=~ 
  \raisebox{-34pt}{
  \begin{picture}(65,75)
   \put(8,8){\scalebox{.75}{\includegraphics{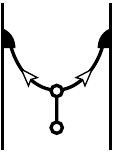}}}
   \put(8,8){
     \setlength{\unitlength}{.75pt}\put(-30,-38){
     \put(25, 28)  {\scriptsize $ M $}
     \put(25,115)  {\scriptsize $ M $}
     \put(79, 28)  {\scriptsize $ N $}
     \put(79,115)  {\scriptsize $ N $}
     \put(12, 87)  {\scriptsize $ \rho_M $}
     \put(86, 87)  {\scriptsize $ \rho_N $}
     \put(41, 62)  {\scriptsize $ A $}
     }\setlength{\unitlength}{1pt}}
  \end{picture}}
\ee
Moreover, there exists morphisms $e_A : M \otimes_A N \rightarrow M \otimes N$ and $r_A : M \otimes N \rightarrow M \otimes_A N$ such that $r_A \circ e_A = \id_{M \otimes_A N}$ and $e_A \circ r_A = P_{\otimes A}$. 
}
\end{defn}

\begin{defn} \label{def:B-over-A}  {\rm
Let $A$ be an algebra in $\EC$. An algebra $B$ is called an algebra over $A$ if there is an algebra homomorphism $f: A \to B$ such that the following diagram commutes: 
$$
\xymatrix{
A\otimes B \ar[r]^{f1} \ar[d]_{c_{B,A}^{-1}} & B\otimes B \ar[r]^{\mu_B} & B \\
B\otimes A \ar[r]^{1f} & B\otimes B \ar[ur]_{\mu_B} & 
}
$$
$B$ is an algebra over $A$ is equivalent to the statement that $B$ is an algebra in $\EC_A$. 
}
\end{defn}

\subsection{Module categories and monoidal centers}

\begin{defn} \label{def:C-mod} {\rm
A left module category over a tensor category $\EC$ (or a left $\EC$-module) is a category $\EM$ equipped with a $\EC$-action: a functor $\otimes: \EC \times \EM \to \EM$ such that there are isomorphisms: 
$$
X \otimes (Y \otimes M) \xrightarrow{\simeq} (X \otimes Y) \otimes M
$$
for $X,Y\in \EC$ and $M\in \EM$ and $\one \otimes M \xrightarrow{\simeq} M$, satisfying some obvious coherence conditions \cite{ostrik}. A $\EC$-module $\EM$ is called {\it semisimple} if every object in $\EM$ is a direct sum of simple objects. $\EM$ is called {\it indecomposable} if it cannot be written as a direct sum of two $\EC$-modules. 
}
\end{defn}

The definition of a right $\EC$-module is similar. For two tensor categories $\EC$ and $\ED$, a $\EC$-$\ED$-bimodule is a category $\EM$ equipped with a left $\EC$-module and a right $\ED$-module structure, such that there are isomorphisms $X\otimes (M\otimes Y) \xrightarrow{\simeq} (X\otimes M) \otimes Y$ satisfying some natural coherence conditions. 

\begin{defn} \label{def:mod-functor} {\rm
A $\EC$ module functor $F: \EM \to \EN$ is a functor from $\EM \to \EN$ together with an isomorphism 
$F(X\otimes M) \xrightarrow{\simeq} X\otimes F(M)$ satisfying some coherence conditions \cite{ostrik}. 
}
\end{defn}

\begin{defn} \label{def:rel-center} {\rm
Let $\EA$ be a fusion category and $\EM$ an $\EA$-bimodule. We define the center $Z_\EA(\EM)$ of $\EM$ by the category of $\EA$-bimodule functors, i.e. $Z_\EA(\EM) := \text{Fun}_{\EA|\EA}(\EM, \EM)$. In the case that $\EA \subset \EB$ as a fusion subcategory, $Z_\EA(\EB)$ is also called the {\it relative center} of $\EB$. If $\EA = \EB$, $Z_\EA(\EA)$ is called {\it monoidal center} of $\EA$, also denoted by $Z(\EA)$. 
}
\end{defn}

\begin{rema}  \label{rema:center} {\rm
An object in $Z(\EA)$ is a pair $(Z, z)$, where $Z$ is an object in $\EA$ and $z$ is a family of isomorphisms $\{ z_X: Z \otimes X \xrightarrow{\simeq} X \otimes Z\}_{X\in \EA}$, satisfying some consistency conditions. There is a forgetful functor $Z(\EA) \to \EA$ defined by $(Z,z) \mapsto Z$. This functor is monoidal. When $\EA$ is modular, $Z(\EA)=\EA \boxtimes \overline{\EA}$, where $\overline{\EA}$ is the same tensor category as $\EA$ but with the braiding given by the anti-braiding of $\EA$. In this case, the forgetful functor coincides with the usual tensor product $a \boxtimes b \mapsto a\otimes b$ for $a,b\in \EA$. 
}
\end{rema}

\smallskip
An important result of M\"{u}ger \cite{mueger} says: 
\begin{thm} \label{thm:mueger} 
If $\EA$ is a spherical fusion category, then the monoidal center $Z(\EA)$ is a MTC. 
\end{thm}

\begin{defn}[\cite{Bez04,dmno}] \label{def:central} {\rm
A functor $F: \EA \to \EB$ from a braided tensor category $A$ to a (not necessarily braided) tensor category $B$, is called {\it central}, or equipped with a structure of a {\it central functor}, if there are natural isomorphisms $B\otimes F(A) \xrightarrow{\simeq} F(A) \otimes B$ satisfying some coherence conditions such that $F$ can be lifted to a braided monoidal functor to the center $Z(\EB)$ of $\EB$. More precisely, there is a functor $\tilde{F}: \EA \to Z(\EB)$ such that the following diagram: 
$$
\xymatrix{
\EA \ar[r]^{\tilde{F}} \ar[rd]_{F} & Z(\EB) \ar[d]^{\text{forget}} \\
& \EB
}
$$
is commutative. 
}
\end{defn}

\begin{defn} \label{def:dominant} {\rm
A functor $F: \EA \to \EB$ is called {\it dominant} if, for any $B \in \EB$, there are $A\in \EA$ such that $\hom_\EB(B, F(A)) \neq 0$. 
}
\end{defn}

\begin{defn} \label{def:morita-cat}  {\rm
Two fusion categories $\EA$ and $\EB$ are called Morita equivalent if there is an indecomposable semisimple $\EA$-module $\EM$ such that $\EB \simeq \text{Fun}_\EA(\EM, \EM)^{\otimes_\op}$ as fusion categories, where $\text{Fun}_\EA(\EM, \EM)^{\otimes_\op}$ is the category of $\EA$-module functors from $\EM$ to $\EM$ but with the tensor product $\otimes_\op$ defined by $f \otimes_\op g = g \otimes f = g\circ f$. 
}
\end{defn}

It was proved in \cite{eno} that two fusion categories $\EA$ and $\EB$ are Morita equivalent if and only if $Z(\EA) \simeq Z(\EB)$ as braided fusion categories.

\end{appendices}

\newpage

\begin{center} \LARGE
Erratum and Addendum:
``Anyon condensation and tensor categories'' [Nucl. Phys. B 886 (2014) 436-482]
\end{center}

\vspace{0.05cm}
\begin{center}
Liang Kong$^{a,b,c}$
%~\footnote{Emails:{\tt  kongl@sustech.edu.cn, tlan@perimeterinstitute.ca, wen@dao.mit.edu, zhengh@sustech.edu.cn}}
\\[1em]
$^a$ Shenzhen Institute for Quantum Science and Engineering, \\
Southern University of Science and Technology, Shenzhen, 518055, China 
\\[0.4em]
$^b$ International Quantum Academy (SIQA), \\
and Shenzhen Branch, Hefei National Laboratory, Futian District, Shenzhen, China 
\\[0.4em]
$^c$ Guangdong Provincial Key Laboratory of Quantum Science and Engineering, \\
Southern University of Science and Technology, Shenzhen, 518055, China
\end{center}

\bigskip
\begin{abstract}
We correct a mistake in the paper [Nucl. Phys. B 886 (2014) 436-482]. The main result of that paper, i.e. Theorem${}^{\mathrm{bs}}$\,4.7, remains correct. We also take the opportunity to simplify the bootstrap analysis in that paper. 
\end{abstract}

\setcounter{section}{0}
\section{Erratum to [Nucl. Phys. B 886 (2014) 436-482]}

We refer to [Nucl. Phys. B 886 (2014) 436-482] (i.e. \cite{Kon14}) for notations and terminology. We first summarize the set-up of the bootstrap analysis in \cite{Kon14}. We consider an anyon condensation occuring in a 2d (spatial dimension) region in a 2d topological order described by a modular tensor category (MTC) $\EC$. Anyons in the condensed phase form a new MTC $\ED$. We assume that 1d domain wall between $\EC$ and $\ED$ is gapped. Then the particles on the wall form a fusion category $\EE$. This set-up is illustrated in the following picture.
$$
\begin{tikzpicture}
\fill[blue!30] (-2,0) rectangle (2,3) ;
\fill[green!30] (0,1.5) circle (1);
\draw[fill=black] (-0.04,2.45) rectangle (0.05,2.55) node[midway,above] {\footnotesize $X$} ;
\node at (-1.5,2.5) {$\EC$} ;
\node at (0,1.2) {$\ED$} ;
\node at (-1.1,1) {$\EE$} ;
\draw [blue, ultra thick] (0,1.5) circle [radius=1] ;
\draw[densely dashed,-stealth] (0,2) .. controls (0.2,2) and (0.3,2.3) ..(0.07,2.45) ;
\draw[densely dashed,-stealth] (0,2) .. controls (-0.2,2) and (-0.3,2.3) .. (-0.07,2.45) ;
\node at (0.2,1.8) {\scriptsize $A=\one_\ED$} ;
\end{tikzpicture}
$$

The main result Theorem${}^{\mathrm{bs}}$\,4.7 in \cite{Kon14} (recalled in Theorem${}^{\mathrm{bs}}$\,\ref{thm:bs-CED-2}) remains correct. However, in Section 3-4 in \cite{Kon14}, I made a mistake in the bootstrap analysis of the domain wall $\EE$. In particular, in Theorem${}^{\mathrm{bs}}$ 3.3 in \cite{Kon14}, the identity ``$\EE=\EC_{B|B}$'' is not correct. By equation (33) and (34) in \cite{Kon14}, the vacuum $B$ on the wall is an algebra over $A$ \cite{DMNO13}, where $A$ is the vacuum of the $\ED$-phase, i.e. $A=\one_\ED$, viewed as an object in $\EC$. Moreover, by \cite{Kon14}, $A$ is a condensable algebra (or a connected commutative normalized-special Frobenius algebra \cite{FS03}) in $\EC$. The ``normalized-specialness'' of a Frobenius algebra $A$ is defined by the conditions: $m_A\circ \Delta_A=\id_A$ and $\epsilon_A\circ\iota_A=\dim A\,\id_\one$ \cite{FS03}. Since $A$ is simple, the map $\iota_B^A: A \hookrightarrow B$ is an embedding of algebras. A particle $X$ on the wall is naturally equipped with two-side actions of both $A$ and $B$. Moreover, the left $A$-action on $X$ should be the one canonically induced from the right $A$-action as illustrated in above picture. Mathematically, this statement says that the following diagram 
\be \label{diag:AXA-1}
\xymatrix{
A\otimes X \ar[r]  \ar[d]_{c_{X,A}^{-1}} &  X \\
X\otimes A \ar[ru] & 
}
\ee
is commutative. Therefore, $X$ is an $A$-$A$-bimodule with the left $A$-action defined by the right $A$-action according to the Diagram (\ref{diag:AXA-1}). Using this bimodule structure, we can see that the category $\EC_A$ of right $A$-modules in $\EC$ has the structure of a fusion category with the tensor product $\otimes_A$ and the tensor unit $A$. Moreover, the two-side $B$-actions on $X$ should be compatible with the $A$-actions on $X$. For example, the following diagram
$$
\xymatrix{
A\otimes X \ar[r]^{\iota_B^A\otimes 1} \ar[d]_{c_{X,A}^{-1}} & B\otimes X \ar[r]^{} & X \\
X\otimes A \ar[r]^{1\otimes \iota_B^A} & X\otimes B \ar[ur]_{} & 
}
$$
should be commutative, and the $B$-actions on $X$ are necessarily morphisms in $\EC_A$. By the same arguments that led to \cite[Lemma${}^{\mathrm{bs}}$\,3.2]{Kon14}, $B$ is a connected normalized-special Frobenius algebra in $\EC_A$. Therefore, we obtain the following correction of Theorem${}^{\mathrm{bs}}$ 3.3 in \cite{Kon14}. 

\begin{pthm}
$\EE=(\EC_A)_{B|B}$ as fusion categories, where $(\EC_A)_{B|B}$ denotes the category of $B$-$B$-bimodules in $\EC_A$. 
\end{pthm}

\begin{rema} {\rm
Note that $(\EC_A)_{B|B}$ is Morita equivalent to $\EC_A$ as fusion categories, and they share the same Drinfeld center, i.e. $Z((\EC_A)_{B|B})\simeq Z(\EC_A)$. 
}
\end{rema}

Moreover, all the ``$\EC_{B|B}$'' appeared in \cite{Kon14} should be replaced by $(\EC_A)_{B|B}$. In particular, Theorem${}^{\mathrm{bs}}$\,4.2 in \cite{Kon14} should be replaced by the following one. 
\begin{pthm} \label{thm:bs-CED}
If a system of anyons, described by a (unitary) modular tensor category $\ED$, is obtained from a condensation of another system of anyons given by a (unitary) modular tensor category $\EC$, and if there is a gapped domain wall between $\EC$ and $\ED$ and the wall excitations form a (unitary) spherical fusion category $\EE$, then we have the following results. 
\bnu
\item The vacuum in $\ED$ can be identified with a connected commutative normalized-special Frobenius algebra $A$ in $\EC$, and $\ED = \EC_A^{loc}$ as (unitary) modular tensor categories.
\item The vacuum in $\EE$ can be identified with a connected normalized-special Frobenius algebra $B$ in the fusion category $\EC_A$, and $\EE = (\EC_A)_{B|B}$ as (unitary) spherical fusion categories.
%\item there is an algebraic homomorphism $\iota_B^A: A \hookrightarrow B$ in $\EC$ such that $B$ is an algebra over $A$;
\item The bulk-to-wall map from the $\EC$-side is given by the monoidal functor 
\be  \label{eq:L-tensor-B-2}
-\otimes B: \EC \to (\EC_A)_{B|B}, \quad\mbox{defined by}\quad C\mapsto C\otimes B, \quad\quad \forall C\in \EC. 
\ee 
\item The bulk-to-wall map from the $\ED$-side is given by the monoidal functor
\be
B\otimes_A -: \EC_A^{loc}\to (\EC_A)_{B|B}, \quad\mbox{defined by}\quad M\mapsto B\otimes_A M, \quad \forall M \in \EC_A^{loc}.
\ee

\enu
\end{pthm}

I apologize for any inconvenience caused by this mistake. Again, when $B=A$, we obtain the main result Theorem${}^{\mathrm{bs}}$\,4.7 in \cite{Kon14}, which remains correct.

\section{Addendum to [Nucl. Phys. B 886 (2014) 436-482]}

It turns out that the bootstrap analysis of the gapped domain wall $\EE$ is entirely unnecessary. It was known to physicists that when a condensation occurs in a 2d (spatical dimension) region, called a {\it 2d condensation}, the condensed 2d $\ED$-phases consists of all the deconfined particles. The work \cite{Kon14} actually provided the mathematical definition of a deconfined particle as a local $A$-module in $\EC$. At the same time, a confined particle can be defined by a ``non-local'' right $A$-module in $\EC$. It was also intuitively clear to physicists that all the confined particles naturally accumulate on the domain wall and are confined to the wall. Moreover, deconfined particles in $\ED$ can move onto the wall and can then move back to $\ED$ freely. Therefore, after the bootstrap analysis of the deconfined particles in $\ED=\EC_A^{loc}$ (see Section 2 in \cite{Kon14}), we can immediately conclude that $\EE = \EC_A$ as the fusion category consisting of all confined particles and deconfined particles. This immediately leads to the main theorem in \cite[Theorem${}^{\mathrm{bs}}$\,4.7]{Kon14} (remains correct), which is rephrased below. 
\begin{pthm} \label{thm:bs-CED-2}
If a system of anyons, described by a (unitary) modular tensor category $\ED$, is obtained from a 2d condensation of another system of anyons given by a (unitary) modular tensor category $\EC$, and if this 2d condensation also produces a gapped domain wall between $\EC$ and $\ED$ consisting of all confined and deconfined particles, then we have the following results. 
\bnu
\item The vacuum in $\ED$ can be identified with a connected commutative normalized-special Frobenius algebra $A$ in $\EC$. The $\ED$-phase consists of all deconfined particles, which form the (unitary) modular tensor category $\EC_A^{loc}$, i.e. $\ED = \EC_A^{loc}$.
\item The confined and deconfined particles naturally accumulate on the domain wall and form the (unitary) spherical fusion categories $\EC_A$.
%\item there is an algebraic homomorphism $\iota_B^A: A \hookrightarrow B$ in $\EC$ such that $B$ is an algebra over $A$;
\item The bulk-to-wall map from the $\EC$-side is given by the monoidal functor 
\be  \label{eq:L-tensor-A-2}
-\otimes A: \EC \to \EC_A, \quad\mbox{defined by}\quad C\mapsto C\otimes A, \quad\quad \forall C\in \EC. 
\ee 
\item The bulk-to-wall map from the $\ED$-side is given by the canonical embedding $\EC_A^{loc}\hookrightarrow \EC_A$, and deconfined particles in $\EC_A$ can move out of the wall to the $\ED$-side freely. 

\enu
\end{pthm}

\begin{rema} {\rm
By \cite{DMNO13}, we have $Z(\EC_A) =\EC\boxtimes \overline{\EC_A^{loc}}$, where $Z(\EC_A)$ denotes the Drinfeld center of $\EC_A$. When $\EC_A^{loc}$ is trivial, i.e. $A$ is Lagrangian, we reobtain the boundary-bulk relation: $\EC=Z(\EE)$ for a gapped boundary $\EE=\EC_A$. The boundary-bulk relation, i.e. ``the bulk is the center of a boundary'' was first discovered in \cite{KK12} based on the Levin-Wen type of lattice models with gapped boundaries. It was noticed model-independently in \cite{FSV13} that the bulk MTC necessarily factors through the Drinfeld center of a boundary fusion category. If we agree that a gapped boundary of a 2d topological order is necessarily the result of a 2d anyon condensation to the trivial phase, then Theorem${}^{\mathrm{bs}}$\,4.7 in \cite{Kon14} provided the first complete model-independent proof of the boundary-bulk relation for 2d topological orders with gapped boundaries. A more general boundary-bulk relation for both gapped/gapless boundaries and for potentially gapless quantum liquid bulk phases in all dimensions was later proposed and proved in \cite{KWZ15,KWZ17}. 
}
\end{rema}

Condensation can also occur in 1d. For example, given a gapped 1d domain wall $\EE$ between two bulk phases $\EC$ and $\ED$, the bootstrap analysis in Section 3 in \cite{Kon14} actually shows that the 1d wall $\EE$ can be condensed to a new 1d gapped phase by condensing a connected normalized-special Frobenius algebra $B$ in $\EE$. The new 1d phase can be described by the fusion category $\EE_{B|B}$, i.e. the category of $B$-$B$-bimodules in $\EE$, with the vacuum given by $B$ and the tensor product given by $\otimes_B$. The new 1d $\EE_{B|B}$-phase share the same bulk as $\EE$, and $\EE_{B|B}$ are Morita equivalent to $\EE$. Such a condensation is called a {\it 1d condensation}.

Using above terminology, it is clear that Theorem${}^{\mathrm{bs}}$\,\ref{thm:bs-CED} can be viewed as a result of two condensations, in which we first do a 2d condensation described by Theorem${}^{\mathrm{bs}}$\,\ref{thm:bs-CED-2}, then do a 1d condensation defined by a connected normalized-special Frobenius algebra $B$ in $\EC_A$. 

When $A=\one_\EC$, $B$ is a connected normalized-special Frobenius algebra in $\EC$. In this case, condensing $B$ produces a defect line in the 2d bulk $\EC$-phase. The particles on the line form the fusion category $\EC_{B|B}$, which is Morita equivalent to $\EC$.

\bigskip
\noindent {\bf Acknowledgement}: I would like to thank Zhi-Hao Zhang for discovering this mistake and for helping me to draw in LaTex. LK is supported by Guangdong Provincial Key Laboratory (Grant No.2019B121203002), by Guangdong Basic and Applied Basic Research Foundation under Grant No. 2020B1515120100 and by NSFC under Grant No. 11971219.


\small
\begin{thebibliography}{XXX}

\bibitem[BJQ13]{bjq}
M.~Barkeshli, C.-M.~Jian, X.-L.~Qi, 
Theory of defects in Abelian topological states, Phys. Rev. B 88, (2013) 235103 %\arXiv{1305.7203}

\bibitem[Bez04]{Bez04}
R. Bezrukavnikov, 
On tensor categories attached to cells in affine Weyl groups, Representation theory of algebraic groups and quantum groups, 69-90, Adv. Stud. Pure Math., 40, Math. Soc. Japan, Tokyo, 2004.

\bibitem[BCKA13]{bcka}
O.~Buerschaper, M.~Christandl, L.~Kong, M.~Aguado,
Electric-magnetic duality of lattice systems with topological order, Nucl. Phys. B. 876 [FS] (2013) 619-636 %\arXiv{1006.5823}

\bibitem[BE98]{be}
J.~B\"{o}ckenhauer, D.~E.~Evans,
Modular invariants, graphs and $\alpha$-induction for nets of subfactors, I. 
Commun. Math. Phys. 197 (1998) 361-386.


\bibitem[BEK99]{bek1}
J.~B\"{o}ckenhauer, D.~E.~Evans and Y.~Kawahigashi, 
On $\alpha$-induction, chiral projectors and modular invariants for subfactors, 
Commun. Math. Phys. 208 (1999) 429-487.

\bibitem[BEK00]{bek2}
J.~B\"{o}ckenhauer, D.~E.~Evans and Y.~Kawahigashi, 
Chiral structure of modular invariants for subfactors. Commun. Math. Phys. 210 (2000), no. 3, 733-784.

\bibitem[BEK01]{bek3}
J.~B\"{o}ckenhauer, D.~E.~Evans and Y.~Kawahigashi, 
Longo-Rehren subfactors arising from $\alpha$-induction. Publ. Res. Inst. Math. Sci. 37 (2001), no. 1, 1-35.

\bibitem[BaK01]{bakalov-kirillov}
B.~Bakalov, A.~Kirillov, Jr., 
Lectures on Tensor Categories and Modular Functors. University
Lecture Series, Vol. 21, Providence, RI: Amer. Math. Soc., 2001

\bibitem[BrK98]{bk}
S.B.~Bravyi, A.Y.~Kitaev,
Quantum codes on a lattice with boundary, [quant-ph/9811052]

\bibitem[BR12]{br}
F. A.~Bais, J. C.~Romers,
The modular S-matrix as order parameter for topological phase transitions,
New J. Phys. 14, (2012) 035024 [arXiv:1108.0683]

\bibitem[BS09]{bs}
F.A.~Bais, J.K.~Slingerland, 
Condensate induced transitions between topologically ordered phases,
Phys. Rev. B 79, 045316 (2009). %[arXiv:0808.0627]

\bibitem[BSH09]{bsh}
F.A. Bais, J.K. Slingerland, S.M. Haaker,
A theory of topological edges and domain walls,
Phys. Rev. Lett. 102 (2009) 220403, \arXiv{0812.4596}

\bibitem[BSS02]{bss1}
F.A.~Bais, B.J.~Schroers, J.K.~Slingerland,
Broken quantum symmetry and confinement phases in planar physics,
Phys. Rev. Lett. 89 (2002) 181601 %\arXiv{hep-th/0205117}

\bibitem[BSS03]{bss2}
F.A.~Bais, B.J.~Schroers, J.K.~Slingerland,
Hopf symmetry breaking and confinement in (2+1)-dimensional gauge theory,
JHEP 0305 (2003) 068 %\arXiv{hep-th/0205114}

\bibitem[BuSS11]{buss}
F. J. Burnell, Steven H. Simon, J. K. Slingerland
Condensation of achiral simple currents in topological lattice models: a Hamiltonian study of topological symmetry breaking, Phys. Rev. B 84 (2011) 125434, %\arXiv{1104.1701} 


\bibitem[BSW11]{bsw}
S.~Beigi, P.~Shor, D.~Whalen,
The quantum double model with boundary: condensations and symmetries,
Commun. Math. Phys. {\bf 306} (3) (2011) 663-694.


\bibitem[BW10]{bw}
M.~Barkeshli, X.-G. Wen,
Anyon Condensation and Continuous Topological Phase Transitions in Non-Abelian Fractional Quantum Hall States, Phys. Rev. Lett.105:216804 (2010) 

\bibitem[BMD08]{bm-d}
H. Bombin, M.A. Martin-Delgado,
A Family of Non-Abelian Kitaev Models on a Lattice: Topological Confinement and Condensation,
Phys. Rev. B 78 (2008 ) 115421

\bibitem[CE08]{ce}
D.~Calaque, P.~Etingof, 
Lectures on tensor categories, 
IRMA Lectures in Mathematics and Theoretical Physics 12 (2008) 1-38, 
%[arXiv:math/0401246]

\bibitem[Dav10a]{Da}
A.~Davydov,
Center of an algebra, 
Adv. Math. {\bf 225} (2010) 319-348 %\arXiv{math.CT/0908.1250}

\bibitem[Dav10b]{da2}
A.~Davydov, 
Modular invariants for group-theoretic modular data I, 
Journal of Algebra {\bf 323} (2010), 1321-1348. [arXiv:0908.1044]

\bibitem[Dav11]{da3}
A.~Davydov,
Anyon condensation and commutative algebras in modular categories, 
Sydney Quantum Information Theory Workshop, 17-20 January 2011

\bibitem[DB12]{db}
A. ~Davydov, T. Booker, 
Commutative algebras in Fibonacci categories,
 J. Algebra {\bf 355} (2012), 176-204. 
 

\bibitem[DMNO10]{dmno}
A. Davydov, M. M\"{u}ger, D.~Nikshych, V.~Ostrik, 
The Witt group of non-degenerate braided fusion categories, [arXiv:1009.2117]

\bibitem[DNO13]{dno}
A. Davydov, D.~Nikshych, V.~Ostrik, 
On the structure of the Witt group of braided fusion categories,
 J. Pure Appl. Algebra {\bf 217} (2013), no. 3, 567-582. %[arXiv:1109.5558]

\bibitem[DGNO10]{drgno}
V.~G.~Drinfeld, S.~Gelaki, D.~Nikshych, V.~Ostrik,
On braided fusion categoriesI,
Selecta Mathematica {\bf 16} (2010) 1-119 %[arXiv:0906.0620]


\bibitem[Eli10]{eliens}
S. Eli\"{e}ns, 
Anyon condensation, University of Amsterdam MSc. Thesis (2010) available at \url{http://www.science.uva.nl/onderwijs/thesis/centraal/files/f1685838870.pdf}.

\bibitem[ERB14]{erb}
I.S.~Eli\"{e}ns, J.C.~Romers, F.A.~Bais,
Diagrammatics for Bose condensation in anyon theories, Phys. Rev. B 90, (2014) 195130 %[arXiv:1310.6001]


\bibitem[EO04]{eo}
P.~Etingof, V.~Ostrik,
Finite tensor categories, 
Moscow Math. Journal {\bf 4} (2004) 627-654. 

\bibitem[ENO05]{eno02}
P.~Etingof, D.~Nikshych, V.~Ostrik,
On fusion categories, 
Ann. Math. 162, (2005) 581-642.

\bibitem[ENO11]{eno}
P.~Etingof, D.~Nikshych, V.~Ostrik,
Weakly group-theoretical and solvable fusion categories, 
Adv. Math. {\bf 226} (2011), 176-205. 

\bibitem[F+07]{ftltkwf}
A.~Feiguin, S.~Trebst, A.~W.~W.~Ludwig, M.~Troyer, A.~Kitaev, Z.~Wang, M.~H.~Freedman,
Interacting anyons in topological quantum liquids: The golden chain,
Phys. Rev. Lett. 98, 160409 (2007). %\arXiv{cond-mat/0612341}

\bibitem[FjFRS08]{unique}
J.~Fjelstad, J.~Fuchs, I.~Runkel, C.~Schweigert,
Uniqueness of open/closed rational CFT with given algebra of open states,
Adv. Theor. Math. Phys. {\bf12} (2008) 1283-1375, %[arXiv:hep-th/0612306]

\bibitem[FRS02]{tft1}
J.~Fuchs, I.~Runkel and C.~Schweigert,
TFT construction of RCFT correlators. I: Partition functions,
Nucl.\ Phys.\ B {\bf 646} (2002) 353--497 %\arXiv{hep-th/0204148}.
%%CITATION = HEP-TH 0204148;%%

\bibitem[FRS04]{tft2}
J.~Fuchs, I.~Runkel and C.~Schweigert,
TFT construction of RCFT correlators II: unoriented world sheets,
Nucl.Phys. B {\bf 678} (2004) 511-637 

\bibitem[FrFRS06]{cor}
J.~Fr\"{o}hlich, J. Fuchs, I.~Runkel, C.~Schweiget, 
Correspondences of ribbon categories,
Adv. Math. {\bf 199} (2006), no. 1, 192-329. 

\bibitem[FrFRS07]{defect-tft}
J.~Fr\"{o}hlich, J. Fuchs, I.~Runkel, C.~Schweiget, 
Nucl.Phys. B {\bf 763} (2007) 354-430. 


\bibitem[FSV13]{fsv}
J.~Fuchs, C.~Schweigert, A.~Valentino
Bicategories for boundary conditions and for surface defects in 3-d TFT, 
Commun. Math. Phys. 321, (2013) 543-575 
%[arXiv:1203.4568]


\bibitem[FSV14]{fsv2}
J.~Fuchs, C.~Schweigert, A.~Valentino, 
A geometric approach to boundaries and surface defects in Dijkgraaf-Witten theories, 
Commun. Math. Phys. 332, (2014) 981-1015
%[arXiv:1307.3632]


\bibitem[GNN09]{gnn}
S.~Gelaki, D.~Naidu, D.~Nikshych,
Centers of graded fusion categories,
Algebra Number Theory 3 (2009), no. 8, 959-990.


\bibitem[Hay99]{hay}
T.~Hayashi, 
A canonical Tannaka duality for finite semisimple tensor categories, \arXiv:{math.QA/9904073}


\bibitem[Hua08]{huang}
Y.-Z.~Huang, 
Ridity and modularity of vertex operator algebras, 
Commun. Contemp. Math. {\bf 10} (2008), suppl. 1, 871-911. 

\bibitem[HKL15]{hkl}
Y.-Z.~Huang, A.~Kirillov, J.~Lepowsky, 
{\it Braided Tensor Categories and Extensions of Vertex Operator Algebras}, 
Commun. Math. Phys. 337 (2015) 1143-1159.


\bibitem[HK04]{osvoa}
Y.-Z.~Huang, L~Kong, 
Open-string vertex algebras, tensor categories and operads, 
Commun. Math. Phys. {\bf 250} (2004) 433-471 %[math.QA/0308248]


\bibitem[HW14]{hw}
L.-Y.~Hung, Y.-D.~Wan, 
Symmetry Enriched Phases via Pseudo Anyon Condensation, International Journal of Modern Physics B, Vol. 28, No. 24, 1450172 (2014)

\bibitem[KS11a]{kas}
A. Kapustin, N. Saulina, 
Topological boundary conditions in abelian Chern-Simons theory, 
Nucl. Phys. B 845 (2011) 393-435 %[math.QA/1008.0654]

\bibitem[KS11b]{kas2}
A. Kapustin, N. Saulina, 
Surface operators in 3d Topological Field Theory and 2d Rational Conformal Field Theory,
Mathematical Foundations of Quantum Field and Perturbative String Theory,  Hisham Sati, Urs Schreiber, 
Proceedings in Symposia in Pure Mathematics, AMS, (2011) \arXiv{1012.0911}


\bibitem[Kap13]{kapustin}
A.~Kapustin, 
Ground-state degeneracy for abelian anyons in the presence of gapped boundaries,
[arXiv:1306.4254]

\bibitem[Kas95]{kassel}
C.~Kassel, 
Quantum Groups, Graduate Texts in Mathematics, 155. Springer-Verlag, New York, 1995.


\bibitem[Ki03]{kitaev}
A.Y.~Kitaev,
Fault-tolerant quantum computation by anyons, 
Ann. Phys. 303, (2003) 2-30 

\bibitem[Ki06]{kitaev1}
A.Y.~Kitaev,
Anyons in an exactly solved model and beyond,
Annals of Physics {\bf 321} (2006) 2-111 

\bibitem[Ki08]{kitaev2}
A.Y.~Kitaev,
unpublished works and private communication in 2008 and 2009. 

\bibitem[KK12]{kk}
A.Y.~Kitaev, L.~Kong, 
Models for gapped boundaries and domain walls, 
Commun. Math. Physics. {\bf 313} (2012) 351-373 %[arXiv:1104.5047]

\bibitem[Kon08]{kong}
L.~Kong, 
Cardy condition for open-closed field algebras, 
Commun. Math. Phys. {\bf 283} (2008) 25-92. %[arXiv:math/0612255]

\bibitem[KO02]{ko}
A.~Kirillov Jr., V.~Ostrik,
On q-analog of Mckay correspondence and ADE classification of $\hat{sl}_2$ conformal field theories, 
Adv. Math. {\bf 171} (2002), no. 2, 183-227. 


\bibitem[KR08]{morita}
L.~Kong, I.~Runkel, 
Morita classes of algebras in modular tensor categories,
Adv. Math. {\bf 219} (2008) 1548-1576. 

\bibitem[KR09]{cardy}
L.~Kong, I.~Runkel, 
Cardy algebras and sewing constraints, I, 
Commun. Math. Phys. {\bf 292} (2009) 871-912 % \arXiv{0807.3356}



\bibitem[Lev13]{levin}
M.~Levin, 
Protected edge modes without symmetry, Phys. Rev. X 3 (2013) 021009 %[arXiv:1301.7355]

\bibitem[LL04]{voa}
J.~Lepowsky, H.-S.~Li,
Introduction to vertex operator algebras and their representations, 
Progress in Mathematics, {\bf 227}, Boston, MA: Birkh\"{a}user Boston, Inc. 2004.

\bibitem[LW05]{lw-mod}
M.~A. Levin, X.-G. Wen,
String-net condensation: A physical mechanism for topological
phases. Phys. Rev. B {\bf 71}, 045110 (2005)

\bibitem[Mue03a]{mueger1}
M.~M\"{u}ger,
From subfactors to categories and topology. I. Frobenius algebras in and Morita equivalence of tensor categories. J. Pure Appl. Algebra 180 (1-2), (2003) 81-157.

\bibitem[Mue03b]{mueger}
M.~M\"{u}ger,
From subfactors to categories and topology. II. The quantum double of tensor categories and subfacts, 
J. Pure Appl. Algebra {\bf 180} (2003), no. 1-2, 159-219. 

\bibitem[Mue08]{mueger2}
M.~M\"{u}ger,
Tensor categories: A selective guided tour, \arXiv{0804.3587}

\bibitem[Ost03a]{ostrik}
V.~Ostrik, 
Module categories, weak Hopf algebras and modular invariants, 
Transform. Group. {\bf 8} (2003) 177-206.

\bibitem[Ost03b]{ostrik2}
V.~Ostrik,
Module categories over the Drinfeld double of a finite group, 
Int. Math. Res. Not. 2003, no. 27, 1507-1520. 

\bibitem[Sch01]{sc}
P.~Schauenburg, 
The monoidal center construction and bimodules, 
J. Pure Appl. Alg. 158 (2001) 325-346. 


\bibitem[Tur94]{turaev}
V.~G.~Turaev, 
Quantum invariant of knots and 3-manifolds, de Gruyter Studies in Mathematics, Vol.
18, Berlin: Walter de Gruyter, 1994.

\bibitem[Wan10]{wang}
Z.~H.~Wang,
Topological Quantum Computation, CBMS Regional Conference Series in Mathematics, vol. 112
American Mathematical Society (2010).

\bibitem[WW]{ww}
J.~Wang, X.-G.~Wen,
Boundary Degeneracy of Topological Order, Phys. Rev. B 91 (2015) 125124 %[arXiv:1212.4863]

\bibitem[Wen13]{wen-private}
X.-G.~Wen,
private communication, 2013.


\end{thebibliography}

\small
\begin{thebibliography}{XXX}

\bibitem[DMNO13]{DMNO13}
A. Davydov, M. M\"{u}ger, D. Nikshych, V. Ostrik, 
{\it The Witt group of non-degenerate braided fusion categories}, 
J. Reine Angew. Math. 677, 135-177 (2013). \arXiv{1009.2117}

\bibitem[FS03]{FS03}
J. Fuchs, C. Schweigert,
{\it Category theory for conformal boundary conditions},
 Fields Institute Communications 39 (2003) 25-71. \arXiv{math/0106050}

\bibitem[FSV13]{FSV13}
J.~Fuchs, C.~Schweigert, A.~Valentino,
{\it Bicategories for boundary conditions and for surface defects in 3-d TFT}, 
Commun. Math. Phys. 321, (2013) 543-575 \arXiv{1203.4568}


\bibitem[KK12]{KK12}
A.~Kitaev, L.~Kong,
{\it Models for Gapped Boundaries and Domain Walls}, Commun. Math. Phys. 313, 351-373 (2012)
\arXiv{1104.5047}

\bibitem[Kon14]{Kon14}
L.~Kong, 
{\it Anyon condensation and tensor categories}, Nucl. Phys. B 886 (2014) 436-482 % \arXiv{1307.8244}. 

\bibitem[KWZ15]{KWZ15}
L. Kong, X.-G. Wen, H. Zheng, 
{\it Boundary-bulk relation for topological orders as the functor mapping higher categories to their centers}, \arXiv{1502.01690}

\bibitem[KWZ17]{KWZ17}
L. Kong, X.-G. Wen, H. Zheng, 
{\it Boundary-bulk relation in topological orders}, Nucl. Phys. B 922 (2017) 62 
\arXiv{1702.00673}

\end{thebibliography}
\end{document}